\documentclass[nofootinbib,pra,twocolumn,showpacs,superscriptaddress,notitlepage,amsmath,amssymb]{revtex4-2}

\usepackage[colorlinks=true,citecolor=blue,linkcolor=blue,urlcolor=blue]{hyperref}

\usepackage{amsmath,amssymb,bm}

\usepackage{tikz}
\usetikzlibrary{external}
\usetikzlibrary{decorations.pathreplacing}
\usetikzlibrary{decorations.text}
\usetikzlibrary{decorations.pathmorphing}
\usetikzlibrary{shapes}

\definecolor{newblue}{RGB}{112,178,255}
\definecolor{neworange}{RGB}{255,204,112}
\definecolor{blue2}{RGB}{120,0,255}
\definecolor{red2}{RGB}{255,0,120}
\definecolor{green2}{RGB}{0,130,130}

\tikzset{snake it/.style={decorate, decoration={snake,segment length=1mm, amplitude=0.5mm}}}

\definecolor{darkred}{RGB}{245,186,183}
\definecolor{lightred}{RGB}{249,217,215}

\usetikzlibrary{arrows,calc,external,shapes.geometric}

\tikzset{
    >=stealth',
    help lines/.style={dashed, thick},
    important line/.style={thick},
    connection/.style={thick, dotted},
}

\tikzstyle{A}=[circle,draw=red!50,fill=red!20,thick]
\tikzstyle{R}=[circle,draw=blue!50,fill=blue!20,thick]
\tikzstyle{U}=[circle,draw=green!50,fill=green!20,thick]
\tikzstyle{V}=[circle,draw=orange!50,fill=orange!20,thick]

\def\bra#1{\mathinner{\langle{#1}|}}
\def\ket#1{\mathinner{|{#1}\rangle}}

\def\inner#1{\mathinner{\langle{#1}\rangle}}
\def\outer#1#2{\mathinner{|{#1}\rangle\langle {#2}|}}
\def\ZZ{\mathbb Z}

\usepackage{framed}
\definecolor{shadecolor}{gray}{0.95}

\begin{document}

\title{Long-Range Entanglement from Measuring Symmetry-Protected Topological Phases}

\author{Nathanan Tantivasadakarn}
\affiliation{Department of Physics, Harvard University, Cambridge, MA 02138, USA}
\author{Ryan Thorngren}
\affiliation{Department of Physics, Harvard University, Cambridge, MA 02138, USA}
\affiliation{Center for Mathematical Sciences and Applications,
Harvard University, Cambridge, MA 02138, USA}
\affiliation{Department of Physics, Massachusetts Institute of Technology, Cambridge, MA 02139, USA}
\author{Ashvin Vishwanath}
\affiliation{Department of Physics, Harvard University, Cambridge, MA 02138, USA}
\author{Ruben Verresen}
\affiliation{Department of Physics, Harvard University, Cambridge, MA 02138, USA}

\begin{abstract}
A fundamental distinction between many-body quantum states are those with short- and long-range entanglement (SRE and LRE). The latter cannot be created by finite-depth circuits, underscoring the nonlocal nature of Schr\"odinger cat states, topological order, and quantum criticality.
Remarkably, examples are known where LRE is obtained by performing single-site measurements on SRE, such as the toric code from measuring a sublattice of a 2D cluster state. However, a systematic understanding of when and how measurements of SRE give rise to LRE is still lacking. Here, we establish that LRE appears upon performing measurements on symmetry-protected topological (SPT) phases---of which the cluster state is one example.
For instance, we show how to implement the Kramers-Wannier transformation by adding a cluster SPT to an input state followed by measurement. This transformation naturally relates states with SRE and LRE. An application is the realization of double-semion order when the input state is the $\mathbb Z_2$ Levin-Gu SPT. Similarly, the addition of fermionic SPTs and measurement leads to an implementation of the Jordan-Wigner transformation of a general state. More generally, we argue that a large class of SPT phases protected by $G \times H$ symmetry gives rise to anomalous LRE upon measuring $G$-charges, and we prove that this persists for generic points in the SPT phase under certain conditions. Our work introduces a new practical tool for using SPT phases as resources for creating LRE, and uncovers the classification result that all states related by sequentially gauging Abelian groups or by Jordan-Wigner transformation are in the same equivalence class, once we augment finite-depth circuits with single-site measurements. In particular, any topological or fracton order with a solvable finite gauge group can be obtained from a product state in this way.
\end{abstract}

\date{\today}

\maketitle

\tableofcontents

\section{Introduction}
Although quantum mechanics exhibits a dichotomy between unitary time evolution and measurement, many-body quantum theory traditionally focuses on unitary aspects. Indeed, the classification of quantum phases of matter at zero temperature takes as its very definition that two states are in the same phase if and only if they can be connected by a unitary time-evolution in a finite time~\cite{BravyiHastingsVerstraete06,Hastings2010,ChenGuWen11A,ChenGuWen11B,ZengWen15,HuangChen2015,Haah2016}. Any state in the same phase as a product state is said to exhibit short-range entanglement (SRE), whereas the other classes have long-range entanglement (LRE)\footnote{Note, this definition of LRE includes some invertible phases like the Kitaev Majorana chain since it cannot be connected to a product state by a finite-depth local unitary circuit.}. Even restricting to gapped phases, the latter contains interesting cases such as intrinsic topological \cite{Read91,Wen_90,Wenbook,Fuchs02,Kitaev_2003,kitaev_anyons_2005} and fracton order \cite{Chamon2005,Haah2011,Yoshida2013,VijayHaahFu2015,VijayHaahFu2016,nandkishore2019fractons,Pretko2020}. States with SRE can also be subdivided into distinct phases of matter if one imposes symmetry constraints on the aforementioned unitaries, giving rise to the notion of symmetry-protected topological (SPT) phases\footnote{Symmetry-broken states can also be regarded as SRE. However, for the purposes of the present work, we will consider their symmetry-preserving cat states, which exhibit LRE.} \cite{Gu09,Pollmann10,Fidkowski_2011,Turner11,Schuch11,ChenGuWen11A,ChenGuWen11B,Chen_2011,Chen_2013,pollmann_symmetry_2012,LuVishwanath12,SenthilLevin13,Levin_2012,VishwanathSenthil2013,Else_2014}.

Recently, there has been growing interest in explicitly incorporating measurements into the study of many-body quantum states. For instance, a multitude of works have studied entanglement reduction from measurements, giving rise to surprising new structures \cite{Li18,Skinner19,Chan19,Li19,Vasseur19,Cao19,Gullans20,Choi20,Tang20,Jian20,Lopez-Piqueres20,Bao20,Rossini20,Piroli20,Fan21,LiChenLudwigFisher21,BenZion20}. However, there are also examples where measurements \emph{increase} the entanglement. For example, it is known that performing single-site measurements on a subset of sites of a cluster state (with SRE) can produce a Greenberger-Horne-Zeilinger (GHZ) cat state \cite{Briegel01}, the toric code \cite{Raussendorf05,Aguado08,Piroli21}, and certain fracton codes via a layered construction \cite{Schmitz19,Williamson21}. In fact, it has been remarked that all states realized by CSS stabilizer codes \cite{Calderbank96,Steane96} (i.e., stabilizers that are of the form $\prod_{\bm i \in \mathcal S} Z_{\bm i}$ or $\prod_{\bm i \in \mathcal S} X_{\bm i}$) can be obtained by measuring an appropriate cluster state \cite{Bolt16}. 

The existence of these examples begs the following question: What is the general framework for when, how, and why one can create LRE from SRE states and single-site measurements? In this work, we argue that the essential fact in the above examples is that the cluster state is an SPT. This deeper understanding confers at least four advantages. First, in contrast to earlier studies, we argue that LRE states are obtained on measuring not just the fixed-point wave function of the SPT but any state within the same phase. Second, the origin of LRE under measurement is tied to a specific anomaly involving the symmetries---related to the anomaly living at the boundary of the original SPT phase---thereby constraining the nature of the resulting LRE. Third, it allows for the preparation of states that are not realized by stabilizer codes, such as topological order described by twisted gauge theories or non-Abelian fracton orders \cite{HW13,SongPremHuangMartinDelgado19,BulmashBarkeshli2019,PremWilliamson2019,AasenBulmashPremSlagleWilliamson20,StephenGarre-RubioDuaWilliamson2020,Wen20,Wang20,WilliamsonCheng20,TJV2}. Fourth, we achieve a new perspective on Kramers-Wannier (KW) \cite{Wegner1971,Kogut1979,CobaneraOrtizNussinov2011,Aasen16,VijayHaahFu2016,Williamson2016,KubicaYoshida2018,Pretko2018,ShirleySlagleChen2019,Radicevic2019} and Jordan-Wigner (JW) \cite{Jordan1928,SMLising,ChenKapustinRadicevic2018,ChenKapustin2019,Chen2019,Tantivasadakarn20,Shirley20,Po21,LiPo21} transformations. Indeed, we show how these nonlocal transformations can be efficiently implemented in a finite time by adding SPT entanglers to arbitrary initial states\footnote{This process is equivalent to stacking SPT states on top of the initial state.} and subsequently performing single-site measurements. In a companion work \cite{Rydberg}, we explain how this general understanding can be utilized to prepare, e.g., $\mathbb Z_3$, $S_3$, and $D_4$ topological order in quantum devices such as Rydberg atom arrays.

This work is structured as follows. In Sec.~\ref{sec:motivation}, we set the stage by reviewing some known examples, explaining how the 1D GHZ and 2D toric code states can be obtained by measuring particular cluster states. In Sec.~\ref{sec:KW}, we generalize these cases by reinterpreting the act of measuring cluster states as effectively implementing a KW transformation. To give illustrative examples, we explain how this allows one to transform the nontrivial $\mathbb Z_2$ SPT in 2D to the double-semion topological order, and to transform the 1D $XY$ chain into two decoupled critical Ising models by using finite-depth circuits and single-site measurements. Moreover, we discuss how certain types of non-Abelian topological order can be obtained by sequential applications of this scheme. Sec.~\ref{sec:JW} generalizes this to the fermionic case, where a similar procedure implements the JW transformation, illustrated by creating the Kitaev chain from a trivial spin chain. Sec.~\ref{sec:generalization} broadens our scope further: First, we argue that this procedure is a robust property of the SPT phase (which we exemplify by obtaining cat states via measuring the spin-$1$ Heisenberg chain), and second we argue that anomalous symmetries and LRE are generically obtained by measuring a broad class of SPT states (which we discuss in detail for the $\mathbb Z_2^3$ SPT in 2D). We conclude with directions for future research in Sec.~\ref{sec:outlook}.

\begin{figure*}[t]
    \centering
 \includegraphics[scale=0.7]{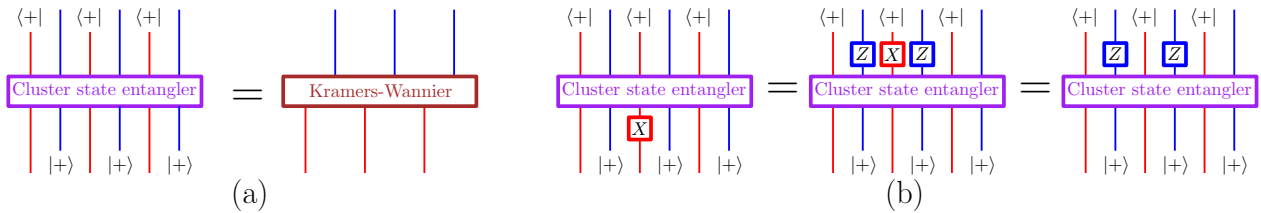}
    \caption{\textbf{From the cluster state entangler to the Kramers-Wannier transformation.} (a) Relation between the cluster state entangler and the Kramers-Wannier duality in arbitrary dimensions, with $A$ legs drawn in red and $B$ legs drawn in blue. Here the entangler is simply a product of controlled-$Z$ on nearest-neighbor sites. (b) Proof of this equality at the level of operators where $X$ on the red sites is interchanged with $ZZ$ on the blue sites.}
    \label{fig:SPTtoKWproof}
\end{figure*}

\section{Motivating examples \label{sec:motivation}}
We begin by reviewing how measuring cluster states in 1D and 2D can produce GHZ states \cite{Briegel01} and the toric code \cite{Raussendorf05}, respectively. Consider a 1D chain with $2N$ qubits. The cluster state $|\psi\rangle$ on this chain is the unique state that satisfies $Z_{n-1} X_{n} Z_{n+1} \ket{\psi} = \ket{\psi}$ for all $n$, where $X,Y,Z$ denote the Pauli matrices. It can be prepared from the product state in the $X$ basis by applying controlled-$Z$ gates on all nearest neighboring qubits:
\begin{align}
    \ket{\psi} = \prod_{n} CZ_{n,n+1} \ket{+}^{\otimes 2N} =: U_{CZ} \ket{+}^{\otimes 2N}.
    \label{eq:clusterstateentangler}
\end{align}
We call the above unitary $U_{CZ}$ the cluster state entangler. Now suppose we measure $X$ on all odd sites, with outcomes $X_{2n+1}=(-1)^{s_{2n+1}}$. Since $Z_{2n-2} X_{2n-1} Z_{2n}$ commutes with the measurement, the state after the measurement $\ket{\psi_\textrm{out}}$ satisfies $Z_{2n-2} Z_{2n} \ket{\psi_\textrm{out}} = (-1)^{s_{2n-1}}\ket{\psi_\textrm{out}}$. On the other hand, the even stabilizers do not commute with the measurement; only their product $\prod_n  Z_{2n-1}X_{2n}Z_{2n+1}  = \prod_n X_{2n} $ commutes, implying $|\psi_\textrm{out}\rangle$ is $\mathbb Z_2$-symmetric. If all the $s_m = 0$, then $\ket{\psi_\textrm{out}}$ is the GHZ state on the even qubits:
\begin{equation}
|\textrm{GHZ}\rangle = \frac{1}{\sqrt{2}} \left( | \uparrow \uparrow \cdots \uparrow \rangle + | \downarrow \downarrow \cdots \downarrow \rangle \right).
\end{equation}
Otherwise, it is the GHZ state  up to single-site spin flips conditioned on the measurement outcomes: $\ket{\textrm{GHZ}} = \prod_{n=1}^N X_{2n}^{\sum_{m=1}^n s_{2m-1}} \ket{\psi_\textrm{out}} $. Thus, regardless of the outcome, $\ket{\psi_\textrm{out}}$ has long-range entanglement, as can, for example, be quantified by quantum Fisher information \cite{Fisher1925,Giovannetti06} (see also Sec.~\ref{subsec:spin1}).

In 2D, we can consider a cluster state on the vertices and edges of the square lattice \cite{Raussendorf05}. The stabilizers of the cluster state for each vertex and edge are $X_v \prod_{e \supset v} Z_e$ and $X_{e} \prod_{v \subset e} Z_v$ respectively, where $e\supset v$ and $v\subset e$ denote edges $e$ that contain the vertex $v$, and vertices $v$ that are contained in $e$, respectively. Measuring $X$ on all the edges will give a GHZ state on the vertices (up to spin flips that depend on measurement outcomes). On the other hand, measuring $X$ on all the vertices gives a state of the toric code: We have the vertex term of the toric code, $\prod_{e \supset v} Z_e = \pm 1$ depending on the measurement outcome, and we have the plaquette operator $\prod_{e \subset p} X_e=1$ coming from a product of four edge stabilizers around a plaquette, which commutes with the measurement. Note that while the topological order of this state is independent of the sign of the aforementioned stabilizers, one can always bring this to a state with $\prod_{e \supset v} Z_e = +1$ by applying string operators that pair up the vertices with $\prod_{e \supset v} Z_e = -1$.

\begin{figure*}
    \centering
    \includegraphics{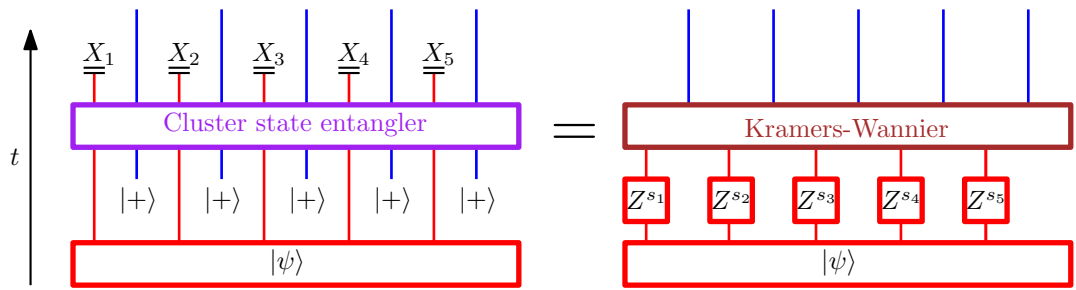}
    \caption{\textbf{The Kramers-Wannier transformation from finite-depth circuit and measurements.} The cluster state entangler can be used to implement Kramers-Wannier duality by measurement. The final state depends on $s_n = 0,1$ corresponding to measurement outcomes $X_n =1,-1$, respectively, which we can express as a product $\prod_n Z^{s_n}$ applied to $\ket{\psi}$ before KW transformation. These operators can be pushed through the KW transformation to obtain a product of $X$ operators on the $B$ sublattice (blue). Hence, by acting with this product on the postmeasurement state, one can obtain the KW transformation of $\ket{\psi}$ without postselection.}
    \label{fig:SPTtoKWmeasure}
\end{figure*}
\section{Kramers-Wannier transformation from measuring cluster state SPT phases\label{sec:KW}}

We have seen that long-range entangled states can be obtained by performing single-site measurements on the cluster state. To explore a deeper reason for this finding, we will show how the cluster state secretly encodes the KW transformation. For simplicity, we will first discuss the 1D case, where the KW transformation is defined as the map $X_n \to Z_n Z_{n+1}$ and $Z_{n-1} Z_n \to X_n$; although this map preserves the locality of $\mathbb Z_2$-symmetric operators, it is a nonlocal mapping, relating SRE to LRE.

A first hint of the connection between the cluster state and the KW transformation is the fact that $Z_{n} Z_{n+2}$ and $X_{n+1}$ act the same way on the cluster state. Moreover, $X_{n+1} U_{CZ} = U_{CZ} Z_n X_{n+1} Z_{n+2}$, where $U_{CZ}$ is the cluster entangler, Eq. \eqref{eq:clusterstateentangler}. Let us divide the sites into the odd and even sublattices, denoted $A$ and $B$, respectively, and define the states $\ket{+}_{A,B}$ on these subspaces. We find that the operator  $\sigma := \bra{+}_A U_{CZ} \ket{+}_B:\mathcal{H}_A \to \mathcal{H}_B$ gives the KW transformation. For example, we show that $X_A$ is correctly mapped to $Z_B Z_B$, i.e., $\sigma X_A = Z_B Z_B \sigma$:

\begin{equation}
\label{eq:KWmapping}
\begin{aligned}
\bra{+}_A U_{CZ} \ket{+}_B X_A
&= \bra{+}_A U_{CZ} X_A \ket{+}_B \\
&= \bra{+}_A Z_B X_A Z_B U_{CZ} \ket{+}_B \\
&= \bra{+}_A Z_B Z_B U_{CZ} \ket{+}_B \\
&= Z_B Z_B \bra{+}_A U_{CZ} \ket{+}_B,
\end{aligned}
\end{equation}
and vice versa. This example is depicted graphically in Fig.~\ref{fig:SPTtoKWproof}. Note that this method works on any bipartite graph using a suitably generalized cluster state in any dimension, in which case, the $Z_B$'s that appear act on the $B$ vertices adjacent to where $X_A$ acts and vice versa.

Eq.~\eqref{eq:KWmapping} suggests a method to apply KW by measurement. We begin with a state in $\mathcal{H}_A$ and then introduce the ancillas $\ket{+}_B$. We then apply $U_{CZ}$ to the combined system and measure the $X$ spins on $A$. If the measurement outcomes are all $+$ spins, then we have exactly implemented the KW duality. Otherwise, we have instead implemented the closely related operator
\begin{equation}
    M = \bra{+}_A \left( \prod_{a \in A} Z^{s_a} \right) U_{CZ} \ket{+}_B = \sigma  \prod_{a \in A} Z^{s_a}
\end{equation} 
where $s_a \in \{0,1\}$ are the measurement outcomes of site $a$. By pushing through the excess operators from the $A$ sites to the $B$ sites using $\sigma$, we can rewrite this formula as
\begin{equation}
M= \left( \prod_{b \in B} X^{s_b} \right) \bra{+}_A U_{CZ} \ket{+}_B = \left( \prod_{b \in B} X^{s_b} \right) \sigma,
\end{equation}
where the $s_b$ are functions of the $s_a$ that depend on the graph. For example, in 1D, where $A$ and $B$ are the odd and even sublattices of the chain respectively, we have $s_b = \sum_{1 < a < b} s_a$. Thus, we see that further applying $\left( \prod_{b \in B} X^{s_b} \right)$ restores the exact KW mapping $\sigma$. See Fig.~\ref{fig:SPTtoKWmeasure}.

This finding explains why the measured 1D cluster state has long-range order---it produces the KW dual of the trivial state $\ket{+}_A$, which is a GHZ state. Likewise in 2D we obtain the KW dual of the trivial state which is a toric code state\footnote{We note that as a by-product, we obtain explicit tensor network representations of these states. This offers an alternative derivation of the 3D toric and fracton code projected entangled pair states (PEPS) \cite{Verstraete06,Cirac2021} obtained in Refs.~\cite{Williamson20,Delcamp20}.}.

We later argue that the long-range order holds for any state in the same SPT phase as the cluster state. Indeed, this fact can be seen by symmetry fractionalization for the two $\ZZ_2$ symmetries $\prod_{a \in A} X_a$ and $\prod_{b \in B} X_b$ (acting on the odd and even sublattices, respectively) protecting the SPT phase. If we act on any state $\ket{\psi}$ in the same SPT phase by the $\ZZ_2^A$ symmetry in a region $\mathcal R$, it will reduce to some $\ZZ_2^B$ charged operators at the boundary of the region: $\prod_{a \in \mathcal R} X_a \ket{\psi} = \mathcal O_L \mathcal O_{R} \ket{\psi}$, where $\mathcal O$ is some operator with finite support situated at the left and right boundaries of $\mathcal R$, which anticommutes with $\ZZ_2^B$. Intuitively, this means that $\ket{\psi}$ has the KW property, exchanging order operators and disorder operators, at long distances. See Sec. \ref{sec:intuitive}.

In higher dimensions, the cluster state is an SPT for higher form or subsystem symmetries that depend on the lattice. For example, if $A$ and $B$ are sites at the vertices and edges of the square lattice, then we have symmetries $\prod_{a \in A} X_a$ and $\prod_{b \in \gamma \subset B} X_b$, where we have a symmetry for each closed curve $\gamma$ drawn along the edges of the direct lattice. The KW so constructed is the duality between the Ising model and Ising gauge theory in 2+1D.

A summary of examples that arise from the KW transformation of various symmetries is given in Table~\ref{tab:examples}.

\begin{table*}[t!]
    \centering
    \begin{tabular}{|c|c|c|c|c|c|}
    \hline
    $D$ &$A$ symmetry & $B$ symmetry & SPT & Product state maps to & See \\
    \hline
    1&$\ZZ_2$ & $\ZZ_2$    &$A B$ & GHZ & Sec.~\ref{sec:motivation}\\
    1& $\ZZ_2$  & $\ZZ_2^F$ & $\eta A$ & Kitaev chain & Sec.~\ref{sec:JW}\\
    2&$\ZZ_2$ & $\ZZ_2[1]$     &$A B$  & Toric code & Sec.~\ref{sec:motivation}\\
    2&$\ZZ_2$ & $\ZZ_2[1]$      &$A^3+ A B$ & Double semion & Sec.~\ref{sec:DS}\\
    2& $\ZZ_2$ (2-foliated line)&$\ZZ_2$ (2-foliated line) &``$A^2 +AB$'' (strong) SSPT & Wen plaquette & Appendix~\ref{app:Wenplaquette}\\
 3 & $\ZZ_2 [1]^2$ & $\ZZ_2 [1]^2$ & $A_1^2 +A_2^2 + A_1A_2 + A_1B_2+A_2 B_1$ & 3-fermion Walker-Wang &Appendix~\ref{sec:3FWW}\\
    3 & $\ZZ_2$ (3-foliated planar)& $\ZZ_2$ (dual subsystem) & ``$AB$" SSPT  & X-cube & Ref.~\cite{Rydberg}\\
    
    3 & $\ZZ_2$ (fractal) &$\ZZ_2$ (dual fractal) &  ``$AB$" fractal SSPT  & Sierpinski fractal spin liquid &Ref.~\cite{Rydberg}\\
 
         \hline
    \end{tabular}
    \caption{Examples of states obtained by measuring SPTs. After evolving the product state with the corresponding SPT entangler, the $A$ sublattice is measured, effectively performing a KW or JW transformation to the product state. All SPTs listed except those that create the Kitaev chain and double semion model are cluster states. Here, $D$ is the space dimension, $\ZZ_2[1]$ denotes a $\ZZ_2$ 1-form symmetry, and $A$, $B$ denote gauge fields defined for the $A$ and $B$ symmetries, respectively. See Sec. \ref{subsecmeasuringgeneralspt} for examples that go beyond this framework.}
    \label{tab:examples}
\end{table*}

\subsection{Twisted gauge theory from measuring cluster + SPT phases} \label{sec:DS}

As a first application, we discuss what happens when we apply this procedure to other states on the $A$ sublattice, such as an SPT. As in Fig.~\ref{fig:SPTtoKWmeasure}, we add $\ket{+}_B$ ancillas, couple $A$ and $B$ with the cluster state entangler, and then perform measurements on the $A$ sublattice. The result of this procedure is equivalent to gauging the SPT phase\footnote{Alternatively, by viewing the SPT and the cluster state as a single state, performing the measurement on this combined SPT can be thought of as a different way of performing the KW duality on the product state. This choice of adding an extra SPT before gauging is also known discrete torsion\cite{Vafa86}, or defectification classes \cite{BarkeshliBondersonChengWang2019} in the literature.}.

To illustrate this procedure, we discuss how beginning with the $A$ sublattice in the pure $\ZZ_2$ or ``Levin-Gu" SPT state $\ket{\psi}$ \cite{Levin_2012} we obtain the double semion topological order \cite{Kitaev_2003} after entangling and measuring. The Levin-Gu SPT is defined on the vertices of the triangular lattice ($A$) and is an eigenstate of the 
following (non-Pauli) stabilizers:
\begin{align}
    X_v \prod_{\inner{vuu'}} e^{\frac{\pi i}{4} Z_uZ_{u'}}=
     \raisebox{-.5\height}{\begin{tikzpicture}[scale=0.5]
\node[label=center:$\color{red}X$] (0) at (1,1.73) {};
\coordinate (1) at (0,0) {};
\coordinate (2) at (2,0) {};
\coordinate (3) at (3,1.73) {};
\coordinate (4) at (2,1.73*2) {};
\coordinate (5) at (0,1.73*2) {};
\coordinate (6) at (-1,1.73) {};
\draw[-,densely dotted,color=gray] (0) -- (1)  {};
\draw[-,densely dotted,color=gray] (0) -- (2)  {};
\draw[-,densely dotted,color=gray] (0) -- (3)  {};
\draw[-,densely dotted,color=gray] (0) -- (4)  {};
\draw[-,densely dotted,color=gray] (0) -- (5)  {};
\draw[-,densely dotted,color=gray] (0) -- (6)  {};
\draw[-,snake it,color=red] (2) -- (1)  {};
\draw[-,snake it,color=red] (2) -- (3)  {};
\draw[-,snake it,color=red] (4) -- (3)  {};
\draw[-,snake it,color=red] (4) -- (5)  {};
\draw[-,snake it,color=red] (6) -- (5)  {};
\draw[-,snake it,color=red] (6) -- (1)  {};
\end{tikzpicture}}
\end{align}
where $\inner{vuu'}$ are the six triangles around $v$, and the wavy lines denote $e^{\frac{\pi i}{4} Z_uZ_{u'}}$ between vertices $u$ and $u'$. Note that this stabilizer is not simply a product of Pauli operators. Let us also stress that since this is an SPT phase, it is possible to prepare this state by a finite-depth circuit\footnote{The unitary that creates the Levin-Gu SPT is given by $e^{\frac{i\pi}{8} (\sum_{\Delta_{uvw}} Z_uZ_vZ_w -2\sum_v Z_v})$ where  $\Delta_{uvw}$ denotes all triangles.}. Following our procedure, we add the $B$ sublattice consisting of edges of the triangular lattice, supporting a product with the trivial stabilizer
\begin{align}
    X_e &=
\raisebox{-.3\height}{\begin{tikzpicture}[scale=0.5]
\coordinate (1) at (0,0) {};
\coordinate (2) at (2,0) {};
\draw[-,densely dotted,color=gray] (1) -- (2) node [midway] {$\color{blue}X$};
\end{tikzpicture}}.
\end{align}
Next, we couple the two sublattices with the cluster state entangler, resulting in the stabilizers
\begin{align}
    X_v \prod_{\inner{vuu'}} e^{\frac{\pi i}{4} Z_uZ_{u'}} \prod_{e \supset v} Z_e &= \raisebox{-.5\height}{\begin{tikzpicture}[scale=0.5]
\node[label=center:$\color{red}X$] (0) at (1,1.73) {};
\coordinate (1) at (0,0) {};
\coordinate (2) at (2,0) {};
\coordinate (3) at (3,1.73) {};
\coordinate (4) at (2,1.73*2) {};
\coordinate (5) at (0,1.73*2) {};
\coordinate (6) at (-1,1.73) {};
\draw[-,densely dotted,color=gray] (0) -- (1) node [midway] {$\color{blue}Z$};
\draw[-,densely dotted,color=gray] (0) -- (2) node [midway] {$\color{blue}Z$};
\draw[-,densely dotted,color=gray] (0) -- (3) node [midway] {$\color{blue}Z$};
\draw[-,densely dotted,color=gray] (0) -- (4) node [midway] {$\color{blue}Z$};
\draw[-,densely dotted,color=gray] (0) -- (5) node [midway] {$\color{blue}Z$};
\draw[-,densely dotted,color=gray] (0) -- (6) node [midway] {$\color{blue}Z$};
\draw[-,snake it,color=red] (2) -- (1)  {};
\draw[-,snake it,color=red] (2) -- (3)  {};
\draw[-,snake it,color=red] (4) -- (3)  {};
\draw[-,snake it,color=red] (4) -- (5)  {};
\draw[-,snake it,color=red] (6) -- (5)  {};
\draw[-,snake it,color=red] (6) -- (1)  {};
\end{tikzpicture}},\\
X_{e} \prod_{v \subset e} Z_v &= 
\raisebox{-.3\height}{\begin{tikzpicture}[scale=0.5]
\node[label=center:$\color{red}Z$] (1) at (0,0) {};
\node[label=center:$\color{red}Z$] (2) at (2,0) {};
\draw[-,densely dotted,color=gray] (1) -- (2) node [midway] {$\color{blue}X$};
\end{tikzpicture}}.
\end{align}
Before we perform the measurements on all $A$ sites (the vertices of the triangular lattice), we note that the vertex stabilizer does not commute with the measurement. Thus, it would not directly give us a useful condition on the postmeasurement state. However, using the fact that
$Z_u Z_{u'}\ket{\psi} = X_{(uu')}\ket{\psi}$, where  $(uu')$ is the edge with $u$ and $u'$ as end points, the following is an equally valid set of stabilizers of $\ket{\psi}$:
\begin{align}
\label{eq:stabilizer}
    X_v \prod_{\inner{vuu'}} R_{(uu')} \prod_{e \supset v} Z_e &=
\raisebox{-.5\height}{\begin{tikzpicture}[scale=0.5]
\node[label=center:$\color{red}X$] (0) at (1,1.73) {};
\coordinate (1) at (0,0) {};
\coordinate (2) at (2,0) {};
\coordinate (3) at (3,1.73) {};
\coordinate (4) at (2,1.73*2) {};
\coordinate (5) at (0,1.73*2) {};
\coordinate (6) at (-1,1.73) {};
\draw[-,densely dotted,color=gray] (0) -- (1) node [midway] {$\color{blue}Z$};
\draw[-,densely dotted,color=gray] (0) -- (2) node [midway] {$\color{blue}Z$};
\draw[-,densely dotted,color=gray] (0) -- (3) node [midway] {$\color{blue}Z$};
\draw[-,densely dotted,color=gray] (0) -- (4) node [midway] {$\color{blue}Z$};
\draw[-,densely dotted,color=gray] (0) -- (5) node [midway] {$\color{blue}Z$};
\draw[-,densely dotted,color=gray] (0) -- (6) node [midway] {$\color{blue}Z$};
\draw[-,densely dotted,color=gray] (2) -- (1) node [midway] {$\color{blue}R$};
\draw[-,densely dotted,color=gray] (2) -- (3)  node [midway] {$\color{blue}R$};
\draw[-,densely dotted,color=gray] (4) -- (3)  node [midway] {$\color{blue}R$};
\draw[-,densely dotted,color=gray] (4) -- (5)  node [midway] {$\color{blue}R$};
\draw[-,densely dotted,color=gray] (6) -- (5)  node [midway] {$\color{blue}R$};
\draw[-,densely dotted,color=gray] (6) -- (1)  node [midway] {$\color{blue}R$};
\end{tikzpicture}},\\
X_{e} \prod_{v \subset e} Z_v &= 
\raisebox{-.3\height}{\begin{tikzpicture}[scale=0.5]
\node[label=center:$\color{red}Z$] (1) at (0,0) {};
\node[label=center:$\color{red}Z$] (2) at (2,0) {};
\draw[-,densely dotted,color=gray] (1) -- (2) node [midway] {$\color{blue}X$};
\end{tikzpicture}},
\end{align}
where $R_e= e^{\frac{\pi i}{4} X_e}$. The vertex stabilizers now commute with the measurement. However, the stabilizers in Eq. \eqref{eq:stabilizer} do not commute for adjacent vertices. However, this problem is cured by restricting to the subspace:
\begin{align}
    \prod_{e \subset \Delta} X_e = \raisebox{-.5\height}{\begin{tikzpicture}[scale=0.5]
\coordinate (0) at (1,1.73) {};
\coordinate (1) at (0,0) {};
\coordinate (2) at (2,0) {};
\draw[-,densely dotted,color=gray] (0) -- (1) node [midway] {$\color{blue}X$};
\draw[-,densely dotted,color=gray] (0) -- (2) node [midway] {$\color{blue}X$};
\draw[-,densely dotted,color=gray] (2) -- (1) node [midway] {$\color{blue}X$};
\end{tikzpicture}}=1.
\end{align}
We can therefore circumvent having non-commuting stabilizers by attaching 
\begin{align}
    O_{vuu'}=\frac{1+X_{(vu)}X_{(vu')}X_{(uu')}}{2} =\raisebox{-.5\height}{\begin{tikzpicture}[scale=0.5]
\coordinate (0) at (1,1.73) {};
\coordinate (1) at (0,0) {};
\coordinate (2) at (2,0) {};
\draw[draw=white, fill=blue,fill opacity=0.2]  (1,1.73) -- (1) -- (2) -- cycle;
\draw[-,densely dotted,color=gray] (0) -- (1) {};
\draw[-,densely dotted,color=gray] (0) -- (2) {};
\draw[-,densely dotted,color=gray] (2) -- (1) {};
\end{tikzpicture}},
\end{align}
 which is a projector into this subspace on each triangle. Finally, $\ket{\psi}$ is identified as the unique state that has eigenvalue $+1$ under the following operators:
\begin{align}
     X_v \prod_{\inner{vuu'}}\left( R_{(uu')} O_{vuu'}\right) \prod_{e \supset v} Z_e &= \raisebox{-.5\height}{\begin{tikzpicture}[scale=0.5]
\coordinate (1) at (0,0) {};
\coordinate (2) at (2,0) {};
\coordinate (3) at (3,1.73) {};
\coordinate (4) at (2,1.73*2) {};
\coordinate (5) at (0,1.73*2) {};
\coordinate (6) at (-1,1.73) {};
\draw[draw=white, fill=blue,fill opacity=0.2]  (1,1.73) -- (1) -- (2) -- cycle;
\draw[draw=white, fill=blue,fill opacity=0.2]  (1,1.73) -- (2) -- (3) -- cycle;
\draw[draw=white, fill=blue,fill opacity=0.2]  (1,1.73) -- (3) -- (4) -- cycle;
\draw[draw=white, fill=blue,fill opacity=0.2]  (1,1.73) -- (4) -- (5) -- cycle;
\draw[draw=white, fill=blue,fill opacity=0.2]  (1,1.73) -- (5) -- (6) -- cycle;
\draw[draw=white, fill=blue,fill opacity=0.2]  (1,1.73) -- (6) -- (1) -- cycle;
\node[label=center:$\color{red}X$] (0) at (1,1.73) {};
\draw[-,densely dotted,color=gray] (0) -- (1) node [midway] {$\color{blue}Z$};
\draw[-,densely dotted,color=gray] (0) -- (2) node [midway] {$\color{blue}Z$};
\draw[-,densely dotted,color=gray] (0) -- (3) node [midway] {$\color{blue}Z$};
\draw[-,densely dotted,color=gray] (0) -- (4) node [midway] {$\color{blue}Z$};
\draw[-,densely dotted,color=gray] (0) -- (5) node [midway] {$\color{blue}Z$};
\draw[-,densely dotted,color=gray] (0) -- (6) node [midway] {$\color{blue}Z$};
\draw[-,densely dotted,color=gray] (2) -- (1) node [midway] {$\color{blue}R$};
\draw[-,densely dotted,color=gray] (2) -- (3)  node [midway] {$\color{blue}R$};
\draw[-,densely dotted,color=gray] (4) -- (3)  node [midway] {$\color{blue}R$};
\draw[-,densely dotted,color=gray] (4) -- (5)  node [midway] {$\color{blue}R$};
\draw[-,densely dotted,color=gray] (6) -- (5)  node [midway] {$\color{blue}R$};
\draw[-,densely dotted,color=gray] (6) -- (1)  node [midway] {$\color{blue}R$};
\end{tikzpicture}},\\
X_{e} \prod_{v \subset e} Z_v &= 
\raisebox{-.3\height}{\begin{tikzpicture}[scale=0.5]
\node[label=center:$\color{red}Z$] (1) at (0,0) {};
\node[label=center:$\color{red}Z$] (2) at (2,0) {};
\draw[-,densely dotted,color=gray] (1) -- (2) node [midway] {$\color{blue}X$};
\end{tikzpicture}}.
\end{align}
Performing the measurement with outcomes $X_v=(-1)^{s_v}$, the postmeasurement state is the unique state that has eigenvalue +1 under the operators:
\begin{align}
      (-1)^{s_v}\prod_{\inner{vuu'}} \left( R_{(uu')} O_{vuu'} \right)\prod_{e \supset v} Z_e  &= (-1)^{s_v}\raisebox{-.5\height}{\begin{tikzpicture}[scale=0.5]
\coordinate (1) at (0,0) {};
\coordinate (2) at (2,0) {};
\coordinate (3) at (3,1.73) {};
\coordinate (4) at (2,1.73*2) {};
\coordinate (5) at (0,1.73*2) {};
\coordinate (6) at (-1,1.73) {};
\draw[draw=white, fill=blue,fill opacity=0.2]  (1,1.73) -- (1) -- (2) -- cycle;
\draw[draw=white, fill=blue,fill opacity=0.2]  (1,1.73) -- (2) -- (3) -- cycle;
\draw[draw=white, fill=blue,fill opacity=0.2]  (1,1.73) -- (3) -- (4) -- cycle;
\draw[draw=white, fill=blue,fill opacity=0.2]  (1,1.73) -- (4) -- (5) -- cycle;
\draw[draw=white, fill=blue,fill opacity=0.2]  (1,1.73) -- (5) -- (6) -- cycle;
\draw[draw=white, fill=blue,fill opacity=0.2]  (1,1.73) -- (6) -- (1) -- cycle;
\coordinate (0) at (1,1.73) {};
\draw[-,densely dotted,color=gray] (0) -- (1) node [midway] {$\color{blue}Z$};
\draw[-,densely dotted,color=gray] (0) -- (2) node [midway] {$\color{blue}Z$};
\draw[-,densely dotted,color=gray] (0) -- (3) node [midway] {$\color{blue}Z$};
\draw[-,densely dotted,color=gray] (0) -- (4) node [midway] {$\color{blue}Z$};
\draw[-,densely dotted,color=gray] (0) -- (5) node [midway] {$\color{blue}Z$};
\draw[-,densely dotted,color=gray] (0) -- (6) node [midway] {$\color{blue}Z$};
\draw[-,densely dotted,color=gray] (2) -- (1) node [midway] {$\color{blue}R$};
\draw[-,densely dotted,color=gray] (2) -- (3)  node [midway] {$\color{blue}R$};
\draw[-,densely dotted,color=gray] (4) -- (3)  node [midway] {$\color{blue}R$};
\draw[-,densely dotted,color=gray] (4) -- (5)  node [midway] {$\color{blue}R$};
\draw[-,densely dotted,color=gray] (6) -- (5)  node [midway] {$\color{blue}R$};
\draw[-,densely dotted,color=gray] (6) -- (1)  node [midway] {$\color{blue}R$};
\end{tikzpicture}},\\
    \prod_{e \subset \Delta} X_e &= \raisebox{-.5\height}{\begin{tikzpicture}[scale=0.5]
\coordinate (0) at (1,1.73) {};
\coordinate (1) at (0,0) {};
\coordinate (2) at (2,0) {};
\draw[-,densely dotted,color=gray] (0) -- (1) node [midway] {$\color{blue}X$};
\draw[-,densely dotted,color=gray] (0) -- (2) node [midway] {$\color{blue}X$};
\draw[-,densely dotted,color=gray] (2) -- (1) node [midway] {$\color{blue}X$};
\end{tikzpicture}},
\end{align}
which is the ground state of the double semion model \cite{Levin_2012} up to single site $X$-rotations on edges that pair up the vertices where $s_v=1$ to remove the signs, and swapping $X_e$ with $Z_e$ to match the choice in Ref. \cite{Levin_2012}.

Our implementation of gauging via combining measurements with a cluster state entangler (including $\ZZ_n$ generalizations) implies that we can produce all twisted quantum double models of a finite Abelian gauge group via stacking general SPTs prior to measuring---which can be prepared by finite-depth circuits \cite{Chen_2013}. Note that these models already contain certain non-Abelian phases, e.g., $D_4$ topological order arises upon gauging the $\mathbb Z_2^3$ symmetry of an SPT phase with a type-III cocycle \cite{P95,Yoshida2016}. (For obtaining non-Abelian topological order associated with any solvable group, see Sec.~\ref{subsec:sequential}.) Similarly, our procedure allows for the creation of twisted fracton phases by gauging 3D subsystem SPT phases \cite{ShirleySlagleChen20,DevakulShirleyWang20,StephenGarre-RubioDuaWilliamson2020,Tantivasadakarn20,Shirley20}. Thus a much wider class of states can be obtained from local unitary circuits and local operations and classical communications (LOCC) ~\cite{Piroli21} than previously established.

\subsection{Physically applying the Kramers-Wannier transformation to a gapless state}

Here, we discuss an example where the input state $|\psi\rangle$ (in Fig.~\ref{fig:SPTtoKWmeasure}) itself has long-range entanglement. In particular, we focus on a well-known example of how the $XY$ chain---an example of a gapless state---can be transformed into two decoupled critical Ising chains by gauging particle-hole symmetry\footnote{Field theoretically, this maps the compact boson to two copies of the Ising CFT \cite{ginsparg_applied_1988}.}. Here, we achieve this gauging by using a finite-depth circuit and single-site measurements.

We place the $XY$ chain on the odd sites ($A$) and initialize with $\ket{+}$ states on the even sites ($B$). The aforementioned state can be considered the ground state of the following Hamiltonian
\begin{align}
    H = \sum_n X_{2n-1}X_{2n+1} + Y_{2n-1}Y_{2n+1} - X_{2n}
\end{align}
Next, we gauge the $\ZZ_2$ subgroup $\prod_n X_{2n-1}$ of the full $U(1)$ symmetry of the XY chain. To do so, we couple the even and odd sites with the cluster state entangler $U=\prod_{n} CZ_{n,n+1}$, resulting in
\begin{align}
    UHU^\dagger =& \sum_n Z_{2n-2}(X_{2n-1}X_{2n+1}+Y_{2n-1}Y_{2n+1})Z_{2n+2}\nonumber\\&- Z_{2n-1}X_{2n}Z_{2n+1}
\end{align}
Note that since $Z_{2n-1}X_{2n}Z_{2n+1}$ is an integral of motion, the following Hamiltonian also has the same wave function as its ground state:
\begin{align}
\sum_n Z_{2n-2}(X_{2n-1}X_{2n+1}-X_{2n})Z_{2n+2}- Z_{2n-1}X_{2n}Z_{2n+1}
\end{align}
Now, we perform a measurement on the odd sites with measurement outcomes $X=(-1)^s$; the state after the measurement is the ground state of the Hamiltonian
\begin{align}
\sum_n (-1)^{s_{2n-1}+s_{2n+1}}Z_{2n-2}Z_{2n+2}-Z_{2n-2}X_{2n}Z_{2n+2}
\end{align}
with the integral of motion $\prod_n X_{2n}$ serving as a global $\ZZ_2$ symmetry. After appropriate spin flips to remove the signs and the circuit $\prod_{n} CZ_{2n,2n+2}$, the Hamiltonian reads
\begin{align}
\sum_n Z_{2n-2}Z_{2n+2}-X_{2n}
\end{align}
which describes two decoupled critical Ising chains. We thus confirm that we have physically implemented the KW transform on a gapless state.

Let us remark that this procedure does not rely on free-fermion solvability of the $XY$ chain and the Ising model. For example, the procedure still works in the presence of the $XXZ$ deformation, which respects the $\ZZ_2$ symmetry (albeit opening up a gap).

\subsection{Non-Abelian topological order from sequentially gauging Abelian groups \label{subsec:sequential}}

Beyond cyclic groups $\ZZ_n$, cluster states and the corresponding KW dualities have been generalized to arbitrary finite groups \cite{Brennen09,Brell2015,Haegeman15}, giving the potential to gauge non-Abelian groups by unitaries and measurement. However, unlike the Abelian case, which produces Abelian anyons depending on the measurement outcome, gauging non-Abelian groups can produce non-Abelian anyons that can only be paired up using linear depth string operators\footnote{We thank David T. Stephen for pointing out this subtlety.}. The intuition for this is that the string operators for moving such anyons consist of noncommuting operators which hence cannot be applied all at once\footnote{We note one potential loophole. If the group is nilpotent, two non-Abelian anyons can be annihilated by first nucleating a whole density of pairs of anyons of the same type along a path connecting the two anyons and subsequently fusing them all at once. This potentially leaves a density of residual anyons all along the path, but the nilpotent sequence ensures that by repeating this process, we obtain simpler and simpler anyons---eventually leading to Abelian anyons which can be efficiently removed. Unfortunately, the known ways of using non-Abelian states for universal quantum computation rely on the group being not nilpotent \cite{Mochon04}.}.

Our implementation of the KW duality avoids this issue by a sequence of circuits and measurements, which can be interpreted as sequentially gauging Abelian groups. In such a method, the measurement outcomes in all intermediate states correspond to Abelian anyons, which can all be paired up in finite depth. In this way, all gauge theories whose gauge group is solvable (i.e., obtained by extending finite Abelian groups) can be constructed efficiently in this manner.
For example, the $S_3$ quantum double can be obtained by gauging a $\ZZ_3$ symmetry (i.e., measuring a $\ZZ_3$ cluster state), which prepares a $\ZZ_3$ toric code, followed by gauging the charge conjugation symmetry that permutes anyons $e \leftrightarrow e^2$ and $m \leftrightarrow m^2$. We note that since $S_3$ is not nilpotent, it can be used for universal quantum computation \cite{Mochon04}. As a second example, the $D_4$ topological order can be obtained by first preparing the 2D color code and gauging the Hadamard symmetry. In our companion paper we provide explicit finite-depth qubit-based circuits for these two examples \cite{Rydberg}.

We note that sequentially gauging Abelian groups can also give rise to states beyond quantum doubles. For instance, the doubled Ising anyon theory can be obtained by gauging the $e\leftrightarrow m$ symmetry of $\mathbb Z_2$ topological order \cite{BarkeshliBondersonChengWang2019}. Such a Kramers-Wannier transformation (implemented using our finite-depth circuit and single-site measurements) can indeed be performed since it is known that the $\mathbb Z_2$ symmetry can be made on-site (for explicit models, see Refs.~\cite{Heinrich16,Cheng17}). By definition, this state can be connected to any other state with $\mathbb Z_2$ topological order through a finite-depth circuit, and we have already described how, e.g., the usual toric code can be obtained from the product state.

\section{Jordan-Wigner transformation from measuring fermionic SPT phases \label{sec:JW}}
\begin{figure}
    \centering
    \includegraphics[scale=0.9]{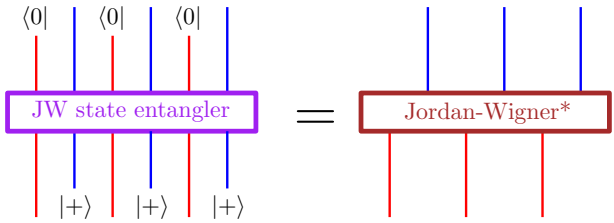}
    \caption{\textbf{Jordan-Wigner transformation from finite-depth circuit and measurements.} We show the process of entangling fermionic (red) and bosonic (blue) degrees of freedom and its relation to the JW transformation. Here $\bra{0}$ corresponds to contracting with the empty state of fermions. We use Jordan-Wigner$^*$ to emphasize that this transformation differs from the usual JW by an additional KW transformation. Similar to Fig.~\ref{fig:SPTtoKWmeasure}, this can be utilized to implement the JW transformation via measurements (see main text).}
    \label{fig:SPTtoJW}
\end{figure}

Analogous to the KW transformation, the Jordan-Wigner (JW) map is a nonlocal transformation which maps between fermionic and bosonic degrees of freedom \cite{Jordan1928,SMLising}. Similar to the KW transformation, here we can prepare and entangle bosonic and fermionic degrees of freedom as shown in Fig.~\ref{fig:SPTtoJW}. We can then perform either bosonization of an arbitrary input fermionic state by measuring the parity of all fermions, or fermionization of an arbitrary input bosonic state by measuring $X$ on all the spins after the entangling step.

\subsection{1+1D bosonization}\label{app:JW1D}

Let us demonstrate this case explicitly by preparing the Kitaev Majorana chain, which cannot be done in finite time with only unitary evolution \cite{HuangChen2015}. We start with $N$ qubits on odd sites initialized in the $\ket{+}$ state and $N$ fermions on even sites initialized in the empty state $P=-i\gamma \gamma'=1$, where $\gamma=c+c^\dagger$ and $\gamma' = -i(c-c^\dagger)$ are Majorana operators. Furthermore, we define the hopping operator $S_{2n} = i\gamma'_{2n-2}\gamma_{2n}$, which hops a fermion from site $2n-2$ to $2n$. We create a $\ZZ_2\times \ZZ_2^F$ SPT \cite{GuWen14,vonKeyserlingkSondhi16,TantivasadakarnVishwanath18,Borla21} with the following circuit:
\begin{align}
    U=\prod_{n=1}^N  CS_{2n-1,2n}
\end{align}
where the operator
\begin{align}
   CS_{2n-1,2n} = \ket{\uparrow}\bra{\uparrow}_{2n-1} + \ket{\downarrow}\bra{\downarrow}_{2n-1} S_{2n} 
\end{align} is a hopping operator controlled by the qubit at $2n-1$. In other words, a fermion is hopped if the spin at site $2n-1$ is down. We also remark that because all gates mutually commute, it can be implemented as a finite -epth circuit. The resulting SPT (which we will call the Jordan-Wigner state) is the $+1$ eigenstate of the stabilizers
\begin{align}
    UX_{2n-1}U^\dagger&=i\gamma'_{2n-2}X_{2n-1}\gamma_{2n},\\
     UP_{2n}U^\dagger&= Z_{2n-1}P_{2n}Z_{2n+1}.
\end{align}
Now, we measure all the spins with outcomes $X_{2n-1} = (-1)^{s_{2n-1}}$. The stabilizers of the measured state are $(-1)^{s_{2n-1}}\gamma'_{2n-2}\gamma_{2n}$ and $\prod_n Z_{2n-1}P_{2n}Z_{2n+1} = \prod_n P_{2n}$, which after applying  $\prod_{n=1}^N P_{2n}^{\sum_{m=1}^n s_{2m-1}}$, gives the ground state of the Kitaev chain. We note that, alternatively, starting with the SPT, measuring the parity of all the fermions gives the GHZ state.

\subsection{2+1D bosonization}\label{app:JW2D}
The recipe above extends to arbitrary dimensions. The generalization of the Jordan-Wigner transformation has been explored in a number of works including \cite{GaiottoKapustin2016,kapustin2017fermionic,ChenKapustinRadicevic2018,ChenKapustin2019,Chen2019,Shukla20,Tantivasadakarn20,Shirley20,Po21,LiPo21}, and can be thought of in the context of this work as gauging the fermion parity symmetry. From this we can construct a particular state of fermions and spins which conserves fermion parity and a higher form $\ZZ_2$ symmetry such that one can perform either bosonization, by measuring the parity of each fermion, or fermionization, by measuring the spins in the $X$-basis. Here, we demonstrate this for the 2D bosonization procedure of Ref.~\onlinecite{ChenKapustinRadicevic2018} on a square lattice.

As with the 2D KW transformation, we consider the square lattice with fermions initialized in the empty state $P_v=1$ on the vertices and spins are initialized in the $\ket{+}$ state on the edges ($X_e=1$). We create an ``SPT" state (see below for caveats) protected by fermion-parity symmetry and a global 1-form symmetry. The stabilizers of this ``JW state" are given by
\begin{align}
    \raisebox{-.5\height}{ \includegraphics[]{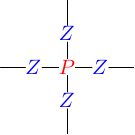} } ,&&  \raisebox{-.5\height}{ \includegraphics[]{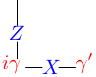} } ,&& \raisebox{-.5\height}{ \includegraphics[]{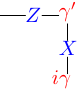} }. \label{eq:JW_SPT}
\end{align}
Upon measuring the fermion parity of all fermions, the resulting state is described by the stabilizers
\begin{align}
     \raisebox{-.5\height}{ \includegraphics[]{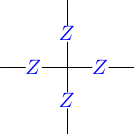} }, &&  \raisebox{-.5\height}{ \includegraphics[]{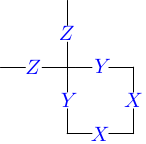} },
\end{align}
which, up to a sign given by measurement outcomes, describe the 2D toric code.

To discuss the circuit required to prepare this SPT, we first define the fermion hopping operator for each edge as
\begin{align}
S_e =  \raisebox{-.5\height}{ \includegraphics[]{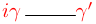} }, &&  \raisebox{-.5\height}{ \includegraphics[]{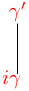}} 
\end{align}
Then, we may define the controlled operator
\begin{align}
    CS_e = \ket{\uparrow}\bra{\uparrow}_{e} + \ket{\downarrow}\bra{\downarrow}_{e} S_{e}.
\end{align}
Here, the only novel subtlety---not present in the bosonic case or the 1D JW transformation--is that not all of the $CS$ gates mutually commute and therefore must be applied sequentially. Nevertheless, it turns out that their ordering is irrelevant: Each choice of ordering gives a valid JW transformation~\cite{Tantivasadakarn20,Shirley20}, and moreover these choices only differ by phase gates. Thus, a given choice determines the spatial anisotropy of the stabilizers.

To obtain the stabilizers of the SPT in Eq.~\ref{eq:JW_SPT}, the unitary that prepares it can be written as
\begin{align}
    U = \prod_v CZ_{e_{N(v)},e_{E(v)}} \prod_{e_x} CS_{e_x} \prod_{e_y} CS_{e_y}.
\end{align}
where $e_{N(v)}$ and $e_{E(v)}$ refer to the edges directly north and east of the vertex $v$, respectively. In other words, we have chosen to apply the control gates on all vertical edges (which mutually commute) followed by those on the horizontal edges; lastly, we apply appropriate $CZ$ gates to obtain the desired form of the stabilizers.

The JW state has the property that if we form the open string operator associated to the 1-form symmetry, by taking a product of stabilizers, we will find a fermion operator at the end. Thus, it looks like a nontrivial SPT for fermion parity and the 1-form symmetry. However, if we consult the cobordism classification, we find there are no nontrivial SPTs in this symmetry class. In fact if we try to construct an SPT class with this property using the Atiyah-Hirzebruch spectral sequence, we find that the relevant class in $H^2(\ZZ_2[1], \Omega^1_{\rm spin})$ has a nonzero differential. It would be a supercohomology class but it does not satisfy the Gu-Wen equation \cite{GuWen14} (also see \cite{anomalousspt}).

The puzzle is resolved by considering the cobordism classification as describing a torsor rather than a group, meaning that with this choice of 1-form symmetry, the associated open string must always end on a fermion, and in that sense there is only one SPT phase, but it is not quite trivial because the 1-form symmetry generator we've chosen is not completely ``on-site".

Indeed, in Refs.\cite{GaiottoKapustin2016,kapustin2017fermionic} it was stressed that the 1-form symmetry in 2+1D bosonization has an anomaly $Sq^2 B$ (unlike in 1+1D bosonization where we obtain an anomaly-free $\ZZ_2$ symmetry upon bosonizing) and the kernel of the bosonization transformation gives a trivialization of this anomaly in the presence of fermions. In simple terms, the $Sq^2 B$ anomaly says that the 1-form symmetry generator needs to obey fermionic statistics. Now, there is no issue with realizing such an anomalous symmetry in a not-on-site fashion, but because of the anomaly, it cannot be screened---there is no end-point operator that will give the open string long-range order. However, if physical fermions are present, we can have a short-range entangled state where the 1-form symmetry generator ends on these fermions, and we interpret this finding as a trivialization of the $Sq^2 B$ anomaly, which is precisely what happens in the JW state. To trivialize the anomaly, the 1-form symmetry generator has to end on a fermion (which is essentially the Gu-Wen equation), so while it looks like a nontrivial SPT, there is really only one option, in harmony with the classification.

Similarly to the KW transformation, we can now apply the JW transformation to arbitrary states by measurements. For example, we can consider preparing the fermions in a 2+1D topological $p+ip$ superconducting state with chiral Majorana edge modes. After coupling to the JW state and measuring fermion parity, the remaining spins will describe a chiral Ising topological order. Similarly, coupling $\nu$ stacks of $p+ip$ superconductors to the SPT and performing the measurement can realize the topological orders in Kitaev's 16-fold way \cite{kitaev_anyons_2005}.

The generalization to higher dimensions~\cite{ChenKapustin2019,Chen2019} and to other types of fermionic gauge theories (including fracton models with fermionic statistics~\cite{Tantivasadakarn20,Shirley20}) is straightforward by taking a sequential product of $CS$ operators that mutually commute within each layer. 

\section{Generalizations \label{sec:generalization}}

Thus far, we have focused on two illustrative cases, where measuring sublattices of the cluster and JW states leads to LRE. In this last section, we generalize this approach in two directions. First, we make the case that the ability to produce LRE from measurements is indeed a property of the whole SPT \emph{phase}, being robust to tuning away from a fixed-point limit. Second, we show that LRE is naturally obtained by measuring a broad class of SPT phases, of which the cluster and JW states are but two examples.

\subsection{LRE generation as stable property of SPT phase}

\subsubsection{Intuition away from fixed-point limit \label{sec:intuitive}}

Let us first consider the 1D cluster SPT phase and ask whether one obtains a cat state upon measuring one of the sublattices starting with an arbitrary state in this phase. We present an intuitive argument, which holds away from the fixed-point limit. A key property of the cluster SPT phase in 1D is that it generically has long-range order for the following string operator \cite{Smacchia11}:
\begin{align}
\lim_{|n-m| \to \infty} \langle Z_{2m} \mathcal S_{2m,2n} Z_{2n} \rangle = C \neq 0, \label{eq:topstring}
\end{align}
where $\mathcal S_{2m,2n} := X_{2m+1} X_{2m+3} \cdots X_{2n-1}$ is a string operator consisting of the $\mathbb Z_2$ symmetry of the odd sites. The SPT invariant \cite{Pollmann12} is encoded in the fact that the string operator for one of the $\mathbb Z_2$ symmetries only has long-range order if one includes an end-point operator that is charged under the \emph{other} $\mathbb Z_2$ symmetry (in this case $Z_{2n}$ which is odd under $\prod_m X_{2m-1}$). Indeed, in the nontrivial SPT phase, one finds that the undressed string does not have long-range order:
\begin{align}
\lim_{|n-m| \to \infty} \langle \mathcal S_{2m,2n} \rangle = 0. \label{eq:trivstring}
\end{align}

We would like to understand what happens if we measure all odd sites in the $X$-basis, which is a rather challenging many-body question, and Secs.~\ref{subsec:statement}-\ref{subsec:spin1} will be devoted to addressing this issue. However, as a first encounter, and to build some intuition, let us imagine that instead of measuring all odd sites, we measure a single global observable, namely the string operator $\mathcal S_{2m,2n}$ for a fixed choice of $m$ and $n$. Since all $X$ measurements commute, we can indeed think of this as a first step in our measurement process, and we find that this first step indeed produces long-range entanglement.

To determine the result of measuring $\mathcal S_{2m,2n}$, first note that Eq.~\eqref{eq:trivstring} tells us that if we choose $n$ and $m$ far enough apart, then $\langle \mathcal S_{2m,2n} \rangle \approx 0$. Hence, both measurement outcomes $\mathcal S_{2m,2n} = \pm 1 = (-1)^s$ are equally likely. The two possible postmeasurement states can thus be written as:
\begin{equation}
|\psi_s\rangle = \frac{1}{\sqrt{2}} \left(1+(-1)^s\mathcal S_{2m,2n} \right) |\psi\rangle.
\end{equation}
Plugging $|\psi\rangle = \frac{1}{\sqrt{2}} \left( |\psi_0\rangle + |\psi_1\rangle \right)$ into Eq.~\eqref{eq:topstring}, we obtain
\begin{equation}
\langle \psi_0| Z_{2m} Z_{2n} |\psi_0\rangle - \langle \psi_1| Z_{2m} Z_{2n} |\psi_1\rangle = 2C.
\end{equation}
Moreover, using the dual string operator, one can prove that $\langle \psi_0| Z_{2m} Z_{2n} |\psi_0\rangle = - \langle \psi_1| Z_{2m} Z_{2n} |\psi_1\rangle$ (see Appendix~\ref{app:string}), such that for either measurement outcome, we have
\begin{equation}
|\langle \psi_i| Z_{2m} Z_{2n} |\psi_i \rangle| = |C| \neq 0. \label{eq:LRO}
\end{equation}
We thus find that measuring the string leads to long-range cat-state-like entanglement between the two end-points! This result is consistent with the notion of SPT entanglement explored in Ref.~\cite{Marvian17}, where the author showed that measuring a large connected block of sites leads to a Bell pair between the two end-points.

The above argument can be extended to higher dimensions. For instance, let us revisit the 2D case mentioned in Sec.~\ref{sec:motivation}: the Lieb lattice with spins on the vertices ($A$ sublattice) and bonds ($B$ sublattice) of the square lattice. The cluster state on this lattice is an SPT phase protected by a global $\mathbb Z_2$ symmetry $U^A = \prod_{a \in A} X_a$, as well as a ``1-form symmetry," $U^B_\gamma = \prod_{b \in \gamma \subset B} X_b$, meaning a symmetry defined for each closed curve $\gamma$ on the bonds of the square lattice \cite{generalizedglobalsymmetries, kapustin2015higher,Yoshida2016}.

In the SPT phase, we have long-range order for the membrane operator $S_{\partial R} \prod_{a \in A \cap R} X_a$ where $R$ is some region and $S_{\partial R}$ is a string operator on the boundary which ``braids" with $U^B_\gamma$, meaning $U^B_\gamma S_{\gamma'} (U^B_\gamma)^\dagger = S_{\gamma'} (-1)^{\gamma \cap \gamma'}$, where the exponent is the number of intersection points between the curves $\gamma$ and $\gamma'$. For the fixed point cluster state, $S_{\gamma'} = \prod_{b \in \gamma'} Z_b$.

Upon measuring the membrane, we are left with long-range order for $S_\gamma$ (see Fig.~\ref{fig:membrane}). This quantity serves as an order parameter for spontaneously breaking the 1-form symmetry, thereby implying topological order. In fact, this point of view naturally generalizes to other SPT phases, as we will discuss in Sec.~\ref{subsecmeasuringgeneralspt}.

However, while the above is intuitive and encouraging, it does not actually prove that the LRE persists upon measuring \emph{all} (or a finite density of) sites. In particular, in the 1D case, we have thus far only measured $\mathcal{S}_{2m,2n}$ and not yet all odd sites. This calculation does not automatically guarantee that the long-range order in Eq.~\eqref{eq:LRO} persists after performing the other measurements\footnote{In fact, in the argument we gave for the 2D case, the unitary string operator $U$ might have overlap with the very sites that we are measuring.} since measurements can reduce entanglement. We now argue that, generically, it does indeed persist.

\subsubsection{Conjecture and theorem: LRE from SPT \label{subsec:statement}}

Having gained the above intuition, let us now try to formalize how and when long-range entanglement is produced by measuring SPT phases. To this end, we state a general conjecture, for which we give plausibility arguments. In addition, we provide a rigorous theorem for a slightly more constrained setting.

We consider a (short-range entangled) wave function $\ket{\psi}$ in a nontrivial SPT phase protected by an Abelian symmetry group $G \times H$. Moreover, we presume that the SPT phase is \emph{mixed}, which means that explicitly breaking either $G$ or $H$ would trivialize the SPT phase. Note that the notion of an on-site symmetry automatically implies the notion of a unit cell, whereby a global symmetry $U \in G\times H$ can be decomposed as a tensor product over the unit cells: $U = \prod_n U_n$. The physical act of measuring the $G$-charge (for a given unit cell $n$) means that, mathematically, we apply a projector
\begin{equation} \label{eq:PG}
P_G(q)_n = \frac{1}{|G|} \sum_{g \in G}\chi_q(g) \;  \left( U_g \right)_n,
\end{equation}
where $q$ is a charge labeling the (random) measurement outcome, and $\chi_q$ is the corresponding character. For a given set of measurement outcomes $\{ q_n \}_n$ (one for each unit cell), we thus obtain the postmeasurement state
\begin{equation} \label{eq:psipost}
\ket{\psi}_{\{ q_n \}} \propto \prod_n P_G(q_n)_n \ket{\psi}.
\end{equation}
The probability of obtaining a given measurement outcome (and thus the corresponding postmeasurement state) is, of course, given by Born's rule. For each given outcome, one can ask whether the postmeasurement state is long-range entangled. We generally expect that this is indeed the case. For concreteness, we will consider the one-dimensional case, although many of the arguments have higher-dimensional analogs. (We will discuss higher-dimensional examples in Sec.~\ref{subsecmeasuringgeneralspt}.)

\begin{shaded}
  \textbf{Conjecture.} If the premeasurement state $\ket{\psi}$ has a conventional SPT string order parameter\footnotemark[11] for a mixed Abelian $G \times H$ SPT phase, then the probability of the postmeasurement state being long-range entangled is unity.
\end{shaded}
\footnotetext[11]{\label{footnote:sop}In other words, the SPT wave function has long-range order in $\langle \mathcal O_m^\dagger U_{m+1} U_{m+2} \cdots U_{n-1} \mathcal O_n\rangle \neq 0$ for a certain $U \in G$ and for a particular choice of end-point operator $\mathcal O$ that is supported on a single unit cell, or at the very least, that commutes with $G$ \emph{in each individual unit cell}. The nontrivial (mixed) SPT class implies that $\mathcal O$ will carry nontrivial charge under $H$. Ref.~\onlinecite{Pollmann12} proved there always exists an $\mathcal O$ that gives long-range order, although it does not guarantee the additional local properties.}
\stepcounter{footnote} 

We will give plausibility arguments for this conjecture in the next subsection. The above claim of unit probability allows for a `measure zero' case where the postmeasurement state can be short-range entangled. Indeed, we will see examples of this in our numerical exploration in Sec.~\ref{subsec:spin1}. However, if we slightly strengthen our assumptions, we can \emph{prove} one \emph{always} obtains long-range entanglement:

\begin{shaded}
  \textbf{Theorem.} Let $\ket{\psi}$ be in a nontrivial mixed SPT phase for Abelian symmetry group $G \times H$. If it admits a finite-bond dimension matrix product state (MPS) description, then there exists a choice of unit cell such that measuring the $G$-charge for each unit cell produces a state with long-range entanglement for \emph{any} measurement outcome. More precisely, the postmeasurement state is a cat state for the (partial) spontaneous symmetry breaking of $H$.
\end{shaded}

To phrase and prove this result, we use the notion of matrix product states (MPS). In fact, this same framework will provide an intuitive justification for our more general conjecture. We thus turn to an MPS-based description of our set-up.

\subsubsection{Proof using matrix product states \label{subsec:MPS}}

For a review of MPS, we point the reader to Refs.~\onlinecite{Cirac2021} or \onlinecite{Hauschild18}. The key idea of MPS is that a wave function is written in terms of finite-dimensional tensors:
\begin{equation}
\ket{\psi} = \sum_{i_1,i_2,\cdots,i_N} \textrm{tr}\left( \prod_{n=1}^N A_{i_1} \right) \ket{i_1,i_2,\cdots,i_N} \label{eq:psiMPS}
\end{equation}
where $N$ labels the number of unit cells, $i=1,\cdots,d$ labels the states in each unit cell, and $A_i$ is a $\chi \times \chi$ matrix. (For convenience, we work with translation-invariant states, where the tensor is identical for all sites.) Here $\chi \in \mathbb N$ is called the bond dimension, with $\chi=1$ corresponding to a product-state wave function. It is known that up to exponentially small errors in local quantities, ground states of gapped Hamiltonians are well approximated by such an MPS \cite{MPSgs,MPSgs2}. In what follows, we will use the graphical notation. For instance, Eq.~\eqref{eq:psiMPS} becomes
\begin{equation}
\ket{\psi} = \raisebox{-.6\height}{
\begin{tikzpicture}
    \draw[dashed] (-1.15,0) -- (-0.5,0);
    \draw[dashed] (3.5,0) -- (4.15,0);
    \draw[-] (-0.5,0) -- (3.5,0);
    \node[R] (R1) at (0, 0) {$A$};
    \node[R] (R2) at (1, 0) {$A$};
    \node[R] (R3) at (2, 0) {$A$};
    \node[R] (R4) at (3, 0) {$A$};
    \draw[-] (R1) -- (0, -0.7);
    \draw[-] (R2) -- (1, -0.7);
    \draw[-] (R3) -- (2, -0.7);
    \draw[-] (R4) -- (3, -0.7);
\end{tikzpicture}
}
\end{equation}
where we ignore boundary conditions, or equivalently, we work in the thermodynamic limit.

A key property that makes MPS such a useful framework, is that global symmetries, such as $U = \prod_n U_n$, imply nice local properties on the MPS tensor. In particular, one can `push' physical symmetries through to the `virtual' level\footnote{The internal indices of the $A$ tensor that are contracted with one another in the wave function are commonly called virtual legs, to distinguish them from the physical legs labeling the spins.} \cite{SymFrac08,Pollmann10,ChenGuWen11A,Schuch11,Cirac2021}: There exists an operator $V_g$ such that
\begin{equation} \label{eq:pushthrough}
\raisebox{-0.82\height}{
\begin{tikzpicture}[inner sep=1mm]
    \node[R] (R) at (0, 0) {$A$};
    \node[U] (U) at (0, -1) {$U_g$};
    \draw[-] (R) -- (U) -- (0, -1.7);
    \draw[-] (R) -- (0.7,0);
    \draw[-] (R) -- (-0.7,0);
    \end{tikzpicture}}
=e^{i\theta_g}
\raisebox{-.6\height}{
    \begin{tikzpicture}[inner sep=1mm]
    \node[R] (R) at (0, 0) {$A$};
    \node[V] (Vr) at (1, 0) {$V_g^\dagger$};
    \node[V] (Vl) at (-1, 0) {$V_g$};
    \draw[-] (R) -- (0, -0.7);
    \draw[-] (R) -- (Vr)-- (1.7,0);
    \draw[-] (R) -- (Vl)-- (-1.7,0);
    \end{tikzpicture}
    }
\end{equation}
In other words, we see that the physical operator $U_g$ is equivalent to acting with $V_g$ and $V_g^\dagger$ at the virtual level. As a sanity check, we indeed see that if we apply $U_g$ on each site, then each $V_g$ is canceled by a $V_g^\dagger$, thereby confirming $\prod_n \left( U_g \right)_n$ is a global symmetry of $\ket{\psi}$.

An interesting property of these virtual symmetry actions $V_g$ is that they only need to form a \emph{projective} representation of the symmetry group. Thus, for any $g,g' \in G \times H$, we have $V_g V_{g'} =  \omega(g,g') V_{gg'}$ with a potentially nontrivial phase factor $\omega(g,g') \in U(1)$. A nontrivial SPT class is then equivalent to the statement that $[\omega] \in H^2(G \times H,U(1))$ is a nontrivial cocycle; the simplest example is when $G \times H = \mathbb Z_2 \times \mathbb Z_2$, where the nontrivial SPT phase corresponds to the projective representation where the two generators anticommute. More generally, a mixed SPT class implies that $\omega(g,h) \neq \omega(h,g)$ for a certain choice of $g\in G$ and $h\in H$, which we will use to derive long-range entanglement in the postmeasurement state.

As discussed, the act of measurement corresponds to applying a projector \eqref{eq:PG}. The MPS tensor for the postmeasurement state \eqref{eq:psipost} is simply:
\begin{equation} \label{eq:MPS_B}
\raisebox{-.6\height}{
\begin{tikzpicture}[inner sep=1mm]
    \node[R] (R) at (0, 0) {$B$};
    \draw[-] (R) -- (0, -0.7);
    \draw[-] (R) -- (0.7,0);
    \draw[-] (R) -- (-0.7,0);
    \end{tikzpicture}}
:=
\raisebox{-.8\height}{
    \begin{tikzpicture}[inner sep=1mm]
    \node[R] (R) at (0, 0) {$A$};
    \node[U] (U) at (0, -1) {$P_G$};
    \draw[-] (R) -- (U) -- (0, -1.7);
    \draw[-] (R) -- (0.7,0);
    \draw[-] (R) -- (-0.7,0);
    \end{tikzpicture}}
\end{equation}
Since for any $g \in G$ we have $U_g P_G = \chi_q(g) P_G$ (i.e., the symmetry \emph{operator} acts like a \emph{number}) we thus have the following local tensor properties:
\begin{align} 
\raisebox{-.6\height}{
    \begin{tikzpicture}[inner sep=1mm]
    \node[R] (R) at (0, 0) {$B$};
    \draw[-] (R) -- (U) -- (0, -0.7);
    \draw[-] (R) -- (0.7,0);
    \draw[-] (R) -- (-0.7,0);
    \end{tikzpicture}}
&=e^{i\theta_g} \chi_q(g)&
\raisebox{-.54\height}{
    \begin{tikzpicture}[inner sep=1mm]
    \node[R] (R) at (0, 0) {$B$};
    \node[V] (Vr) at (1, 0) {$V_g^\dagger$};
    \node[V] (Vl) at (-1, 0) {$V_g$};
    \draw[-] (R) -- (0, -0.7);
    \draw[-] (R) -- (Vr)-- (1.7,0);
    \draw[-] (R) -- (Vl)-- (-1.7,0);
    \end{tikzpicture}},
\label{eq:B_G} \\
\raisebox{-0.8\height}{
\begin{tikzpicture}[inner sep=1mm]
    \node[R] (R) at (0, 0) {$B$};
    \node[U] (U) at (0, -1) {$U_h$};
    \draw[-] (R) -- (U) -- (0, -1.7);
    \draw[-] (R) -- (0.7,0);
    \draw[-] (R) -- (-0.7,0);
    \end{tikzpicture}}
&=e^{i\theta_h}&
\raisebox{-.54\height}{
    \begin{tikzpicture}[inner sep=1mm]
    \node[R] (R) at (0, 0) {$B$};
    \node[V] (Vr) at (1, 0) {$V_h^\dagger$};
    \node[V] (Vl) at (-1, 0) {$V_h$};
    \draw[-] (R) -- (0, -0.7);
    \draw[-] (R) -- (Vr)-- (1.7,0);
    \draw[-] (R) -- (Vl)-- (-1.7,0);
    \end{tikzpicture}},
 \label{eq:B_H}
\end{align}
for $g \in G$ and $h \in H$. Eq.~\eqref{eq:B_H} tells us that $H$ still acts like a physical symmetry on the postmeasurement state; however, Eq.~\eqref{eq:B_G} tells us that $G$ now only acts on the virtual degrees of freedom, which we can interpret as a sort of higher symmetry. More concretely, as we will now argue, $V_g$ acts as an order parameter for the spontaneous breaking of $H$ symmetry, such that the postmeasurement state is a long-range entangled cat state for symmetry breaking.

The key identity we will need is the ability to push $V_g$ from the virtual level to the physical level. In particular, the question is whether there exists an operator $\mathcal O_g$ such that
\begin{equation} \label{eq:Og}
\boxed{ \raisebox{-.8\height}{
    \begin{tikzpicture}[inner sep=1mm]
    \node[R] (R) at (0, 0) {$B$};
    \node[U] (U) at (0, -1) {$\mathcal O_g$};
    \draw[-] (R) -- (U) -- (0, -1.7);
    \draw[-] (R) -- (0.7,0);
    \draw[-] (R) -- (-0.7,0);
    \end{tikzpicture}
    }
    \overset{?}{=}
 \raisebox{-.6\height}{
    \begin{tikzpicture}[inner sep=1mm]
    \node[R] (R) at (0, 0) {$B$};
    \node[V] (Vl) at (-1, 0) {$V_g$};
    \draw[-] (R) -- (0, -0.7);
    \draw[-] (R) -- (0.7,0);
    \draw[-] (R) -- (Vl)-- (-1.7,0);
    \end{tikzpicture}
    }
    }
\end{equation}
Let us temporarily earmark the question of \emph{whether} $\mathcal O_g$ exists and first explain how its existence is sufficient to prove that the postmeasurement state is long-range entangled.

From the projective group relations $V_g V_{g'} = \frac{\omega(g,g')}{\omega(g',g)} V_{g'} V_g$, one can straightforwardly prove that if $\mathcal O_g$ exists, it must carry charge under $H$. In particular, in Appendix~\ref{app:symcharge} we prove that Eq.~\eqref{eq:Og} implies
\begin{equation} \label{eq:Og_charge}
U_h \mathcal O_g U_h^\dagger = \underbrace{\omega(g,h) \overline{ \omega(h,g)}}_{\equiv \alpha_{g,h}} \; \mathcal O_g.
\end{equation}
Since we are considering a mixed SPT phase, we know that this phase factor is nontrivial for certain $g \in G$ and $h \in H$; let us henceforth fix those elements, such that $\alpha_{g,h}  \neq 1$.

One consequence of Eq.~\eqref{eq:Og} is that in the postmeasurement state, the expectation value of $\mathcal O_g$ must vanish. Indeed, taking the expectation value of both sides of Eq.~\eqref{eq:Og_charge} and using that $U_h$ is a symmetry, we obtain
\begin{equation}
\langle \mathcal O_g \rangle_{\textrm{postmeas}} = \alpha_{g,h} \langle \mathcal O_g \rangle_{\textrm{postmeas}}.
\end{equation}
Since $\alpha_{g,h} \neq 1$, this implies that $\langle \mathcal O_g \rangle_{\textrm{postmeas}} = 0$. However, the \emph{two-point} correlation is nonzero. Indeed, combining Eq.~\eqref{eq:Og} with Eq.~\eqref{eq:B_G} directly implies that
\begin{equation}
\left|\left\langle \left( \mathcal O_g\right)_m^\dagger \left( \mathcal O_g\right)_n \right\rangle_{\textrm{postmeas}} \right| = 1.
\end{equation}
We thus have long-range mutual information and thus long-range entanglement. In more physical terms, we see that the postmeasurement state can be interpreted as a cat state for the (partial) spontaneous symmetry breaking of $H$.

We have thus proven that the existence of $\mathcal O_g$, as defined in Eq.~\eqref{eq:Og}, is sufficient to prove long-range entanglement. The final issue is when we expect this to hold. One scenario where we can show that $\mathcal O_g$ exists is when the conditions of the theorem in Sec.~\ref{subsec:statement} are met. Indeed, it is known that short-range entangled MPS satisfy a certain injectivity condition \cite{Cirac2021} which means that after potentially blocking sites a \emph{finite} number of times, the MPS tensor defines an injective map where we consider the virtual legs to be its input and the physical leg its output. Equivalently, there exists a tensor\footnote{If the wave function has zero correlation length, then $C$ is simply the complex conjugate of $A$. However, $C$ exists even for nonzero correlation length.} $C$ that functions as an inverse for $A$:
\begin{equation} \label{eq:inj}
\raisebox{-.4\height}{
    \begin{tikzpicture}[inner sep=1mm]
    \node[R] (R) at (0, 0) {$A$};
    \node[R] (Rb) at (0, -1) {$C$};
    \draw[-] (R) -- (Rb);
    \draw[-] (R) -- (0.7,0);
    \draw[-] (R) -- (-0.7,0);
    \draw[-] (Rb) -- (0.7,-1);
    \draw[-] (Rb) -- (-0.7,-1);
    \end{tikzpicture}
    }
=
\raisebox{-.4\height}{
    \begin{tikzpicture}[inner sep=1mm]
    \draw (-0.3,0) -- (0,0) to [in=0, out=0] (0,-1) -- (-0.3,-1);
    \draw (1.3,0) -- (1,0) to [in=180, out=180] (1,-1) -- (1.3,-1);
    \end{tikzpicture}
    }
\end{equation}
where we will henceforth presume one has blocked the unit cell to achieve the injectivity condition.
Using this, we can define the physical operator $\mathcal O_g$ as follows:
\begin{equation}
\raisebox{-.45\height}{
    \begin{tikzpicture}[inner sep=1mm]
    \node[U] (U) at (0,0) {$\mathcal O_g$};
    \draw[-] (0,1) -- (U) -- (0, -1);
    \end{tikzpicture}}
\equiv
\raisebox{-.45\height}{
    \begin{tikzpicture}[inner sep=1mm]
    \node[R] (C) at (0, 0) {$C$};
    \node[R] (A) at (0, -1) {$A$};
    \node[V] (V) at (-1, -1) {$V_g$};
    \draw[-] (V) -- (A) -- (0.5,-1) to[in=0, out=0] (0.5,0) -- (C) -- (-1.6,0) to[in=180,out=180] (-1.6,-1) -- (V);
    \draw[-] (C) -- (0, 0.7);
    \draw[-] (A) -- (0, -1.7);
    \end{tikzpicture}
    }
\end{equation}
Using Eq.~\eqref{eq:inj}, one sees that this operator satisfies Eq.~\eqref{eq:Og} for the $A$ tensor. Moreover, one can prove that $\mathcal O_g$ commutes with the $G$ symmetry, i.e., $U_g \mathcal O_g U_g^\dagger = \mathcal O_g$ for any $g \in G$ (see Appendix~\ref{app:symcharge}). Hence, $\mathcal O_g$ commutes with the projection $P_G$, such that we obtain Eq.~\eqref{eq:Og} also for the $B$ tensor. This concludes the proof of the theorem in Sec.~\ref{subsec:statement}.

Thus, if we are willing to block unit cells a finite\footnote{We emphasize the \emph{finiteness} since if one is willing to block an \emph{unbounded} number of times, we can effectively appeal to an RG-based argument whereby one flows to the fixed-point state with zero correlation length, which would be less interesting.} number of times, we can prove that LRE is obtained for \emph{any} measurement outcome. In the absence of such blocking, we believe one can only make a probabilistic statement. In fact, while we do not offer a proof of the conjecture stated in Sec.~\ref{subsec:statement}, the above MPS arguments provide an intuitive justification. To see this case, let us first remark that to make probabilistic arguments, one only needs a weaker version of Eq.~\eqref{eq:Og}, namely, that there exists an $\mathcal O_g$ such that one has finite overlap with the virtual $V_g$ action, i.e.,
\begin{equation} \label{eq:Ogweak}
\raisebox{-.79\height}{
    \begin{tikzpicture}[inner sep=1mm]
    \node[R] (R) at (0, 0) {$B$};
    \node[U] (U) at (0, -1) {$\mathcal O_g$};
    \draw[-] (R) -- (U) -- (0, -1.7);
    \draw[-] (R) -- (0.7,0);
    \draw[-] (R) -- (-0.7,0);
    \end{tikzpicture}
    }
= \lambda
\raisebox{-.56\height}{
    \begin{tikzpicture}[inner sep=1mm]
    \node[R] (R) at (0, 0) {$B$};
    \node[V] (Vl) at (-1, 0) {$V_g$};
    \draw[-] (R) -- (0, -0.7);
    \draw[-] (R) -- (0.7,0);
    \draw[-] (R) -- (Vl)-- (-1.7,0);
    \end{tikzpicture}}
    + \cdots
\end{equation}
for some $\lambda \neq 0$. Indeed, one can again show that this implies $\mathcal O_g$ carries nontrivial charge under $H$. Moreover, the same argument as above still implies that one expects $\mathcal O_g$ to have a long-range two-point function, since it picks up on the long-range order of $V_g$ (see Eq.~\eqref{eq:B_G}). The only way this case can fail is if the \emph{multiple} terms on the right-hand side of Eq.~\eqref{eq:Ogweak} conspire to exactly \emph{cancel out} the long-range contributions, which this certainly \emph{can} happen (we will give an example in the next subsection); however this requires a delicate balancing of terms and is thus likely a \emph{measure zero} case over the ensemble of all possible measurement outcomes. Lastly, we note that Eq.~\ref{eq:Ogweak} can be expected to hold for SPT phases which admit a conventional SPT order parameter, as defined in footnote \ref{footnote:sop}. Indeed, the very reason the string order parameters have nontrivial end-point operators is because they are able to cancel out the virtual $V_g$ action of the symmetry string or disorder operator \cite{Pollmann12}. Commonly used string order operators have an end-point $\mathcal O_g$ supported on a single unit cell and commute with the corresponding symmetry generator $U_g$, such that if Eq.~\eqref{eq:Ogweak} applies to the $A$ tensor it also automatically carriers over to the postmeasurement $B$ tensor. In conclusion, for these reasons, we conjecture that only a measure zero of measurement outcomes can fail to give long-range entanglement. It would be interesting to sharpen this intuition into a rigorous proof of our conjecture.

\subsubsection{Analytics: Cat state from the deformed cluster state and AKLT state \label{subsec:examples}}

Let us illustrate our general theorem with two MPS-based examples. Both examples will be SPT phases with nonzero correlation length, i.e., away from the simple fixed-point cases studied in the earlier sections of this work.

First, we consider a deformation of the cluster state:
\begin{equation}
\ket{\psi(\beta)} \propto e^{ \beta \sum_n X_n} \ket{\textrm{cluster}}.
\end{equation}
Here $\ket{\psi(0)}$ is the cluster state of Eq.~\eqref{eq:clusterstateentangler}. For any $\beta$, this state admits a $\chi=2$ MPS representation \cite{Wolf06} and one can show that for any finite $\beta$, this state is in the nontrivial SPT phase protected by $\mathbb Z_2 \times \mathbb Z_2$ symmetry. Its correlation length $\xi$ increases monotonically with $\beta$ and diverges as $\beta \to \infty$. The MPS tensor turns out to be injective without blocking, meaning that our theorem implies that measuring, say, $X_{2n+1}$ on odd sites produces a long-range entangled state on the remaining qubits---for any possible measurement outcome.

As a second example, we consider the paradigmatic spin-1 AKLT state \cite{AKLT1987}, which is known to be described by a $\chi=2$ MPS and is an SPT phase protected by the $\mathbb Z_2 \times \mathbb Z_2$ symmetry of $\pi$-rotations. As generators, we can choose $R^x = \prod_n e^{i \pi S^x_n}$ and $R^z = \prod_n e^{i \pi S^z_n}$. If we block the spin-1s into two-site unit cells, then the MPS satisfies the aforementioned injectivity property. Hence, our MPS-based arguments prove that if one measures, say, $R^z_{2n-1} R^z_{2n} \in \{-1,1\}$ charge on each two-site unit cell, then the postmeasurement state will always have long-range entanglement.

What if we did not block in the last example? If we measure $R^z_n \in \{-1,1\}$ in each single-site unit cell, then there is a measure-zero chance that we obtain $R^z_n = 1$ for all sites. In this case, the postmeasurement state is simply the product state $\ket{0}^N$, where $\ket{0}$ is the unique $+1$ eigenstate of $R^z= e^{i\pi S^z}$. However, as long as a finite density of sites projects onto the $-1$ eigenstate of $R^z$, the postmeasurement state is a long-range entangled of GHZ type, capturing the spontaneous symmetry-breaking of $R^x$. This example is thus consistent with our conjecture and it illustrates the importance of making probabilistic statements in the cases where one does not block unit cells.

While both examples are illustrative, by definition they are analytically tractable. One might wonder about SPT phases of ground states that are not exactly solvable. For this reason, we now turn to a numerical exploration.

\subsubsection{Numerics: Cat state from the spin-1 Heisenberg chain \label{subsec:spin1}}

To emphasize the generality of our claim that SPT phases can be used to generated LRE upon measurement, we consider the incarnation of the Haldane SPT phase in the spin-1 Heisenberg chain. Its Hamiltonian is a just nearest-neighbor antiferromagnetic coupling:
\begin{equation}
H = \sum_n \bm S_{n} \bm{\cdot S}_{n+1}. \label{eq:spin1}
\end{equation}
This spin chain is known to be gapped \cite{Haldane1983}, forming a nontrivial SPT phase for the $\mathbb Z_2 \times \mathbb Z_2$ group of $\pi$-rotations generated by $R^\gamma = \prod_n e^{i \pi S^\gamma_n}$ with $\gamma=x,y,z$ \cite{AKLT1987,dennijs89,Gu09,pollmann_entanglement_2010}. Indeed, it has been argued to be in the same phase as the tractable AKLT state encountered in the previous section \cite{AKLT1987}.

By our general proposal, we expect that measuring, say, the $R^z$ charge for every site, should result in a cat state for the remaining $\mathbb Z_2$ symmetry. An interesting difference from the cluster chain is that the symmetries do not act on distinct sites. We thus measure $R^z_n = e^{i \pi S^z_n}$ on \emph{every single} site. Effectively, this process comes down to measuring whether $\left( S^z_n \right)^2$ is 0 or 1. For the first outcome, the site has no degree of freedom left, whereas for the latter, we still have a remaining qubit ($S^z_n = \pm 1$) which is toggled by $R^x$.
Hence, with the exception of there being no qubits left (which is of measure zero in the thermodynamic limit), we expect a cat state for the remaining chain of qubits. This is similar to the AKLT discussion in Sec.~\ref{subsec:examples}, although now we cannot rely on an exact solution.

\begin{figure}[t!]
    \centering
    \includegraphics[scale=0.35]{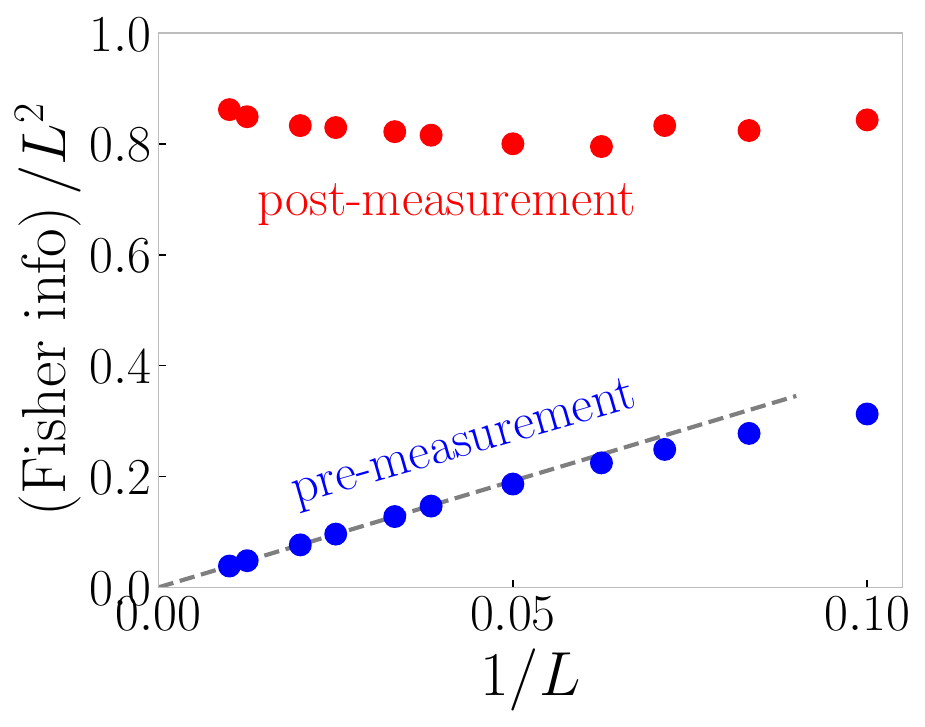}
    \caption{\textbf{Cat state from measuring the Haldane SPT phase.} We consider the ground state of the spin-$1$ Heisenberg chain, which is in a nontrivial SPT phase for the $\mathbb Z_2\times \mathbb Z_2$ symmetry of $\pi$-rotations. In accordance with its short-range entanglement, we find that the Fisher information scales linearly with system size (blue dots). In contrast, if we measure the $R^z_n=e^{i \pi S^z_n}$-charge on every site, the remaining state has Fisher information $F \sim L^2$ (red dots), signaling long-range entanglement in the post-meaurement state (here we have chosen different random measurement outcomes for each $L$). This finding confirms that measuring one $\mathbb Z_2$ symmetry of the Haldane SPT phase creates a cat state for the remaining $\mathbb Z_2$ symmetry, even if one is not at a fine-tuned fixed-point limit.}
    \label{fig:spin1}
\end{figure}

To test this prediction, we numerically obtain the ground state of Eq.~\eqref{eq:spin1} using the density matrix renormalization group (DMRG) \cite{White92,White93,Hauschild18} for a variable system size $L$ with periodic boundary conditions. We then project each site into $\left( S^z_n \right)^2 = 0$ with probability $1/3$ or $\left( S^z_n \right)^2 = 1$ with probability $2/3$. As a robust way of detecting whether the resulting state is a cat state, we calculate the Fisher information, which in this case is simply the variance of the total (staggered) magnetization:
\begin{equation}
F = \left\langle \left( \sum_{n=1}^L  (-1)^n S^z_n \right)^2 \right\rangle -   \left\langle\sum_{n=1}^L (-1)^n   S^z_n  \right\rangle^2.
\end{equation}
This Fisher information is a quantitative measure for the use of the state for quantum metrology purposes \cite{Fisher1925,Giovannetti06}. While SRE states obey a scaling $F \sim L$, only nonlocal cat states have $F \sim L^2$. Our numerical results\footnote{We went up to system sizes of $L=100$, where we found that $\chi \approx 500$ was sufficient to guarantee convergence of the Fisher information.} are shown in Fig.~\ref{fig:spin1}. While the original ground state has $F\sim L$, we find that the postmeasurement state indeed has $F\sim L^2$, confirming that it is a cat state. In addition, it is interesting to see that $F(L)$ varies relatively continuously with $L$, despite each system size having a completely random measurement outcome (each red dot is computed for only a single measurement shot).

The above emergence of a cat state can actually be linked to the original interpretation of the Haldane SPT phase. Indeed, when the topological string order parameter was first introduced in 1989 \cite{dennijs89}, it was designed to pick up the `hidden symmetry-breaking' of the state, where it was observed that if one imagines removing all $S^z_n =0$ states, then the remaining $S^z_n = \pm 1$ states form long-range N\'eel order. However, since the $S^z_n =0$ states are interspersed within the $S^z_n = \pm 1$ states and are allowed to have quantum fluctuations, they disorder this local order (which can now only be picked up with a string order parameter). Our above procedure can be interpreted as making this hidden order manifest: The measurement pins the location of $S^z_n= 0$, preventing them from disordering the N\'eel state.

\subsection{Measuring general SPT phases}\label{subsecmeasuringgeneralspt}

Here, we discuss how LRE arises upon measuring more general SPT states, even beyond 1D. As a natural starting point, we consider one of the simplest SPT phases (beyond 1D) which are protected by more than a single cyclic group---such that it is meaningful to measure one symmetry and preserve the other. Let us thus consider the $\ZZ_2^3$ ``cubic" SPT in 2+1D. One model for this phase \cite{yoshida_topological_2015,Yoshida2016} is given by placing spins on the sites of a triangular lattice, with each $\ZZ_2$ acting as $\prod_{j \in A,B,C} X_j$ on each of three triangular sublattices $A,B,C$. For each site $j$, there is a stabilizer given by 
\begin{equation}
S_j = X_j \prod_{\langle jqq' \rangle} CZ_{q,q'},
\end{equation}
where the product is over triangles $\langle jqq' \rangle$ with vertices $j,q,q'$. When we measure $X_j$ on the $A$ sublattice, we are left with a state on a honeycomb sublattice with
\begin{equation}
\prod_{\langle jqq' \rangle} CZ_{q,q'} = (-1)^{s_j}
\label{eq:CZring}
\end{equation} around each hexagon, for some fixed signs (determined by our measurement outcome $s_j$).

The loop operators $\prod_{ij \in \gamma} CZ_{i,j}$ along a closed path $\gamma$ of vertices can be considered as a $\ZZ_2$ 1-form symmetry of this state. Note that this acts as the cluster SPT entangler for $\ZZ_2^{B,C}$ along $\gamma$, which implies there is a mixed anomaly; therefore, the resulting state obtained from measurement cannot be short-range entangled. Note that this anomaly can be realized on the boundary of a lattice model of a 3D SPT protected by $\ZZ_2^2 \times \ZZ_2[1]$ as studied in Ref.~\onlinecite{Yoshida2016}.

We believe that a similar conclusion holds generally when we measure SPT states, at least when the corresponding topological term is linear in the gauge field associated with the measured charge.
Let $G$ and $H$ be $(p-1)$- and $(q-1)$-form symmetries where $G$ and $H$ are onsite symmetries that act only on subsystems $A$ and $B$ respectively. Denote the background gauge fields of $G$ and $H$, $A_p$ and $B_q$, respectively. Now, consider an SPT associated with the cohomology class $A_p F(B_q) \in H^{d+1}(G \times H, U(1))$, where $d$ is the space dimension and $F(B_q)\in H^{d+1-p}( H,G^\star)$  describes a topological $G$ current made from $B_q$ where $G^\star = {\rm Hom}(G,U(1))$. Physically, $F(B_q)$ can be understood as an $H$ SPT in $d-p+1$ spatial dimensions, and the SPT $A_p F(B_q)$ corresponds to decorating fluctuating $G$-domain walls with this $H$ SPT\cite{decorateddomainwalls}.

\begin{figure}
    \centering
    \includegraphics[scale=0.6]{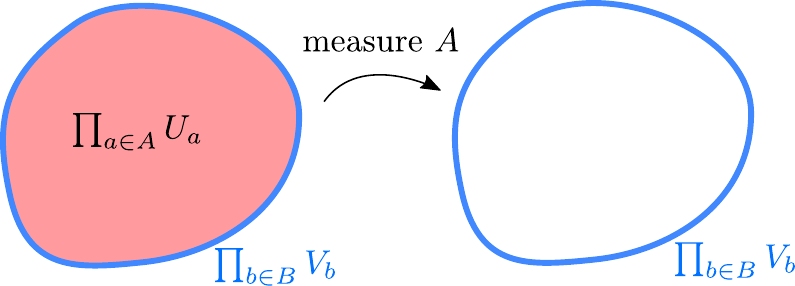}
    \caption{\textbf{Anomalous symmetry from measuring an SPT phase.} In an SPT phase, applying the symmetry in a region is equivalent to applying a unitary operator just near the boundary of that region; equivalently, the membrane operator has long-range order if we include the appropriate unitary operator along its boundary. In the $G \times H$ SPT fixed point models of the linear form $A_p F(B_q)$, $G$ acts only on the $A$ sublattice and the boundary operator acts only on the $B$ sublattice. If we then measure the spins of the $A$ sublattice, this boundary operator remains as a symmetry, now locally defined along the boundary. Because the boundary is codimension $p$, this defines a $G$ $p$-form symmetry, which acts as the entangler for a nontrivial $H$ SPT phase. This implies that the $G$ $p$-form symmetry in the post-measured state has a mixed anomaly with $H$, implying that the state cannot be short-range entangled.}
    \label{fig:membrane}
\end{figure}
In this fixed-point model, if we now measure the $G$ charges, we essentially project out the topological current $F(B_q)$. Analogously to the $CZ$ ring in Eq.\eqref{eq:CZring}, we similarly obtain a $p$-form symmetry, the remnant of the $G$ symmetry action by symmetry fractionalization of the parent SPT phase before measurement---applying the $G$ symmetry in a region is equivalent to acting on the boundary of that region with the entangler of the $H$-SPT (see Fig. \ref{fig:membrane}).

 This anomaly can also be seen from studying the topological response of the $G \times H$ SPT. Projecting out $G$ charges is equivalent to making the $G$ gauge field $A_p$ dynamical. Measuring the $G$ charges can be thought of as making $A_p$ dynamical with a charge background fixed by the measurement outcome. Since we began with a gapped phase, there are no fluctuating $G$ charges at low energies. As a result, there is an emergent $p$-form symmetry that acts as $A_p \mapsto A_p + \lambda$, known as the center, or electric symmetry \cite{generalizedglobalsymmetries}. This symmetry is the same as the $p$-form symmetry we defined above. From the form of the topological response, assumed to be $A_p F(B_q)$, we see that this global symmetry is broken when there is a nontrivial $B_q$ since it produces a variation of the effective action, namely $\int \lambda F(B_q)$. This variation is characteristic of an anomaly associated with a $d+1$-dimensional topological response $\tilde A_{p+1} F(B_q)$ \cite{Komargodski_2019}, where $\tilde A_{p+1}$ is the background $p+1$-form gauge field (note the shift) associated with the center symmetry.

When the SPT class is not linear in $A_p$, we will not be able to fractionalize the $G$ symmetry so that the boundary operator commutes with the $G$ charges \cite{Else_2014}. However, if it is the form $F_1(A_p) F_2(B_q)$, where $F_1(A_p) \in H^{j}(G,K)$ and $F_2(B_q) \in H^{d+1-j}(G,K^*)$, for some Abelian group $K$, then there will be a codimension $j + 1$ defect Poincar\'e dual to $d F_1(A_p)$ that can factorize, defining a $j+1$-form symmetry in the fixed point model postmeasurement corresponding to a field $C_{j+2}$. The anomaly will then be $C_{j+2} F_2(B_q) \in H^{d+2}(K[j+1] \times H,U(1))$. For example, if we measure both $\ZZ_2^{(1)}$ and $\ZZ_2^{(2)}$ in the cubic SPT, then in this case, we identify  $A_p = (A_1^{(1)},A_1^{(2)})$, $B_q = A_1^{(3)}$,  $F(A_p) = \frac{1}{2} A_{1}^{(1)} A_{1}^{(2)}$ and $F(B_q) =  A_{1}^{(3)}$. Thus, the anomaly postmeasurement is  $\frac{1}{2} C_{3} A^{(3)}_1$ for a $\ZZ_2$ 2-form symmetry associated with $C_3$.

\section{Outlook \label{sec:outlook}}

In this work, we have presented a general framework for which performing measurements of short-range entangled states produces long-range entanglement. We have given some intuitive arguments that this is a stable property of the SPT phase, as well as proven that this always holds if the measurements are performed in an appropriately large enough unit cell. We would also like to determine the nature of the long-range entangled states which appear.

We have also described how non-local transformations including Kramers-Wannier and Jordan-Wigner arise from coupling an arbitrary state with a symmetry to a cluster-like SPT and performing measurements. It would be interesting to see whether other SPTs define useful transformations this way. If so, what family of MPOs do they define? We note that, given a general MPO, it is not obvious how to implement it from finite-depth unitaries and measurements.

Sequential applications of our procedure even lead to non-Abelian topological order, including quantum doubles for solvable groups. A natural question is to find an analogue for nonsolvable groups---or to prove a no-go theorem. We also argued that non-Abelian states beyond quantum doubles can be obtained, such as the doubled Ising anyon theory, although we have left an explicit prescription of a circuit to future work.

Another feature of our method is that it can be performed in an arbitrary region, producing a duality defect on its boundary. We expect this defect to be topological \cite{frohlichKW,aasen_topological_2016,tdl}. It might even be natural to consider moving it by measurements?

It is also interesting to note the similarities to quantum teleportation \cite{Teleportation} and measurement-based quantum computation (MBQC) \cite{GottesmanChuang1999,BriegelRaussendorf2001,RaussendorfBriegel2001,RaussendorfBrowneBriegel2003}, where measurement effectively performs unitary operations on the input state. Here, the act of measurement instead performs a non-local transformation on the initial state. It would be interesting to make contact with similar notions of ``computational phases of matter" in MBQC \cite{Raussendorfetal2017,Raussendorfetal2018,Stephenetal2017,DevakulWilliamson2018,DanielAlexanderMiyake20}
 . Exploring connections to the topological bootstrap \cite{LuHsieh} is also a promising future direction. 

It may also be interesting to ``soften" the projectors, considering either weak measurements or an open system weakly interacting with the environment by a subset of its degrees of freedom.

\vspace{10pt}



\emph{Note added}---While finalizing the preprint of the present manuscript, Ref.~\cite{Ashkenazi21} appeared, which overlaps with our section of KW duality. Moreover, after our preprint appeared, we learned of a parallel work preparing quantum double topological order via measurements \cite{Bravyi22}. Our results agree with both of these works where they intersect.

\begin{acknowledgments}
We thank David T. Stephen for insightful observations about measurement-based preparation of non-Abelian topological order. DMRG simulations were performed using the TeNPy Library \cite{Hauschild18}. N.T. is supported by NSERC. R.V. and A.V. are supported by the Simons Collaboration on Ultra-Quantum Matter, which is a grant from the Simons Foundation (651440, A.V.). R.V. is supported by the Harvard Quantum Initiative Postdoctoral Fellowship in Science and Engineering.
\end{acknowledgments}

\bibliography{bib.bib}

\begin{thebibliography}{162}%
\makeatletter
\providecommand \@ifxundefined [1]{%
 \@ifx{#1\undefined}
}%
\providecommand \@ifnum [1]{%
 \ifnum #1\expandafter \@firstoftwo
 \else \expandafter \@secondoftwo
 \fi
}%
\providecommand \@ifx [1]{%
 \ifx #1\expandafter \@firstoftwo
 \else \expandafter \@secondoftwo
 \fi
}%
\providecommand \natexlab [1]{#1}%
\providecommand \enquote  [1]{``#1''}%
\providecommand \bibnamefont  [1]{#1}%
\providecommand \bibfnamefont [1]{#1}%
\providecommand \citenamefont [1]{#1}%
\providecommand \href@noop [0]{\@secondoftwo}%
\providecommand \href [0]{\begingroup \@sanitize@url \@href}%
\providecommand \@href[1]{\@@startlink{#1}\@@href}%
\providecommand \@@href[1]{\endgroup#1\@@endlink}%
\providecommand \@sanitize@url [0]{\catcode `\\12\catcode `\$12\catcode
  `\&12\catcode `\#12\catcode `\^12\catcode `\_12\catcode `\%12\relax}%
\providecommand \@@startlink[1]{}%
\providecommand \@@endlink[0]{}%
\providecommand \url  [0]{\begingroup\@sanitize@url \@url }%
\providecommand \@url [1]{\endgroup\@href {#1}{\urlprefix }}%
\providecommand \urlprefix  [0]{URL }%
\providecommand \Eprint [0]{\href }%
\providecommand \doibase [0]{https://doi.org/}%
\providecommand \selectlanguage [0]{\@gobble}%
\providecommand \bibinfo  [0]{\@secondoftwo}%
\providecommand \bibfield  [0]{\@secondoftwo}%
\providecommand \translation [1]{[#1]}%
\providecommand \BibitemOpen [0]{}%
\providecommand \bibitemStop [0]{}%
\providecommand \bibitemNoStop [0]{.\EOS\space}%
\providecommand \EOS [0]{\spacefactor3000\relax}%
\providecommand \BibitemShut  [1]{\csname bibitem#1\endcsname}%
\let\auto@bib@innerbib\@empty
\bibitem [{\citenamefont {Bravyi}\ \emph {et~al.}(2006)\citenamefont {Bravyi},
  \citenamefont {Hastings},\ and\ \citenamefont
  {Verstraete}}]{BravyiHastingsVerstraete06}%
  \BibitemOpen
  \bibfield  {author} {\bibinfo {author} {\bibfnamefont {S.}~\bibnamefont
  {Bravyi}}, \bibinfo {author} {\bibfnamefont {M.~B.}\ \bibnamefont
  {Hastings}},\ and\ \bibinfo {author} {\bibfnamefont {F.}~\bibnamefont
  {Verstraete}},\ }\bibfield  {title} {\bibinfo {title} {Lieb-robinson bounds
  and the generation of correlations and topological quantum order},\ }\href
  {https://doi.org/10.1103/PhysRevLett.97.050401} {\bibfield  {journal}
  {\bibinfo  {journal} {Phys. Rev. Lett.}\ }\textbf {\bibinfo {volume} {97}},\
  \bibinfo {pages} {050401} (\bibinfo {year} {2006})}\BibitemShut {NoStop}%
\bibitem [{\citenamefont {Hastings}(2012)}]{Hastings2010}%
  \BibitemOpen
  \bibfield  {author} {\bibinfo {author} {\bibfnamefont {M.~B.}\ \bibnamefont
  {Hastings}},\ }\bibfield  {title} {\bibinfo {title} {Locality in quantum
  systems},\ }in\ \href@noop {} {\emph {\bibinfo {booktitle} {Quantum Theory
  from Small to Large Scales}}},\ \bibinfo {series and number} {Les Houches
  2010, Session 95}\ (\bibinfo  {publisher} {Oxford University Press},\
  \bibinfo {year} {2012})\ pp.\ \bibinfo {pages} {171--212}\BibitemShut
  {NoStop}%
\bibitem [{\citenamefont {Chen}\ \emph
  {et~al.}(2011{\natexlab{a}})\citenamefont {Chen}, \citenamefont {Gu},\ and\
  \citenamefont {Wen}}]{ChenGuWen11A}%
  \BibitemOpen
  \bibfield  {author} {\bibinfo {author} {\bibfnamefont {X.}~\bibnamefont
  {Chen}}, \bibinfo {author} {\bibfnamefont {Z.-C.}\ \bibnamefont {Gu}},\ and\
  \bibinfo {author} {\bibfnamefont {X.-G.}\ \bibnamefont {Wen}},\ }\bibfield
  {title} {\bibinfo {title} {Classification of gapped symmetric phases in
  one-dimensional spin systems},\ }\href
  {https://doi.org/10.1103/PhysRevB.83.035107} {\bibfield  {journal} {\bibinfo
  {journal} {Phys. Rev. B}\ }\textbf {\bibinfo {volume} {83}},\ \bibinfo
  {pages} {035107} (\bibinfo {year} {2011}{\natexlab{a}})}\BibitemShut
  {NoStop}%
\bibitem [{\citenamefont {Chen}\ \emph
  {et~al.}(2011{\natexlab{b}})\citenamefont {Chen}, \citenamefont {Gu},\ and\
  \citenamefont {Wen}}]{ChenGuWen11B}%
  \BibitemOpen
  \bibfield  {author} {\bibinfo {author} {\bibfnamefont {X.}~\bibnamefont
  {Chen}}, \bibinfo {author} {\bibfnamefont {Z.-C.}\ \bibnamefont {Gu}},\ and\
  \bibinfo {author} {\bibfnamefont {X.-G.}\ \bibnamefont {Wen}},\ }\bibfield
  {title} {\bibinfo {title} {Complete classification of one-dimensional gapped
  quantum phases in interacting spin systems},\ }\href
  {https://doi.org/10.1103/PhysRevB.84.235128} {\bibfield  {journal} {\bibinfo
  {journal} {Phys. Rev. B}\ }\textbf {\bibinfo {volume} {84}},\ \bibinfo
  {pages} {235128} (\bibinfo {year} {2011}{\natexlab{b}})}\BibitemShut
  {NoStop}%
\bibitem [{\citenamefont {Zeng}\ and\ \citenamefont {Wen}(2015)}]{ZengWen15}%
  \BibitemOpen
  \bibfield  {author} {\bibinfo {author} {\bibfnamefont {B.}~\bibnamefont
  {Zeng}}\ and\ \bibinfo {author} {\bibfnamefont {X.-G.}\ \bibnamefont {Wen}},\
  }\bibfield  {title} {\bibinfo {title} {Gapped quantum liquids and topological
  order, stochastic local transformations and emergence of unitarity},\ }\href
  {https://doi.org/10.1103/PhysRevB.91.125121} {\bibfield  {journal} {\bibinfo
  {journal} {Phys. Rev. B}\ }\textbf {\bibinfo {volume} {91}},\ \bibinfo
  {pages} {125121} (\bibinfo {year} {2015})}\BibitemShut {NoStop}%
\bibitem [{\citenamefont {Huang}\ and\ \citenamefont
  {Chen}(2015)}]{HuangChen2015}%
  \BibitemOpen
  \bibfield  {author} {\bibinfo {author} {\bibfnamefont {Y.}~\bibnamefont
  {Huang}}\ and\ \bibinfo {author} {\bibfnamefont {X.}~\bibnamefont {Chen}},\
  }\bibfield  {title} {\bibinfo {title} {Quantum circuit complexity of
  one-dimensional topological phases},\ }\href
  {https://doi.org/10.1103/PhysRevB.91.195143} {\bibfield  {journal} {\bibinfo
  {journal} {Phys. Rev. B}\ }\textbf {\bibinfo {volume} {91}},\ \bibinfo
  {pages} {195143} (\bibinfo {year} {2015})}\BibitemShut {NoStop}%
\bibitem [{\citenamefont {Haah}(2016)}]{Haah2016}%
  \BibitemOpen
  \bibfield  {author} {\bibinfo {author} {\bibfnamefont {J.}~\bibnamefont
  {Haah}},\ }\bibfield  {title} {\bibinfo {title} {An invariant of
  topologically ordered states under local unitary transformations},\ }\href
  {https://doi.org/10.1007/s00220-016-2594-y} {\bibfield  {journal} {\bibinfo
  {journal} {Communications in Mathematical Physics}\ }\textbf {\bibinfo
  {volume} {342}},\ \bibinfo {pages} {771} (\bibinfo {year}
  {2016})}\BibitemShut {NoStop}%
\bibitem [{\citenamefont {Read}\ and\ \citenamefont {Sachdev}(1991)}]{Read91}%
  \BibitemOpen
  \bibfield  {author} {\bibinfo {author} {\bibfnamefont {N.}~\bibnamefont
  {Read}}\ and\ \bibinfo {author} {\bibfnamefont {S.}~\bibnamefont {Sachdev}},\
  }\bibfield  {title} {\bibinfo {title} {Large-n expansion for frustrated
  quantum antiferromagnets},\ }\href
  {https://doi.org/10.1103/PhysRevLett.66.1773} {\bibfield  {journal} {\bibinfo
   {journal} {Phys. Rev. Lett.}\ }\textbf {\bibinfo {volume} {66}},\ \bibinfo
  {pages} {1773} (\bibinfo {year} {1991})}\BibitemShut {NoStop}%
\bibitem [{\citenamefont {Wen}(1990)}]{Wen_90}%
  \BibitemOpen
  \bibfield  {author} {\bibinfo {author} {\bibfnamefont {X.-G.}\ \bibnamefont
  {Wen}},\ }\bibfield  {title} {\bibinfo {title} {Topological orders in rigid
  states},\ }\href {https://doi.org/10.1142/S0217979290000139} {\bibfield
  {journal} {\bibinfo  {journal} {International Journal of Modern Physics B}\
  }\textbf {\bibinfo {volume} {04}},\ \bibinfo {pages} {239} (\bibinfo {year}
  {1990})}\BibitemShut {NoStop}%
\bibitem [{\citenamefont {Wen}(2004)}]{Wenbook}%
  \BibitemOpen
  \bibfield  {author} {\bibinfo {author} {\bibfnamefont {X.}~\bibnamefont
  {Wen}},\ }\href
  {https://global.oup.com/academic/product/quantum-field-theory-of-many-body-systems-9780198530947?cc=de&lang=en&}
  {\emph {\bibinfo {title} {Quantum Field Theory of Many-body Systems}}},\
  Oxford graduate texts\ (\bibinfo  {publisher} {Oxford University Press},\
  \bibinfo {year} {2004})\BibitemShut {NoStop}%
\bibitem [{\citenamefont {Fuchs}\ \emph {et~al.}(2002)\citenamefont {Fuchs},
  \citenamefont {Runkel},\ and\ \citenamefont {Schweigert}}]{Fuchs02}%
  \BibitemOpen
  \bibfield  {author} {\bibinfo {author} {\bibfnamefont {J.}~\bibnamefont
  {Fuchs}}, \bibinfo {author} {\bibfnamefont {I.}~\bibnamefont {Runkel}},\ and\
  \bibinfo {author} {\bibfnamefont {C.}~\bibnamefont {Schweigert}},\ }\bibfield
   {title} {\bibinfo {title} {Tft construction of rcft correlators i: partition
  functions},\ }\href {https://doi.org/10.1016/s0550-3213(02)00744-7}
  {\bibfield  {journal} {\bibinfo  {journal} {Nuclear Physics B}\ }\textbf
  {\bibinfo {volume} {646}},\ \bibinfo {pages} {353–497} (\bibinfo {year}
  {2002})}\BibitemShut {NoStop}%
\bibitem [{\citenamefont {Kitaev}(2003)}]{Kitaev_2003}%
  \BibitemOpen
  \bibfield  {author} {\bibinfo {author} {\bibfnamefont {A.}~\bibnamefont
  {Kitaev}},\ }\bibfield  {title} {\bibinfo {title} {Fault-tolerant quantum
  computation by anyons},\ }\href
  {https://doi.org/10.1016/s0003-4916(02)00018-0} {\bibfield  {journal}
  {\bibinfo  {journal} {Annals of Physics}\ }\textbf {\bibinfo {volume}
  {303}},\ \bibinfo {pages} {2–30} (\bibinfo {year} {2003})}\BibitemShut
  {NoStop}%
\bibitem [{\citenamefont {Kitaev}(2006)}]{kitaev_anyons_2005}%
  \BibitemOpen
  \bibfield  {author} {\bibinfo {author} {\bibfnamefont {A.}~\bibnamefont
  {Kitaev}},\ }\bibfield  {title} {\bibinfo {title} {Anyons in an exactly
  solved model and beyond},\ }\href {https://doi.org/10.1016/j.aop.2005.10.005}
  {\bibfield  {journal} {\bibinfo  {journal} {Annals of Physics}\ }\textbf
  {\bibinfo {volume} {321}},\ \bibinfo {pages} {2} (\bibinfo {year}
  {2006})}\BibitemShut {NoStop}%
\bibitem [{\citenamefont {Chamon}(2005)}]{Chamon2005}%
  \BibitemOpen
  \bibfield  {author} {\bibinfo {author} {\bibfnamefont {C.}~\bibnamefont
  {Chamon}},\ }\bibfield  {title} {\bibinfo {title} {Quantum glassiness in
  strongly correlated clean systems: An example of topological
  overprotection},\ }\href {https://doi.org/10.1103/PhysRevLett.94.040402}
  {\bibfield  {journal} {\bibinfo  {journal} {Phys. Rev. Lett.}\ }\textbf
  {\bibinfo {volume} {94}},\ \bibinfo {pages} {040402} (\bibinfo {year}
  {2005})}\BibitemShut {NoStop}%
\bibitem [{\citenamefont {Haah}(2011)}]{Haah2011}%
  \BibitemOpen
  \bibfield  {author} {\bibinfo {author} {\bibfnamefont {J.}~\bibnamefont
  {Haah}},\ }\bibfield  {title} {\bibinfo {title} {Local stabilizer codes in
  three dimensions without string logical operators},\ }\href
  {https://doi.org/10.1103/PhysRevA.83.042330} {\bibfield  {journal} {\bibinfo
  {journal} {Phys. Rev. A}\ }\textbf {\bibinfo {volume} {83}},\ \bibinfo
  {pages} {042330} (\bibinfo {year} {2011})}\BibitemShut {NoStop}%
\bibitem [{\citenamefont {Yoshida}(2013)}]{Yoshida2013}%
  \BibitemOpen
  \bibfield  {author} {\bibinfo {author} {\bibfnamefont {B.}~\bibnamefont
  {Yoshida}},\ }\bibfield  {title} {\bibinfo {title} {Exotic topological order
  in fractal spin liquids},\ }\href
  {https://doi.org/10.1103/PhysRevB.88.125122} {\bibfield  {journal} {\bibinfo
  {journal} {Phys. Rev. B}\ }\textbf {\bibinfo {volume} {88}},\ \bibinfo
  {pages} {125122} (\bibinfo {year} {2013})}\BibitemShut {NoStop}%
\bibitem [{\citenamefont {Vijay}\ \emph {et~al.}(2015)\citenamefont {Vijay},
  \citenamefont {Haah},\ and\ \citenamefont {Fu}}]{VijayHaahFu2015}%
  \BibitemOpen
  \bibfield  {author} {\bibinfo {author} {\bibfnamefont {S.}~\bibnamefont
  {Vijay}}, \bibinfo {author} {\bibfnamefont {J.}~\bibnamefont {Haah}},\ and\
  \bibinfo {author} {\bibfnamefont {L.}~\bibnamefont {Fu}},\ }\bibfield
  {title} {\bibinfo {title} {A new kind of topological quantum order: A
  dimensional hierarchy of quasiparticles built from stationary excitations},\
  }\href {https://doi.org/10.1103/PhysRevB.92.235136} {\bibfield  {journal}
  {\bibinfo  {journal} {Phys. Rev. B}\ }\textbf {\bibinfo {volume} {92}},\
  \bibinfo {pages} {235136} (\bibinfo {year} {2015})}\BibitemShut {NoStop}%
\bibitem [{\citenamefont {Vijay}\ \emph {et~al.}(2016)\citenamefont {Vijay},
  \citenamefont {Haah},\ and\ \citenamefont {Fu}}]{VijayHaahFu2016}%
  \BibitemOpen
  \bibfield  {author} {\bibinfo {author} {\bibfnamefont {S.}~\bibnamefont
  {Vijay}}, \bibinfo {author} {\bibfnamefont {J.}~\bibnamefont {Haah}},\ and\
  \bibinfo {author} {\bibfnamefont {L.}~\bibnamefont {Fu}},\ }\bibfield
  {title} {\bibinfo {title} {Fracton topological order, generalized lattice
  gauge theory, and duality},\ }\href
  {https://doi.org/10.1103/PhysRevB.94.235157} {\bibfield  {journal} {\bibinfo
  {journal} {Phys. Rev. B}\ }\textbf {\bibinfo {volume} {94}},\ \bibinfo
  {pages} {235157} (\bibinfo {year} {2016})}\BibitemShut {NoStop}%
\bibitem [{\citenamefont {Nandkishore}\ and\ \citenamefont
  {Hermele}(2019)}]{nandkishore2019fractons}%
  \BibitemOpen
  \bibfield  {author} {\bibinfo {author} {\bibfnamefont {R.~M.}\ \bibnamefont
  {Nandkishore}}\ and\ \bibinfo {author} {\bibfnamefont {M.}~\bibnamefont
  {Hermele}},\ }\bibfield  {title} {\bibinfo {title} {Fractons},\ }\href
  {https://doi.org/10.1146/annurev-conmatphys-031218-013604} {\bibfield
  {journal} {\bibinfo  {journal} {Annual Review of Condensed Matter Physics}\
  }\textbf {\bibinfo {volume} {10}},\ \bibinfo {pages} {295} (\bibinfo {year}
  {2019})}\BibitemShut {NoStop}%
\bibitem [{\citenamefont {Pretko}\ \emph {et~al.}(2020)\citenamefont {Pretko},
  \citenamefont {Chen},\ and\ \citenamefont {You}}]{Pretko2020}%
  \BibitemOpen
  \bibfield  {author} {\bibinfo {author} {\bibfnamefont {M.}~\bibnamefont
  {Pretko}}, \bibinfo {author} {\bibfnamefont {X.}~\bibnamefont {Chen}},\ and\
  \bibinfo {author} {\bibfnamefont {Y.}~\bibnamefont {You}},\ }\bibfield
  {title} {\bibinfo {title} {Fracton phases of matter},\ }\href
  {https://doi.org/10.1142/S0217751X20300033} {\bibfield  {journal} {\bibinfo
  {journal} {International Journal of Modern Physics A}\ }\textbf {\bibinfo
  {volume} {35}},\ \bibinfo {pages} {2030003} (\bibinfo {year}
  {2020})}\BibitemShut {NoStop}%
\bibitem [{\citenamefont {Gu}\ and\ \citenamefont {Wen}(2009)}]{Gu09}%
  \BibitemOpen
  \bibfield  {author} {\bibinfo {author} {\bibfnamefont {Z.-C.}\ \bibnamefont
  {Gu}}\ and\ \bibinfo {author} {\bibfnamefont {X.-G.}\ \bibnamefont {Wen}},\
  }\bibfield  {title} {\bibinfo {title} {Tensor-entanglement-filtering
  renormalization approach and symmetry-protected topological order},\ }\href
  {https://doi.org/10.1103/PhysRevB.80.155131} {\bibfield  {journal} {\bibinfo
  {journal} {Phys. Rev. B}\ }\textbf {\bibinfo {volume} {80}},\ \bibinfo
  {pages} {155131} (\bibinfo {year} {2009})}\BibitemShut {NoStop}%
\bibitem [{\citenamefont {Pollmann}\ \emph
  {et~al.}(2010{\natexlab{a}})\citenamefont {Pollmann}, \citenamefont {Turner},
  \citenamefont {Berg},\ and\ \citenamefont {Oshikawa}}]{Pollmann10}%
  \BibitemOpen
  \bibfield  {author} {\bibinfo {author} {\bibfnamefont {F.}~\bibnamefont
  {Pollmann}}, \bibinfo {author} {\bibfnamefont {A.~M.}\ \bibnamefont
  {Turner}}, \bibinfo {author} {\bibfnamefont {E.}~\bibnamefont {Berg}},\ and\
  \bibinfo {author} {\bibfnamefont {M.}~\bibnamefont {Oshikawa}},\ }\bibfield
  {title} {\bibinfo {title} {Entanglement spectrum of a topological phase in
  one dimension},\ }\href {https://doi.org/10.1103/PhysRevB.81.064439}
  {\bibfield  {journal} {\bibinfo  {journal} {Phys. Rev. B}\ }\textbf {\bibinfo
  {volume} {81}},\ \bibinfo {pages} {064439} (\bibinfo {year}
  {2010}{\natexlab{a}})}\BibitemShut {NoStop}%
\bibitem [{\citenamefont {Fidkowski}\ and\ \citenamefont
  {Kitaev}(2011)}]{Fidkowski_2011}%
  \BibitemOpen
  \bibfield  {author} {\bibinfo {author} {\bibfnamefont {L.}~\bibnamefont
  {Fidkowski}}\ and\ \bibinfo {author} {\bibfnamefont {A.}~\bibnamefont
  {Kitaev}},\ }\bibfield  {title} {\bibinfo {title} {Topological phases of
  fermions in one dimension},\ }\href
  {https://doi.org/10.1103/PhysRevB.83.075103} {\bibfield  {journal} {\bibinfo
  {journal} {Phys. Rev. B}\ }\textbf {\bibinfo {volume} {83}},\ \bibinfo
  {pages} {075103} (\bibinfo {year} {2011})}\BibitemShut {NoStop}%
\bibitem [{\citenamefont {Turner}\ \emph {et~al.}(2011)\citenamefont {Turner},
  \citenamefont {Pollmann},\ and\ \citenamefont {Berg}}]{Turner11}%
  \BibitemOpen
  \bibfield  {author} {\bibinfo {author} {\bibfnamefont {A.~M.}\ \bibnamefont
  {Turner}}, \bibinfo {author} {\bibfnamefont {F.}~\bibnamefont {Pollmann}},\
  and\ \bibinfo {author} {\bibfnamefont {E.}~\bibnamefont {Berg}},\ }\bibfield
  {title} {\bibinfo {title} {Topological phases of one-dimensional fermions: An
  entanglement point of view},\ }\href
  {https://doi.org/10.1103/PhysRevB.83.075102} {\bibfield  {journal} {\bibinfo
  {journal} {Phys. Rev. B}\ }\textbf {\bibinfo {volume} {83}},\ \bibinfo
  {pages} {075102} (\bibinfo {year} {2011})}\BibitemShut {NoStop}%
\bibitem [{\citenamefont {Schuch}\ \emph {et~al.}(2011)\citenamefont {Schuch},
  \citenamefont {P\'erez-Garc\'{\i}a},\ and\ \citenamefont {Cirac}}]{Schuch11}%
  \BibitemOpen
  \bibfield  {author} {\bibinfo {author} {\bibfnamefont {N.}~\bibnamefont
  {Schuch}}, \bibinfo {author} {\bibfnamefont {D.}~\bibnamefont
  {P\'erez-Garc\'{\i}a}},\ and\ \bibinfo {author} {\bibfnamefont
  {I.}~\bibnamefont {Cirac}},\ }\bibfield  {title} {\bibinfo {title}
  {Classifying quantum phases using matrix product states and projected
  entangled pair states},\ }\href {https://doi.org/10.1103/PhysRevB.84.165139}
  {\bibfield  {journal} {\bibinfo  {journal} {Phys. Rev. B}\ }\textbf {\bibinfo
  {volume} {84}},\ \bibinfo {pages} {165139} (\bibinfo {year}
  {2011})}\BibitemShut {NoStop}%
\bibitem [{\citenamefont {Chen}\ \emph
  {et~al.}(2011{\natexlab{c}})\citenamefont {Chen}, \citenamefont {Liu},\ and\
  \citenamefont {Wen}}]{Chen_2011}%
  \BibitemOpen
  \bibfield  {author} {\bibinfo {author} {\bibfnamefont {X.}~\bibnamefont
  {Chen}}, \bibinfo {author} {\bibfnamefont {Z.-X.}\ \bibnamefont {Liu}},\ and\
  \bibinfo {author} {\bibfnamefont {X.-G.}\ \bibnamefont {Wen}},\ }\bibfield
  {title} {\bibinfo {title} {Two-dimensional symmetry-protected topological
  orders and their protected gapless edge excitations},\ }\href
  {https://doi.org/10.1103/PhysRevB.84.235141} {\bibfield  {journal} {\bibinfo
  {journal} {Phys. Rev. B}\ }\textbf {\bibinfo {volume} {84}},\ \bibinfo
  {pages} {235141} (\bibinfo {year} {2011}{\natexlab{c}})}\BibitemShut
  {NoStop}%
\bibitem [{\citenamefont {Chen}\ \emph {et~al.}(2013)\citenamefont {Chen},
  \citenamefont {Gu}, \citenamefont {Liu},\ and\ \citenamefont
  {Wen}}]{Chen_2013}%
  \BibitemOpen
  \bibfield  {author} {\bibinfo {author} {\bibfnamefont {X.}~\bibnamefont
  {Chen}}, \bibinfo {author} {\bibfnamefont {Z.-C.}\ \bibnamefont {Gu}},
  \bibinfo {author} {\bibfnamefont {Z.-X.}\ \bibnamefont {Liu}},\ and\ \bibinfo
  {author} {\bibfnamefont {X.-G.}\ \bibnamefont {Wen}},\ }\bibfield  {title}
  {\bibinfo {title} {Symmetry protected topological orders and the group
  cohomology of their symmetry group},\ }\href
  {https://doi.org/10.1103/PhysRevB.87.155114} {\bibfield  {journal} {\bibinfo
  {journal} {Phys. Rev. B}\ }\textbf {\bibinfo {volume} {87}},\ \bibinfo
  {pages} {155114} (\bibinfo {year} {2013})}\BibitemShut {NoStop}%
\bibitem [{\citenamefont {Pollmann}\ \emph {et~al.}(2012)\citenamefont
  {Pollmann}, \citenamefont {Berg}, \citenamefont {Turner},\ and\ \citenamefont
  {Oshikawa}}]{pollmann_symmetry_2012}%
  \BibitemOpen
  \bibfield  {author} {\bibinfo {author} {\bibfnamefont {F.}~\bibnamefont
  {Pollmann}}, \bibinfo {author} {\bibfnamefont {E.}~\bibnamefont {Berg}},
  \bibinfo {author} {\bibfnamefont {A.~M.}\ \bibnamefont {Turner}},\ and\
  \bibinfo {author} {\bibfnamefont {M.}~\bibnamefont {Oshikawa}},\ }\bibfield
  {title} {\bibinfo {title} {Symmetry protection of topological phases in
  one-dimensional quantum spin systems},\ }\href
  {https://doi.org/10.1103/PhysRevB.85.075125} {\bibfield  {journal} {\bibinfo
  {journal} {Phys. Rev. B}\ }\textbf {\bibinfo {volume} {85}},\ \bibinfo
  {pages} {075125} (\bibinfo {year} {2012})}\BibitemShut {NoStop}%
\bibitem [{\citenamefont {Lu}\ and\ \citenamefont
  {Vishwanath}(2012)}]{LuVishwanath12}%
  \BibitemOpen
  \bibfield  {author} {\bibinfo {author} {\bibfnamefont {Y.-M.}\ \bibnamefont
  {Lu}}\ and\ \bibinfo {author} {\bibfnamefont {A.}~\bibnamefont
  {Vishwanath}},\ }\bibfield  {title} {\bibinfo {title} {Theory and
  classification of interacting integer topological phases in two dimensions: A
  chern-simons approach},\ }\href {https://doi.org/10.1103/PhysRevB.86.125119}
  {\bibfield  {journal} {\bibinfo  {journal} {Phys. Rev. B}\ }\textbf {\bibinfo
  {volume} {86}},\ \bibinfo {pages} {125119} (\bibinfo {year}
  {2012})}\BibitemShut {NoStop}%
\bibitem [{\citenamefont {Senthil}\ and\ \citenamefont
  {Levin}(2013)}]{SenthilLevin13}%
  \BibitemOpen
  \bibfield  {author} {\bibinfo {author} {\bibfnamefont {T.}~\bibnamefont
  {Senthil}}\ and\ \bibinfo {author} {\bibfnamefont {M.}~\bibnamefont
  {Levin}},\ }\bibfield  {title} {\bibinfo {title} {Integer quantum hall effect
  for bosons},\ }\href {https://doi.org/10.1103/PhysRevLett.110.046801}
  {\bibfield  {journal} {\bibinfo  {journal} {Phys. Rev. Lett.}\ }\textbf
  {\bibinfo {volume} {110}},\ \bibinfo {pages} {046801} (\bibinfo {year}
  {2013})}\BibitemShut {NoStop}%
\bibitem [{\citenamefont {Levin}\ and\ \citenamefont {Gu}(2012)}]{Levin_2012}%
  \BibitemOpen
  \bibfield  {author} {\bibinfo {author} {\bibfnamefont {M.}~\bibnamefont
  {Levin}}\ and\ \bibinfo {author} {\bibfnamefont {Z.-C.}\ \bibnamefont {Gu}},\
  }\bibfield  {title} {\bibinfo {title} {Braiding statistics approach to
  symmetry-protected topological phases},\ }\href
  {https://doi.org/10.1103/PhysRevB.86.115109} {\bibfield  {journal} {\bibinfo
  {journal} {Phys. Rev. B}\ }\textbf {\bibinfo {volume} {86}},\ \bibinfo
  {pages} {115109} (\bibinfo {year} {2012})}\BibitemShut {NoStop}%
\bibitem [{\citenamefont {Vishwanath}\ and\ \citenamefont
  {Senthil}(2013)}]{VishwanathSenthil2013}%
  \BibitemOpen
  \bibfield  {author} {\bibinfo {author} {\bibfnamefont {A.}~\bibnamefont
  {Vishwanath}}\ and\ \bibinfo {author} {\bibfnamefont {T.}~\bibnamefont
  {Senthil}},\ }\bibfield  {title} {\bibinfo {title} {Physics of
  three-dimensional bosonic topological insulators: Surface-deconfined
  criticality and quantized magnetoelectric effect},\ }\href
  {https://doi.org/10.1103/PhysRevX.3.011016} {\bibfield  {journal} {\bibinfo
  {journal} {Phys. Rev. X}\ }\textbf {\bibinfo {volume} {3}},\ \bibinfo {pages}
  {011016} (\bibinfo {year} {2013})}\BibitemShut {NoStop}%
\bibitem [{\citenamefont {Else}\ and\ \citenamefont {Nayak}(2014)}]{Else_2014}%
  \BibitemOpen
  \bibfield  {author} {\bibinfo {author} {\bibfnamefont {D.~V.}\ \bibnamefont
  {Else}}\ and\ \bibinfo {author} {\bibfnamefont {C.}~\bibnamefont {Nayak}},\
  }\bibfield  {title} {\bibinfo {title} {Classifying symmetry-protected
  topological phases through the anomalous action of the symmetry on the
  edge},\ }\href {https://doi.org/10.1103/PhysRevB.90.235137} {\bibfield
  {journal} {\bibinfo  {journal} {Phys. Rev. B}\ }\textbf {\bibinfo {volume}
  {90}},\ \bibinfo {pages} {235137} (\bibinfo {year} {2014})}\BibitemShut
  {NoStop}%
\bibitem [{\citenamefont {Li}\ \emph {et~al.}(2018)\citenamefont {Li},
  \citenamefont {Chen},\ and\ \citenamefont {Fisher}}]{Li18}%
  \BibitemOpen
  \bibfield  {author} {\bibinfo {author} {\bibfnamefont {Y.}~\bibnamefont
  {Li}}, \bibinfo {author} {\bibfnamefont {X.}~\bibnamefont {Chen}},\ and\
  \bibinfo {author} {\bibfnamefont {M.~P.~A.}\ \bibnamefont {Fisher}},\
  }\bibfield  {title} {\bibinfo {title} {Quantum zeno effect and the many-body
  entanglement transition},\ }\href
  {https://doi.org/10.1103/PhysRevB.98.205136} {\bibfield  {journal} {\bibinfo
  {journal} {Phys. Rev. B}\ }\textbf {\bibinfo {volume} {98}},\ \bibinfo
  {pages} {205136} (\bibinfo {year} {2018})}\BibitemShut {NoStop}%
\bibitem [{\citenamefont {Skinner}\ \emph {et~al.}(2019)\citenamefont
  {Skinner}, \citenamefont {Ruhman},\ and\ \citenamefont {Nahum}}]{Skinner19}%
  \BibitemOpen
  \bibfield  {author} {\bibinfo {author} {\bibfnamefont {B.}~\bibnamefont
  {Skinner}}, \bibinfo {author} {\bibfnamefont {J.}~\bibnamefont {Ruhman}},\
  and\ \bibinfo {author} {\bibfnamefont {A.}~\bibnamefont {Nahum}},\ }\bibfield
   {title} {\bibinfo {title} {Measurement-induced phase transitions in the
  dynamics of entanglement},\ }\href
  {https://doi.org/10.1103/PhysRevX.9.031009} {\bibfield  {journal} {\bibinfo
  {journal} {Phys. Rev. X}\ }\textbf {\bibinfo {volume} {9}},\ \bibinfo {pages}
  {031009} (\bibinfo {year} {2019})}\BibitemShut {NoStop}%
\bibitem [{\citenamefont {Chan}\ \emph {et~al.}(2019)\citenamefont {Chan},
  \citenamefont {Nandkishore}, \citenamefont {Pretko},\ and\ \citenamefont
  {Smith}}]{Chan19}%
  \BibitemOpen
  \bibfield  {author} {\bibinfo {author} {\bibfnamefont {A.}~\bibnamefont
  {Chan}}, \bibinfo {author} {\bibfnamefont {R.~M.}\ \bibnamefont
  {Nandkishore}}, \bibinfo {author} {\bibfnamefont {M.}~\bibnamefont
  {Pretko}},\ and\ \bibinfo {author} {\bibfnamefont {G.}~\bibnamefont
  {Smith}},\ }\bibfield  {title} {\bibinfo {title} {Unitary-projective
  entanglement dynamics},\ }\href {https://doi.org/10.1103/PhysRevB.99.224307}
  {\bibfield  {journal} {\bibinfo  {journal} {Phys. Rev. B}\ }\textbf {\bibinfo
  {volume} {99}},\ \bibinfo {pages} {224307} (\bibinfo {year}
  {2019})}\BibitemShut {NoStop}%
\bibitem [{\citenamefont {Li}\ \emph {et~al.}(2019)\citenamefont {Li},
  \citenamefont {Chen},\ and\ \citenamefont {Fisher}}]{Li19}%
  \BibitemOpen
  \bibfield  {author} {\bibinfo {author} {\bibfnamefont {Y.}~\bibnamefont
  {Li}}, \bibinfo {author} {\bibfnamefont {X.}~\bibnamefont {Chen}},\ and\
  \bibinfo {author} {\bibfnamefont {M.~P.~A.}\ \bibnamefont {Fisher}},\
  }\bibfield  {title} {\bibinfo {title} {Measurement-driven entanglement
  transition in hybrid quantum circuits},\ }\href
  {https://doi.org/10.1103/PhysRevB.100.134306} {\bibfield  {journal} {\bibinfo
   {journal} {Phys. Rev. B}\ }\textbf {\bibinfo {volume} {100}},\ \bibinfo
  {pages} {134306} (\bibinfo {year} {2019})}\BibitemShut {NoStop}%
\bibitem [{\citenamefont {Vasseur}\ \emph {et~al.}(2019)\citenamefont
  {Vasseur}, \citenamefont {Potter}, \citenamefont {You},\ and\ \citenamefont
  {Ludwig}}]{Vasseur19}%
  \BibitemOpen
  \bibfield  {author} {\bibinfo {author} {\bibfnamefont {R.}~\bibnamefont
  {Vasseur}}, \bibinfo {author} {\bibfnamefont {A.~C.}\ \bibnamefont {Potter}},
  \bibinfo {author} {\bibfnamefont {Y.-Z.}\ \bibnamefont {You}},\ and\ \bibinfo
  {author} {\bibfnamefont {A.~W.~W.}\ \bibnamefont {Ludwig}},\ }\bibfield
  {title} {\bibinfo {title} {Entanglement transitions from holographic random
  tensor networks},\ }\href {https://doi.org/10.1103/PhysRevB.100.134203}
  {\bibfield  {journal} {\bibinfo  {journal} {Phys. Rev. B}\ }\textbf {\bibinfo
  {volume} {100}},\ \bibinfo {pages} {134203} (\bibinfo {year}
  {2019})}\BibitemShut {NoStop}%
\bibitem [{\citenamefont {Cao}\ \emph {et~al.}(2019)\citenamefont {Cao},
  \citenamefont {Tilloy},\ and\ \citenamefont {Luca}}]{Cao19}%
  \BibitemOpen
  \bibfield  {author} {\bibinfo {author} {\bibfnamefont {X.}~\bibnamefont
  {Cao}}, \bibinfo {author} {\bibfnamefont {A.}~\bibnamefont {Tilloy}},\ and\
  \bibinfo {author} {\bibfnamefont {A.~D.}\ \bibnamefont {Luca}},\ }\bibfield
  {title} {\bibinfo {title} {{Entanglement in a fermion chain under continuous
  monitoring}},\ }\href {https://doi.org/10.21468/SciPostPhys.7.2.024}
  {\bibfield  {journal} {\bibinfo  {journal} {SciPost Phys.}\ }\textbf
  {\bibinfo {volume} {7}},\ \bibinfo {pages} {24} (\bibinfo {year}
  {2019})}\BibitemShut {NoStop}%
\bibitem [{\citenamefont {Gullans}\ and\ \citenamefont
  {Huse}(2020)}]{Gullans20}%
  \BibitemOpen
  \bibfield  {author} {\bibinfo {author} {\bibfnamefont {M.~J.}\ \bibnamefont
  {Gullans}}\ and\ \bibinfo {author} {\bibfnamefont {D.~A.}\ \bibnamefont
  {Huse}},\ }\bibfield  {title} {\bibinfo {title} {Dynamical purification phase
  transition induced by quantum measurements},\ }\href
  {https://doi.org/10.1103/PhysRevX.10.041020} {\bibfield  {journal} {\bibinfo
  {journal} {Phys. Rev. X}\ }\textbf {\bibinfo {volume} {10}},\ \bibinfo
  {pages} {041020} (\bibinfo {year} {2020})}\BibitemShut {NoStop}%
\bibitem [{\citenamefont {Choi}\ \emph {et~al.}(2020)\citenamefont {Choi},
  \citenamefont {Bao}, \citenamefont {Qi},\ and\ \citenamefont
  {Altman}}]{Choi20}%
  \BibitemOpen
  \bibfield  {author} {\bibinfo {author} {\bibfnamefont {S.}~\bibnamefont
  {Choi}}, \bibinfo {author} {\bibfnamefont {Y.}~\bibnamefont {Bao}}, \bibinfo
  {author} {\bibfnamefont {X.-L.}\ \bibnamefont {Qi}},\ and\ \bibinfo {author}
  {\bibfnamefont {E.}~\bibnamefont {Altman}},\ }\bibfield  {title} {\bibinfo
  {title} {Quantum error correction in scrambling dynamics and
  measurement-induced phase transition},\ }\href
  {https://doi.org/10.1103/PhysRevLett.125.030505} {\bibfield  {journal}
  {\bibinfo  {journal} {Phys. Rev. Lett.}\ }\textbf {\bibinfo {volume} {125}},\
  \bibinfo {pages} {030505} (\bibinfo {year} {2020})}\BibitemShut {NoStop}%
\bibitem [{\citenamefont {Tang}\ and\ \citenamefont {Zhu}(2020)}]{Tang20}%
  \BibitemOpen
  \bibfield  {author} {\bibinfo {author} {\bibfnamefont {Q.}~\bibnamefont
  {Tang}}\ and\ \bibinfo {author} {\bibfnamefont {W.}~\bibnamefont {Zhu}},\
  }\bibfield  {title} {\bibinfo {title} {Measurement-induced phase transition:
  A case study in the nonintegrable model by density-matrix renormalization
  group calculations},\ }\href
  {https://doi.org/10.1103/PhysRevResearch.2.013022} {\bibfield  {journal}
  {\bibinfo  {journal} {Phys. Rev. Research}\ }\textbf {\bibinfo {volume}
  {2}},\ \bibinfo {pages} {013022} (\bibinfo {year} {2020})}\BibitemShut
  {NoStop}%
\bibitem [{\citenamefont {Jian}\ \emph {et~al.}(2020)\citenamefont {Jian},
  \citenamefont {You}, \citenamefont {Vasseur},\ and\ \citenamefont
  {Ludwig}}]{Jian20}%
  \BibitemOpen
  \bibfield  {author} {\bibinfo {author} {\bibfnamefont {C.-M.}\ \bibnamefont
  {Jian}}, \bibinfo {author} {\bibfnamefont {Y.-Z.}\ \bibnamefont {You}},
  \bibinfo {author} {\bibfnamefont {R.}~\bibnamefont {Vasseur}},\ and\ \bibinfo
  {author} {\bibfnamefont {A.~W.~W.}\ \bibnamefont {Ludwig}},\ }\bibfield
  {title} {\bibinfo {title} {Measurement-induced criticality in random quantum
  circuits},\ }\href {https://doi.org/10.1103/PhysRevB.101.104302} {\bibfield
  {journal} {\bibinfo  {journal} {Phys. Rev. B}\ }\textbf {\bibinfo {volume}
  {101}},\ \bibinfo {pages} {104302} (\bibinfo {year} {2020})}\BibitemShut
  {NoStop}%
\bibitem [{\citenamefont {Lopez-Piqueres}\ \emph {et~al.}(2020)\citenamefont
  {Lopez-Piqueres}, \citenamefont {Ware},\ and\ \citenamefont
  {Vasseur}}]{Lopez-Piqueres20}%
  \BibitemOpen
  \bibfield  {author} {\bibinfo {author} {\bibfnamefont {J.}~\bibnamefont
  {Lopez-Piqueres}}, \bibinfo {author} {\bibfnamefont {B.}~\bibnamefont
  {Ware}},\ and\ \bibinfo {author} {\bibfnamefont {R.}~\bibnamefont
  {Vasseur}},\ }\bibfield  {title} {\bibinfo {title} {Mean-field entanglement
  transitions in random tree tensor networks},\ }\href
  {https://doi.org/10.1103/PhysRevB.102.064202} {\bibfield  {journal} {\bibinfo
   {journal} {Phys. Rev. B}\ }\textbf {\bibinfo {volume} {102}},\ \bibinfo
  {pages} {064202} (\bibinfo {year} {2020})}\BibitemShut {NoStop}%
\bibitem [{\citenamefont {Bao}\ \emph {et~al.}(2020)\citenamefont {Bao},
  \citenamefont {Choi},\ and\ \citenamefont {Altman}}]{Bao20}%
  \BibitemOpen
  \bibfield  {author} {\bibinfo {author} {\bibfnamefont {Y.}~\bibnamefont
  {Bao}}, \bibinfo {author} {\bibfnamefont {S.}~\bibnamefont {Choi}},\ and\
  \bibinfo {author} {\bibfnamefont {E.}~\bibnamefont {Altman}},\ }\bibfield
  {title} {\bibinfo {title} {Theory of the phase transition in random unitary
  circuits with measurements},\ }\href
  {https://doi.org/10.1103/PhysRevB.101.104301} {\bibfield  {journal} {\bibinfo
   {journal} {Phys. Rev. B}\ }\textbf {\bibinfo {volume} {101}},\ \bibinfo
  {pages} {104301} (\bibinfo {year} {2020})}\BibitemShut {NoStop}%
\bibitem [{\citenamefont {Rossini}\ and\ \citenamefont
  {Vicari}(2020)}]{Rossini20}%
  \BibitemOpen
  \bibfield  {author} {\bibinfo {author} {\bibfnamefont {D.}~\bibnamefont
  {Rossini}}\ and\ \bibinfo {author} {\bibfnamefont {E.}~\bibnamefont
  {Vicari}},\ }\bibfield  {title} {\bibinfo {title} {Measurement-induced
  dynamics of many-body systems at quantum criticality},\ }\href
  {https://doi.org/10.1103/PhysRevB.102.035119} {\bibfield  {journal} {\bibinfo
   {journal} {Phys. Rev. B}\ }\textbf {\bibinfo {volume} {102}},\ \bibinfo
  {pages} {035119} (\bibinfo {year} {2020})}\BibitemShut {NoStop}%
\bibitem [{\citenamefont {Piroli}\ \emph {et~al.}(2020)\citenamefont {Piroli},
  \citenamefont {S{\"u}nderhauf},\ and\ \citenamefont {Qi}}]{Piroli20}%
  \BibitemOpen
  \bibfield  {author} {\bibinfo {author} {\bibfnamefont {L.}~\bibnamefont
  {Piroli}}, \bibinfo {author} {\bibfnamefont {C.}~\bibnamefont
  {S{\"u}nderhauf}},\ and\ \bibinfo {author} {\bibfnamefont {X.-L.}\
  \bibnamefont {Qi}},\ }\bibfield  {title} {\bibinfo {title} {A random unitary
  circuit model for black hole evaporation},\ }\href
  {https://doi.org/10.1007/JHEP04(2020)063} {\bibfield  {journal} {\bibinfo
  {journal} {Journal of High Energy Physics}\ }\textbf {\bibinfo {volume}
  {2020}},\ \bibinfo {pages} {63} (\bibinfo {year} {2020})}\BibitemShut
  {NoStop}%
\bibitem [{\citenamefont {Fan}\ \emph {et~al.}(2021)\citenamefont {Fan},
  \citenamefont {Vijay}, \citenamefont {Vishwanath},\ and\ \citenamefont
  {You}}]{Fan21}%
  \BibitemOpen
  \bibfield  {author} {\bibinfo {author} {\bibfnamefont {R.}~\bibnamefont
  {Fan}}, \bibinfo {author} {\bibfnamefont {S.}~\bibnamefont {Vijay}}, \bibinfo
  {author} {\bibfnamefont {A.}~\bibnamefont {Vishwanath}},\ and\ \bibinfo
  {author} {\bibfnamefont {Y.-Z.}\ \bibnamefont {You}},\ }\bibfield  {title}
  {\bibinfo {title} {Self-organized error correction in random unitary circuits
  with measurement},\ }\href {https://doi.org/10.1103/PhysRevB.103.174309}
  {\bibfield  {journal} {\bibinfo  {journal} {Phys. Rev. B}\ }\textbf {\bibinfo
  {volume} {103}},\ \bibinfo {pages} {174309} (\bibinfo {year}
  {2021})}\BibitemShut {NoStop}%
\bibitem [{\citenamefont {Li}\ \emph {et~al.}(2021)\citenamefont {Li},
  \citenamefont {Chen}, \citenamefont {Ludwig},\ and\ \citenamefont
  {Fisher}}]{LiChenLudwigFisher21}%
  \BibitemOpen
  \bibfield  {author} {\bibinfo {author} {\bibfnamefont {Y.}~\bibnamefont
  {Li}}, \bibinfo {author} {\bibfnamefont {X.}~\bibnamefont {Chen}}, \bibinfo
  {author} {\bibfnamefont {A.~W.~W.}\ \bibnamefont {Ludwig}},\ and\ \bibinfo
  {author} {\bibfnamefont {M.~P.~A.}\ \bibnamefont {Fisher}},\ }\bibfield
  {title} {\bibinfo {title} {Conformal invariance and quantum nonlocality in
  critical hybrid circuits},\ }\href
  {https://doi.org/10.1103/PhysRevB.104.104305} {\bibfield  {journal} {\bibinfo
   {journal} {Phys. Rev. B}\ }\textbf {\bibinfo {volume} {104}},\ \bibinfo
  {pages} {104305} (\bibinfo {year} {2021})}\BibitemShut {NoStop}%
\bibitem [{\citenamefont {Ben-Zion}\ \emph {et~al.}(2020)\citenamefont
  {Ben-Zion}, \citenamefont {McGreevy},\ and\ \citenamefont
  {Grover}}]{BenZion20}%
  \BibitemOpen
  \bibfield  {author} {\bibinfo {author} {\bibfnamefont {D.}~\bibnamefont
  {Ben-Zion}}, \bibinfo {author} {\bibfnamefont {J.}~\bibnamefont {McGreevy}},\
  and\ \bibinfo {author} {\bibfnamefont {T.}~\bibnamefont {Grover}},\
  }\bibfield  {title} {\bibinfo {title} {Disentangling quantum matter with
  measurements},\ }\href {https://doi.org/10.1103/PhysRevB.101.115131}
  {\bibfield  {journal} {\bibinfo  {journal} {Phys. Rev. B}\ }\textbf {\bibinfo
  {volume} {101}},\ \bibinfo {pages} {115131} (\bibinfo {year}
  {2020})}\BibitemShut {NoStop}%
\bibitem [{\citenamefont {Briegel}\ and\ \citenamefont
  {Raussendorf}(2001{\natexlab{a}})}]{Briegel01}%
  \BibitemOpen
  \bibfield  {author} {\bibinfo {author} {\bibfnamefont {H.~J.}\ \bibnamefont
  {Briegel}}\ and\ \bibinfo {author} {\bibfnamefont {R.}~\bibnamefont
  {Raussendorf}},\ }\bibfield  {title} {\bibinfo {title} {Persistent
  entanglement in arrays of interacting particles},\ }\href
  {https://doi.org/10.1103/PhysRevLett.86.910} {\bibfield  {journal} {\bibinfo
  {journal} {Phys. Rev. Lett.}\ }\textbf {\bibinfo {volume} {86}},\ \bibinfo
  {pages} {910} (\bibinfo {year} {2001}{\natexlab{a}})}\BibitemShut {NoStop}%
\bibitem [{\citenamefont {Raussendorf}\ \emph
  {et~al.}(2005{\natexlab{a}})\citenamefont {Raussendorf}, \citenamefont
  {Bravyi},\ and\ \citenamefont {Harrington}}]{Raussendorf05}%
  \BibitemOpen
  \bibfield  {author} {\bibinfo {author} {\bibfnamefont {R.}~\bibnamefont
  {Raussendorf}}, \bibinfo {author} {\bibfnamefont {S.}~\bibnamefont
  {Bravyi}},\ and\ \bibinfo {author} {\bibfnamefont {J.}~\bibnamefont
  {Harrington}},\ }\bibfield  {title} {\bibinfo {title} {Long-range quantum
  entanglement in noisy cluster states},\ }\href
  {https://doi.org/10.1103/PhysRevA.71.062313} {\bibfield  {journal} {\bibinfo
  {journal} {Phys. Rev. A}\ }\textbf {\bibinfo {volume} {71}},\ \bibinfo
  {pages} {062313} (\bibinfo {year} {2005}{\natexlab{a}})}\BibitemShut
  {NoStop}%
\bibitem [{\citenamefont {Aguado}\ \emph {et~al.}(2008)\citenamefont {Aguado},
  \citenamefont {Brennen}, \citenamefont {Verstraete},\ and\ \citenamefont
  {Cirac}}]{Aguado08}%
  \BibitemOpen
  \bibfield  {author} {\bibinfo {author} {\bibfnamefont {M.}~\bibnamefont
  {Aguado}}, \bibinfo {author} {\bibfnamefont {G.~K.}\ \bibnamefont {Brennen}},
  \bibinfo {author} {\bibfnamefont {F.}~\bibnamefont {Verstraete}},\ and\
  \bibinfo {author} {\bibfnamefont {J.~I.}\ \bibnamefont {Cirac}},\ }\bibfield
  {title} {\bibinfo {title} {Creation, manipulation, and detection of abelian
  and non-abelian anyons in optical lattices},\ }\href
  {https://doi.org/10.1103/PhysRevLett.101.260501} {\bibfield  {journal}
  {\bibinfo  {journal} {Phys. Rev. Lett.}\ }\textbf {\bibinfo {volume} {101}},\
  \bibinfo {pages} {260501} (\bibinfo {year} {2008})}\BibitemShut {NoStop}%
\bibitem [{\citenamefont {Piroli}\ \emph {et~al.}(2021)\citenamefont {Piroli},
  \citenamefont {Styliaris},\ and\ \citenamefont {Cirac}}]{Piroli21}%
  \BibitemOpen
  \bibfield  {author} {\bibinfo {author} {\bibfnamefont {L.}~\bibnamefont
  {Piroli}}, \bibinfo {author} {\bibfnamefont {G.}~\bibnamefont {Styliaris}},\
  and\ \bibinfo {author} {\bibfnamefont {J.~I.}\ \bibnamefont {Cirac}},\
  }\bibfield  {title} {\bibinfo {title} {Quantum circuits assisted by local
  operations and classical communication: Transformations and phases of
  matter},\ }\href {https://doi.org/10.1103/PhysRevLett.127.220503} {\bibfield
  {journal} {\bibinfo  {journal} {Phys. Rev. Lett.}\ }\textbf {\bibinfo
  {volume} {127}},\ \bibinfo {pages} {220503} (\bibinfo {year}
  {2021})}\BibitemShut {NoStop}%
\bibitem [{\citenamefont {Schmitz}(2019)}]{Schmitz19}%
  \BibitemOpen
  \bibfield  {author} {\bibinfo {author} {\bibfnamefont {A.~T.}\ \bibnamefont
  {Schmitz}},\ }\bibfield  {title} {\bibinfo {title} {Distilling fractons from
  layered subsystem-symmetry protected phases},\ }\href@noop {} {\bibfield
  {journal} {\bibinfo  {journal} {arXiv preprint arXiv:1910.04765}\ } (\bibinfo
  {year} {2019})}\BibitemShut {NoStop}%
\bibitem [{\citenamefont {Williamson}\ and\ \citenamefont
  {Devakul}(2021)}]{Williamson21}%
  \BibitemOpen
  \bibfield  {author} {\bibinfo {author} {\bibfnamefont {D.~J.}\ \bibnamefont
  {Williamson}}\ and\ \bibinfo {author} {\bibfnamefont {T.}~\bibnamefont
  {Devakul}},\ }\bibfield  {title} {\bibinfo {title} {Type-ii fractons from
  coupled spin chains and layers},\ }\href
  {https://doi.org/10.1103/PhysRevB.103.155140} {\bibfield  {journal} {\bibinfo
   {journal} {Phys. Rev. B}\ }\textbf {\bibinfo {volume} {103}},\ \bibinfo
  {pages} {155140} (\bibinfo {year} {2021})}\BibitemShut {NoStop}%
\bibitem [{\citenamefont {Calderbank}\ and\ \citenamefont
  {Shor}(1996)}]{Calderbank96}%
  \BibitemOpen
  \bibfield  {author} {\bibinfo {author} {\bibfnamefont {A.~R.}\ \bibnamefont
  {Calderbank}}\ and\ \bibinfo {author} {\bibfnamefont {P.~W.}\ \bibnamefont
  {Shor}},\ }\bibfield  {title} {\bibinfo {title} {Good quantum
  error-correcting codes exist},\ }\href
  {https://doi.org/10.1103/PhysRevA.54.1098} {\bibfield  {journal} {\bibinfo
  {journal} {Phys. Rev. A}\ }\textbf {\bibinfo {volume} {54}},\ \bibinfo
  {pages} {1098} (\bibinfo {year} {1996})}\BibitemShut {NoStop}%
\bibitem [{\citenamefont {Steane}(1996)}]{Steane96}%
  \BibitemOpen
  \bibfield  {author} {\bibinfo {author} {\bibfnamefont {A.}~\bibnamefont
  {Steane}},\ }\bibfield  {title} {\bibinfo {title} {Multiple-particle
  interference and quantum error correction},\ }\href
  {https://doi.org/10.1098/rspa.1996.0136} {\bibfield  {journal} {\bibinfo
  {journal} {Proceedings of the Royal Society of London. Series A:
  Mathematical, Physical and Engineering Sciences}\ }\textbf {\bibinfo {volume}
  {452}},\ \bibinfo {pages} {2551} (\bibinfo {year} {1996})}\BibitemShut
  {NoStop}%
\bibitem [{\citenamefont {Bolt}\ \emph {et~al.}(2016)\citenamefont {Bolt},
  \citenamefont {Duclos-Cianci}, \citenamefont {Poulin},\ and\ \citenamefont
  {Stace}}]{Bolt16}%
  \BibitemOpen
  \bibfield  {author} {\bibinfo {author} {\bibfnamefont {A.}~\bibnamefont
  {Bolt}}, \bibinfo {author} {\bibfnamefont {G.}~\bibnamefont {Duclos-Cianci}},
  \bibinfo {author} {\bibfnamefont {D.}~\bibnamefont {Poulin}},\ and\ \bibinfo
  {author} {\bibfnamefont {T.~M.}\ \bibnamefont {Stace}},\ }\bibfield  {title}
  {\bibinfo {title} {Foliated quantum error-correcting codes},\ }\href
  {https://doi.org/10.1103/PhysRevLett.117.070501} {\bibfield  {journal}
  {\bibinfo  {journal} {Phys. Rev. Lett.}\ }\textbf {\bibinfo {volume} {117}},\
  \bibinfo {pages} {070501} (\bibinfo {year} {2016})}\BibitemShut {NoStop}%
\bibitem [{\citenamefont {Hu}\ \emph {et~al.}(2013)\citenamefont {Hu},
  \citenamefont {Wan},\ and\ \citenamefont {Wu}}]{HW13}%
  \BibitemOpen
  \bibfield  {author} {\bibinfo {author} {\bibfnamefont {Y.}~\bibnamefont
  {Hu}}, \bibinfo {author} {\bibfnamefont {Y.}~\bibnamefont {Wan}},\ and\
  \bibinfo {author} {\bibfnamefont {Y.-S.}\ \bibnamefont {Wu}},\ }\bibfield
  {title} {\bibinfo {title} {Twisted quantum double model of topological phases
  in two dimensions},\ }\href {https://doi.org/10.1103/PhysRevB.87.125114}
  {\bibfield  {journal} {\bibinfo  {journal} {Phys. Rev. B}\ }\textbf {\bibinfo
  {volume} {87}},\ \bibinfo {pages} {125114} (\bibinfo {year}
  {2013})}\BibitemShut {NoStop}%
\bibitem [{\citenamefont {Song}\ \emph {et~al.}(2019)\citenamefont {Song},
  \citenamefont {Prem}, \citenamefont {Huang},\ and\ \citenamefont
  {Martin-Delgado}}]{SongPremHuangMartinDelgado19}%
  \BibitemOpen
  \bibfield  {author} {\bibinfo {author} {\bibfnamefont {H.}~\bibnamefont
  {Song}}, \bibinfo {author} {\bibfnamefont {A.}~\bibnamefont {Prem}}, \bibinfo
  {author} {\bibfnamefont {S.-J.}\ \bibnamefont {Huang}},\ and\ \bibinfo
  {author} {\bibfnamefont {M.~A.}\ \bibnamefont {Martin-Delgado}},\ }\bibfield
  {title} {\bibinfo {title} {Twisted fracton models in three dimensions},\
  }\href {https://doi.org/10.1103/PhysRevB.99.155118} {\bibfield  {journal}
  {\bibinfo  {journal} {Phys. Rev. B}\ }\textbf {\bibinfo {volume} {99}},\
  \bibinfo {pages} {155118} (\bibinfo {year} {2019})}\BibitemShut {NoStop}%
\bibitem [{\citenamefont {Bulmash}\ and\ \citenamefont
  {Barkeshli}(2019)}]{BulmashBarkeshli2019}%
  \BibitemOpen
  \bibfield  {author} {\bibinfo {author} {\bibfnamefont {D.}~\bibnamefont
  {Bulmash}}\ and\ \bibinfo {author} {\bibfnamefont {M.}~\bibnamefont
  {Barkeshli}},\ }\bibfield  {title} {\bibinfo {title} {Gauging fractons:
  Immobile non-abelian quasiparticles, fractals, and position-dependent
  degeneracies},\ }\href {https://doi.org/10.1103/PhysRevB.100.155146}
  {\bibfield  {journal} {\bibinfo  {journal} {Phys. Rev. B}\ }\textbf {\bibinfo
  {volume} {100}},\ \bibinfo {pages} {155146} (\bibinfo {year}
  {2019})}\BibitemShut {NoStop}%
\bibitem [{\citenamefont {Prem}\ and\ \citenamefont
  {Williamson}(2019)}]{PremWilliamson2019}%
  \BibitemOpen
  \bibfield  {author} {\bibinfo {author} {\bibfnamefont {A.}~\bibnamefont
  {Prem}}\ and\ \bibinfo {author} {\bibfnamefont {D.~J.}\ \bibnamefont
  {Williamson}},\ }\bibfield  {title} {\bibinfo {title} {{Gauging permutation
  symmetries as a route to non-Abelian fractons}},\ }\href
  {https://doi.org/10.21468/SciPostPhys.7.5.068} {\bibfield  {journal}
  {\bibinfo  {journal} {SciPost Phys.}\ }\textbf {\bibinfo {volume} {7}},\
  \bibinfo {pages} {68} (\bibinfo {year} {2019})}\BibitemShut {NoStop}%
\bibitem [{\citenamefont {Aasen}\ \emph {et~al.}(2020)\citenamefont {Aasen},
  \citenamefont {Bulmash}, \citenamefont {Prem}, \citenamefont {Slagle},\ and\
  \citenamefont {Williamson}}]{AasenBulmashPremSlagleWilliamson20}%
  \BibitemOpen
  \bibfield  {author} {\bibinfo {author} {\bibfnamefont {D.}~\bibnamefont
  {Aasen}}, \bibinfo {author} {\bibfnamefont {D.}~\bibnamefont {Bulmash}},
  \bibinfo {author} {\bibfnamefont {A.}~\bibnamefont {Prem}}, \bibinfo {author}
  {\bibfnamefont {K.}~\bibnamefont {Slagle}},\ and\ \bibinfo {author}
  {\bibfnamefont {D.~J.}\ \bibnamefont {Williamson}},\ }\bibfield  {title}
  {\bibinfo {title} {Topological defect networks for fractons of all types},\
  }\href {https://doi.org/10.1103/PhysRevResearch.2.043165} {\bibfield
  {journal} {\bibinfo  {journal} {Phys. Rev. Research}\ }\textbf {\bibinfo
  {volume} {2}},\ \bibinfo {pages} {043165} (\bibinfo {year}
  {2020})}\BibitemShut {NoStop}%
\bibitem [{\citenamefont {Stephen}\ \emph {et~al.}(2020)\citenamefont
  {Stephen}, \citenamefont {Garre-Rubio}, \citenamefont {Dua},\ and\
  \citenamefont {Williamson}}]{StephenGarre-RubioDuaWilliamson2020}%
  \BibitemOpen
  \bibfield  {author} {\bibinfo {author} {\bibfnamefont {D.~T.}\ \bibnamefont
  {Stephen}}, \bibinfo {author} {\bibfnamefont {J.}~\bibnamefont
  {Garre-Rubio}}, \bibinfo {author} {\bibfnamefont {A.}~\bibnamefont {Dua}},\
  and\ \bibinfo {author} {\bibfnamefont {D.~J.}\ \bibnamefont {Williamson}},\
  }\bibfield  {title} {\bibinfo {title} {Subsystem symmetry enriched
  topological order in three dimensions},\ }\href
  {https://doi.org/10.1103/PhysRevResearch.2.033331} {\bibfield  {journal}
  {\bibinfo  {journal} {Phys. Rev. Research}\ }\textbf {\bibinfo {volume}
  {2}},\ \bibinfo {pages} {033331} (\bibinfo {year} {2020})}\BibitemShut
  {NoStop}%
\bibitem [{\citenamefont {Wen}(2020)}]{Wen20}%
  \BibitemOpen
  \bibfield  {author} {\bibinfo {author} {\bibfnamefont {X.-G.}\ \bibnamefont
  {Wen}},\ }\bibfield  {title} {\bibinfo {title} {Systematic construction of
  gapped nonliquid states},\ }\href
  {https://doi.org/10.1103/PhysRevResearch.2.033300} {\bibfield  {journal}
  {\bibinfo  {journal} {Phys. Rev. Research}\ }\textbf {\bibinfo {volume}
  {2}},\ \bibinfo {pages} {033300} (\bibinfo {year} {2020})}\BibitemShut
  {NoStop}%
\bibitem [{\citenamefont {Wang}(2022)}]{Wang20}%
  \BibitemOpen
  \bibfield  {author} {\bibinfo {author} {\bibfnamefont {J.}~\bibnamefont
  {Wang}},\ }\bibfield  {title} {\bibinfo {title} {Nonliquid cellular states:
  Gluing gauge-higher-symmetry-breaking versus gauge-higher-symmetry-extension
  interfacial defects},\ }\href
  {https://doi.org/10.1103/PhysRevResearch.4.023258} {\bibfield  {journal}
  {\bibinfo  {journal} {Phys. Rev. Res.}\ }\textbf {\bibinfo {volume} {4}},\
  \bibinfo {pages} {023258} (\bibinfo {year} {2022})}\BibitemShut {NoStop}%
\bibitem [{\citenamefont {Williamson}\ and\ \citenamefont
  {Cheng}(2023)}]{WilliamsonCheng20}%
  \BibitemOpen
  \bibfield  {author} {\bibinfo {author} {\bibfnamefont {D.~J.}\ \bibnamefont
  {Williamson}}\ and\ \bibinfo {author} {\bibfnamefont {M.}~\bibnamefont
  {Cheng}},\ }\bibfield  {title} {\bibinfo {title} {Designer non-abelian
  fractons from topological layers},\ }\href
  {https://doi.org/10.1103/PhysRevB.107.035103} {\bibfield  {journal} {\bibinfo
   {journal} {Phys. Rev. B}\ }\textbf {\bibinfo {volume} {107}},\ \bibinfo
  {pages} {035103} (\bibinfo {year} {2023})}\BibitemShut {NoStop}%
\bibitem [{\citenamefont {Tantivasadakarn}\ \emph {et~al.}(2021)\citenamefont
  {Tantivasadakarn}, \citenamefont {Ji},\ and\ \citenamefont {Vijay}}]{TJV2}%
  \BibitemOpen
  \bibfield  {author} {\bibinfo {author} {\bibfnamefont {N.}~\bibnamefont
  {Tantivasadakarn}}, \bibinfo {author} {\bibfnamefont {W.}~\bibnamefont
  {Ji}},\ and\ \bibinfo {author} {\bibfnamefont {S.}~\bibnamefont {Vijay}},\
  }\bibfield  {title} {\bibinfo {title} {Non-abelian hybrid fracton orders},\
  }\href {https://doi.org/10.1103/PhysRevB.104.115117} {\bibfield  {journal}
  {\bibinfo  {journal} {Phys. Rev. B}\ }\textbf {\bibinfo {volume} {104}},\
  \bibinfo {pages} {115117} (\bibinfo {year} {2021})}\BibitemShut {NoStop}%
\bibitem [{\citenamefont {Wegner}(1971)}]{Wegner1971}%
  \BibitemOpen
  \bibfield  {author} {\bibinfo {author} {\bibfnamefont {F.~J.}\ \bibnamefont
  {Wegner}},\ }\bibfield  {title} {\bibinfo {title} {Duality in generalized
  ising models and phase transitions without local order parameters},\ }\href
  {https://doi.org/10.1063/1.1665530} {\bibfield  {journal} {\bibinfo
  {journal} {J. Math. Phys.}\ }\textbf {\bibinfo {volume} {12}},\ \bibinfo
  {pages} {2259} (\bibinfo {year} {1971})}\BibitemShut {NoStop}%
\bibitem [{\citenamefont {Kogut}(1979)}]{Kogut1979}%
  \BibitemOpen
  \bibfield  {author} {\bibinfo {author} {\bibfnamefont {J.~B.}\ \bibnamefont
  {Kogut}},\ }\bibfield  {title} {\bibinfo {title} {An introduction to lattice
  gauge theory and spin systems},\ }\href
  {https://doi.org/10.1103/RevModPhys.51.659} {\bibfield  {journal} {\bibinfo
  {journal} {Rev. Mod. Phys.}\ }\textbf {\bibinfo {volume} {51}},\ \bibinfo
  {pages} {659} (\bibinfo {year} {1979})}\BibitemShut {NoStop}%
\bibitem [{\citenamefont {Cobanera}\ \emph {et~al.}(2011)\citenamefont
  {Cobanera}, \citenamefont {Ortiz},\ and\ \citenamefont
  {Nussinov}}]{CobaneraOrtizNussinov2011}%
  \BibitemOpen
  \bibfield  {author} {\bibinfo {author} {\bibfnamefont {E.}~\bibnamefont
  {Cobanera}}, \bibinfo {author} {\bibfnamefont {G.}~\bibnamefont {Ortiz}},\
  and\ \bibinfo {author} {\bibfnamefont {Z.}~\bibnamefont {Nussinov}},\
  }\bibfield  {title} {\bibinfo {title} {The bond-algebraic approach to
  dualities},\ }\href {https://doi.org/10.1080/00018732.2011.619814} {\bibfield
   {journal} {\bibinfo  {journal} {Advances in physics}\ }\textbf {\bibinfo
  {volume} {60}},\ \bibinfo {pages} {679} (\bibinfo {year} {2011})}\BibitemShut
  {NoStop}%
\bibitem [{\citenamefont {Aasen}\ \emph
  {et~al.}(2016{\natexlab{a}})\citenamefont {Aasen}, \citenamefont {Mong},\
  and\ \citenamefont {Fendley}}]{Aasen16}%
  \BibitemOpen
  \bibfield  {author} {\bibinfo {author} {\bibfnamefont {D.}~\bibnamefont
  {Aasen}}, \bibinfo {author} {\bibfnamefont {R.~S.}\ \bibnamefont {Mong}},\
  and\ \bibinfo {author} {\bibfnamefont {P.}~\bibnamefont {Fendley}},\
  }\bibfield  {title} {\bibinfo {title} {Topological defects on the lattice: I.
  the ising model},\ }\href {https://doi.org/10.1088/1751-8113/49/35/354001}
  {\bibfield  {journal} {\bibinfo  {journal} {Journal of Physics A:
  Mathematical and Theoretical}\ }\textbf {\bibinfo {volume} {49}},\ \bibinfo
  {pages} {354001} (\bibinfo {year} {2016}{\natexlab{a}})}\BibitemShut
  {NoStop}%
\bibitem [{\citenamefont {Williamson}(2016)}]{Williamson2016}%
  \BibitemOpen
  \bibfield  {author} {\bibinfo {author} {\bibfnamefont {D.~J.}\ \bibnamefont
  {Williamson}},\ }\bibfield  {title} {\bibinfo {title} {Fractal symmetries:
  Ungauging the cubic code},\ }\href
  {https://doi.org/10.1103/PhysRevB.94.155128} {\bibfield  {journal} {\bibinfo
  {journal} {Phys. Rev. B}\ }\textbf {\bibinfo {volume} {94}},\ \bibinfo
  {pages} {155128} (\bibinfo {year} {2016})}\BibitemShut {NoStop}%
\bibitem [{\citenamefont {Kubica}\ and\ \citenamefont
  {Yoshida}(2018)}]{KubicaYoshida2018}%
  \BibitemOpen
  \bibfield  {author} {\bibinfo {author} {\bibfnamefont {A.}~\bibnamefont
  {Kubica}}\ and\ \bibinfo {author} {\bibfnamefont {B.}~\bibnamefont
  {Yoshida}},\ }\bibfield  {title} {\bibinfo {title} {Ungauging quantum
  error-correcting codes},\ }\href@noop {} {\bibfield  {journal} {\bibinfo
  {journal} {arXiv preprint arXiv:1805.01836}\ } (\bibinfo {year}
  {2018})}\BibitemShut {NoStop}%
\bibitem [{\citenamefont {Pretko}(2018)}]{Pretko2018}%
  \BibitemOpen
  \bibfield  {author} {\bibinfo {author} {\bibfnamefont {M.}~\bibnamefont
  {Pretko}},\ }\bibfield  {title} {\bibinfo {title} {The fracton gauge
  principle},\ }\href {https://doi.org/10.1103/PhysRevB.98.115134} {\bibfield
  {journal} {\bibinfo  {journal} {Phys. Rev. B}\ }\textbf {\bibinfo {volume}
  {98}},\ \bibinfo {pages} {115134} (\bibinfo {year} {2018})}\BibitemShut
  {NoStop}%
\bibitem [{\citenamefont {Shirley}\ \emph {et~al.}(2019)\citenamefont
  {Shirley}, \citenamefont {Slagle},\ and\ \citenamefont
  {Chen}}]{ShirleySlagleChen2019}%
  \BibitemOpen
  \bibfield  {author} {\bibinfo {author} {\bibfnamefont {W.}~\bibnamefont
  {Shirley}}, \bibinfo {author} {\bibfnamefont {K.}~\bibnamefont {Slagle}},\
  and\ \bibinfo {author} {\bibfnamefont {X.}~\bibnamefont {Chen}},\ }\bibfield
  {title} {\bibinfo {title} {{Foliated fracton order from gauging subsystem
  symmetries}},\ }\href {https://doi.org/10.21468/SciPostPhys.6.4.041}
  {\bibfield  {journal} {\bibinfo  {journal} {SciPost Phys.}\ }\textbf
  {\bibinfo {volume} {6}},\ \bibinfo {pages} {41} (\bibinfo {year}
  {2019})}\BibitemShut {NoStop}%
\bibitem [{\citenamefont {Radicevic}(2019)}]{Radicevic2019}%
  \BibitemOpen
  \bibfield  {author} {\bibinfo {author} {\bibfnamefont {D.}~\bibnamefont
  {Radicevic}},\ }\bibfield  {title} {\bibinfo {title} {Systematic
  constructions of fracton theories},\ }\href@noop {} {\bibfield  {journal}
  {\bibinfo  {journal} {arXiv preprint arXiv:1910.06336}\ } (\bibinfo {year}
  {2019})}\BibitemShut {NoStop}%
\bibitem [{\citenamefont {Jordan}\ and\ \citenamefont
  {Wigner}(1928)}]{Jordan1928}%
  \BibitemOpen
  \bibfield  {author} {\bibinfo {author} {\bibfnamefont {P.}~\bibnamefont
  {Jordan}}\ and\ \bibinfo {author} {\bibfnamefont {E.}~\bibnamefont
  {Wigner}},\ }\bibfield  {title} {\bibinfo {title} {Uber das paulische
  equivalenzverbot},\ }\href {https://doi.org/10.1007/bf01331938} {\bibfield
  {journal} {\bibinfo  {journal} {Zeitschrift fur Physik}\ }\textbf {\bibinfo
  {volume} {47}},\ \bibinfo {pages} {631} (\bibinfo {year} {1928})}\BibitemShut
  {NoStop}%
\bibitem [{\citenamefont {Schultz}\ \emph {et~al.}(1964)\citenamefont
  {Schultz}, \citenamefont {Mattis},\ and\ \citenamefont {Lieb}}]{SMLising}%
  \BibitemOpen
  \bibfield  {author} {\bibinfo {author} {\bibfnamefont {T.~D.}\ \bibnamefont
  {Schultz}}, \bibinfo {author} {\bibfnamefont {D.~C.}\ \bibnamefont
  {Mattis}},\ and\ \bibinfo {author} {\bibfnamefont {E.~H.}\ \bibnamefont
  {Lieb}},\ }\bibfield  {title} {\bibinfo {title} {Two-dimensional ising model
  as a soluble problem of many fermions},\ }\href
  {https://doi.org/10.1103/RevModPhys.36.856} {\bibfield  {journal} {\bibinfo
  {journal} {Rev. Mod. Phys.}\ }\textbf {\bibinfo {volume} {36}},\ \bibinfo
  {pages} {856} (\bibinfo {year} {1964})}\BibitemShut {NoStop}%
\bibitem [{\citenamefont {Chen}\ \emph {et~al.}(2018)\citenamefont {Chen},
  \citenamefont {Kapustin},\ and\ \citenamefont
  {Radicevic}}]{ChenKapustinRadicevic2018}%
  \BibitemOpen
  \bibfield  {author} {\bibinfo {author} {\bibfnamefont {Y.-A.}\ \bibnamefont
  {Chen}}, \bibinfo {author} {\bibfnamefont {A.}~\bibnamefont {Kapustin}},\
  and\ \bibinfo {author} {\bibfnamefont {D.}~\bibnamefont {Radicevic}},\
  }\bibfield  {title} {\bibinfo {title} {Exact bosonization in two spatial
  dimensions and a new class of lattice gauge theories},\ }\href
  {https://doi.org/https://doi.org/10.1016/j.aop.2018.03.024} {\bibfield
  {journal} {\bibinfo  {journal} {Annals of Physics}\ }\textbf {\bibinfo
  {volume} {393}},\ \bibinfo {pages} {234 } (\bibinfo {year}
  {2018})}\BibitemShut {NoStop}%
\bibitem [{\citenamefont {Chen}\ and\ \citenamefont
  {Kapustin}(2019)}]{ChenKapustin2019}%
  \BibitemOpen
  \bibfield  {author} {\bibinfo {author} {\bibfnamefont {Y.-A.}\ \bibnamefont
  {Chen}}\ and\ \bibinfo {author} {\bibfnamefont {A.}~\bibnamefont
  {Kapustin}},\ }\bibfield  {title} {\bibinfo {title} {Bosonization in three
  spatial dimensions and a 2-form gauge theory},\ }\href
  {https://doi.org/10.1103/PhysRevB.100.245127} {\bibfield  {journal} {\bibinfo
   {journal} {Phys. Rev. B}\ }\textbf {\bibinfo {volume} {100}},\ \bibinfo
  {pages} {245127} (\bibinfo {year} {2019})}\BibitemShut {NoStop}%
\bibitem [{\citenamefont {Chen}(2020)}]{Chen2019}%
  \BibitemOpen
  \bibfield  {author} {\bibinfo {author} {\bibfnamefont {Y.-A.}\ \bibnamefont
  {Chen}},\ }\bibfield  {title} {\bibinfo {title} {Exact bosonization in
  arbitrary dimensions},\ }\href
  {https://doi.org/10.1103/PhysRevResearch.2.033527} {\bibfield  {journal}
  {\bibinfo  {journal} {Phys. Rev. Research}\ }\textbf {\bibinfo {volume}
  {2}},\ \bibinfo {pages} {033527} (\bibinfo {year} {2020})}\BibitemShut
  {NoStop}%
\bibitem [{\citenamefont {Tantivasadakarn}(2020)}]{Tantivasadakarn20}%
  \BibitemOpen
  \bibfield  {author} {\bibinfo {author} {\bibfnamefont {N.}~\bibnamefont
  {Tantivasadakarn}},\ }\bibfield  {title} {\bibinfo {title} {Jordan-wigner
  dualities for translation-invariant hamiltonians in any dimension: Emergent
  fermions in fracton topological order},\ }\href
  {https://doi.org/10.1103/PhysRevResearch.2.023353} {\bibfield  {journal}
  {\bibinfo  {journal} {Phys. Rev. Research}\ }\textbf {\bibinfo {volume}
  {2}},\ \bibinfo {pages} {023353} (\bibinfo {year} {2020})}\BibitemShut
  {NoStop}%
\bibitem [{\citenamefont {Shirley}(2020)}]{Shirley20}%
  \BibitemOpen
  \bibfield  {author} {\bibinfo {author} {\bibfnamefont {W.}~\bibnamefont
  {Shirley}},\ }\bibfield  {title} {\bibinfo {title} {Fractonic order and
  emergent fermionic gauge theory},\ }\href@noop {} {\bibfield  {journal}
  {\bibinfo  {journal} {arXiv preprint arXiv:2002.12026}\ } (\bibinfo {year}
  {2020})}\BibitemShut {NoStop}%
\bibitem [{\citenamefont {Po}(2021)}]{Po21}%
  \BibitemOpen
  \bibfield  {author} {\bibinfo {author} {\bibfnamefont {H.~C.}\ \bibnamefont
  {Po}},\ }\bibfield  {title} {\bibinfo {title} {Symmetric jordan-wigner
  transformation in higher dimensions},\ }\href@noop {} {\bibfield  {journal}
  {\bibinfo  {journal} {arXiv preprint arXiv:2107.10842}\ } (\bibinfo {year}
  {2021})}\BibitemShut {NoStop}%
\bibitem [{\citenamefont {Li}\ and\ \citenamefont {Po}(2022)}]{LiPo21}%
  \BibitemOpen
  \bibfield  {author} {\bibinfo {author} {\bibfnamefont {K.}~\bibnamefont
  {Li}}\ and\ \bibinfo {author} {\bibfnamefont {H.~C.}\ \bibnamefont {Po}},\
  }\bibfield  {title} {\bibinfo {title} {Higher-dimensional jordan-wigner
  transformation and auxiliary majorana fermions},\ }\href
  {https://doi.org/10.1103/PhysRevB.106.115109} {\bibfield  {journal} {\bibinfo
   {journal} {Phys. Rev. B}\ }\textbf {\bibinfo {volume} {106}},\ \bibinfo
  {pages} {115109} (\bibinfo {year} {2022})}\BibitemShut {NoStop}%
\bibitem [{\citenamefont {Verresen}\ \emph {et~al.}(2021)\citenamefont
  {Verresen}, \citenamefont {Tantivasadakarn},\ and\ \citenamefont
  {Vishwanath}}]{Rydberg}%
  \BibitemOpen
  \bibfield  {author} {\bibinfo {author} {\bibfnamefont {R.}~\bibnamefont
  {Verresen}}, \bibinfo {author} {\bibfnamefont {N.}~\bibnamefont
  {Tantivasadakarn}},\ and\ \bibinfo {author} {\bibfnamefont {A.}~\bibnamefont
  {Vishwanath}},\ }\bibfield  {title} {\bibinfo {title} {Efficiently preparing
  schr\"odinger's cat, fractons and non-abelian topological order in quantum
  devices},\ }\href@noop {} {\bibfield  {journal} {\bibinfo  {journal} {arXiv
  preprint arXiv:2112.03061}\ } (\bibinfo {year} {2021})}\BibitemShut {NoStop}%
\bibitem [{\citenamefont {Fisher}(1925)}]{Fisher1925}%
  \BibitemOpen
  \bibfield  {author} {\bibinfo {author} {\bibfnamefont {R.~A.}\ \bibnamefont
  {Fisher}},\ }\bibfield  {title} {\bibinfo {title} {Theory of statistical
  estimation},\ }\href {https://doi.org/10.1017/S0305004100009580} {\bibfield
  {journal} {\bibinfo  {journal} {Mathematical Proceedings of the Cambridge
  Philosophical Society}\ }\textbf {\bibinfo {volume} {22}},\ \bibinfo {pages}
  {700–725} (\bibinfo {year} {1925})}\BibitemShut {NoStop}%
\bibitem [{\citenamefont {Giovannetti}\ \emph {et~al.}(2006)\citenamefont
  {Giovannetti}, \citenamefont {Lloyd},\ and\ \citenamefont
  {Maccone}}]{Giovannetti06}%
  \BibitemOpen
  \bibfield  {author} {\bibinfo {author} {\bibfnamefont {V.}~\bibnamefont
  {Giovannetti}}, \bibinfo {author} {\bibfnamefont {S.}~\bibnamefont {Lloyd}},\
  and\ \bibinfo {author} {\bibfnamefont {L.}~\bibnamefont {Maccone}},\
  }\bibfield  {title} {\bibinfo {title} {Quantum metrology},\ }\href
  {https://doi.org/10.1103/PhysRevLett.96.010401} {\bibfield  {journal}
  {\bibinfo  {journal} {Phys. Rev. Lett.}\ }\textbf {\bibinfo {volume} {96}},\
  \bibinfo {pages} {010401} (\bibinfo {year} {2006})}\BibitemShut {NoStop}%
\bibitem [{\citenamefont {Verstraete}\ \emph {et~al.}(2006)\citenamefont
  {Verstraete}, \citenamefont {Wolf}, \citenamefont {Perez-Garcia},\ and\
  \citenamefont {Cirac}}]{Verstraete06}%
  \BibitemOpen
  \bibfield  {author} {\bibinfo {author} {\bibfnamefont {F.}~\bibnamefont
  {Verstraete}}, \bibinfo {author} {\bibfnamefont {M.~M.}\ \bibnamefont
  {Wolf}}, \bibinfo {author} {\bibfnamefont {D.}~\bibnamefont {Perez-Garcia}},\
  and\ \bibinfo {author} {\bibfnamefont {J.~I.}\ \bibnamefont {Cirac}},\
  }\bibfield  {title} {\bibinfo {title} {Criticality, the area law, and the
  computational power of projected entangled pair states},\ }\href
  {https://doi.org/10.1103/PhysRevLett.96.220601} {\bibfield  {journal}
  {\bibinfo  {journal} {Phys. Rev. Lett.}\ }\textbf {\bibinfo {volume} {96}},\
  \bibinfo {pages} {220601} (\bibinfo {year} {2006})}\BibitemShut {NoStop}%
\bibitem [{\citenamefont {Cirac}\ \emph {et~al.}(2021)\citenamefont {Cirac},
  \citenamefont {P\'erez-Garc\'{\i}a}, \citenamefont {Schuch},\ and\
  \citenamefont {Verstraete}}]{Cirac2021}%
  \BibitemOpen
  \bibfield  {author} {\bibinfo {author} {\bibfnamefont {J.~I.}\ \bibnamefont
  {Cirac}}, \bibinfo {author} {\bibfnamefont {D.}~\bibnamefont
  {P\'erez-Garc\'{\i}a}}, \bibinfo {author} {\bibfnamefont {N.}~\bibnamefont
  {Schuch}},\ and\ \bibinfo {author} {\bibfnamefont {F.}~\bibnamefont
  {Verstraete}},\ }\bibfield  {title} {\bibinfo {title} {Matrix product states
  and projected entangled pair states: Concepts, symmetries, theorems},\ }\href
  {https://doi.org/10.1103/RevModPhys.93.045003} {\bibfield  {journal}
  {\bibinfo  {journal} {Rev. Mod. Phys.}\ }\textbf {\bibinfo {volume} {93}},\
  \bibinfo {pages} {045003} (\bibinfo {year} {2021})}\BibitemShut {NoStop}%
\bibitem [{\citenamefont {Williamson}\ \emph {et~al.}(2021)\citenamefont
  {Williamson}, \citenamefont {Delcamp}, \citenamefont {Verstraete},\ and\
  \citenamefont {Schuch}}]{Williamson20}%
  \BibitemOpen
  \bibfield  {author} {\bibinfo {author} {\bibfnamefont {D.~J.}\ \bibnamefont
  {Williamson}}, \bibinfo {author} {\bibfnamefont {C.}~\bibnamefont {Delcamp}},
  \bibinfo {author} {\bibfnamefont {F.}~\bibnamefont {Verstraete}},\ and\
  \bibinfo {author} {\bibfnamefont {N.}~\bibnamefont {Schuch}},\ }\bibfield
  {title} {\bibinfo {title} {On the stability of topological order in tensor
  network states},\ }\href {https://doi.org/10.1103/PhysRevB.104.235151}
  {\bibfield  {journal} {\bibinfo  {journal} {Phys. Rev. B}\ }\textbf {\bibinfo
  {volume} {104}},\ \bibinfo {pages} {235151} (\bibinfo {year}
  {2021})}\BibitemShut {NoStop}%
\bibitem [{\citenamefont {Delcamp}\ and\ \citenamefont
  {Schuch}(2021)}]{Delcamp20}%
  \BibitemOpen
  \bibfield  {author} {\bibinfo {author} {\bibfnamefont {C.}~\bibnamefont
  {Delcamp}}\ and\ \bibinfo {author} {\bibfnamefont {N.}~\bibnamefont
  {Schuch}},\ }\bibfield  {title} {\bibinfo {title} {On tensor network
  representations of the (3+ 1) d toric code},\ }\href
  {https://doi.org/10.22331/q-2021-12-16-604} {\bibfield  {journal} {\bibinfo
  {journal} {Quantum}\ }\textbf {\bibinfo {volume} {5}},\ \bibinfo {pages}
  {604} (\bibinfo {year} {2021})}\BibitemShut {NoStop}%
\bibitem [{\citenamefont {Vafa}(1986)}]{Vafa86}%
  \BibitemOpen
  \bibfield  {author} {\bibinfo {author} {\bibfnamefont {C.}~\bibnamefont
  {Vafa}},\ }\bibfield  {title} {\bibinfo {title} {Modular invariance and
  discrete torsion on orbifolds},\ }\href
  {https://doi.org/10.1016/0550-3213(86)90379-2} {\bibfield  {journal}
  {\bibinfo  {journal} {Nuclear Physics B}\ }\textbf {\bibinfo {volume}
  {273}},\ \bibinfo {pages} {592} (\bibinfo {year} {1986})}\BibitemShut
  {NoStop}%
\bibitem [{\citenamefont {Barkeshli}\ \emph {et~al.}(2019)\citenamefont
  {Barkeshli}, \citenamefont {Bonderson}, \citenamefont {Cheng},\ and\
  \citenamefont {Wang}}]{BarkeshliBondersonChengWang2019}%
  \BibitemOpen
  \bibfield  {author} {\bibinfo {author} {\bibfnamefont {M.}~\bibnamefont
  {Barkeshli}}, \bibinfo {author} {\bibfnamefont {P.}~\bibnamefont
  {Bonderson}}, \bibinfo {author} {\bibfnamefont {M.}~\bibnamefont {Cheng}},\
  and\ \bibinfo {author} {\bibfnamefont {Z.}~\bibnamefont {Wang}},\ }\bibfield
  {title} {\bibinfo {title} {Symmetry fractionalization, defects, and gauging
  of topological phases},\ }\href {https://doi.org/10.1103/PhysRevB.100.115147}
  {\bibfield  {journal} {\bibinfo  {journal} {Phys. Rev. B}\ }\textbf {\bibinfo
  {volume} {100}},\ \bibinfo {pages} {115147} (\bibinfo {year}
  {2019})}\BibitemShut {NoStop}%
\bibitem [{\citenamefont {{de Wild Propitius}}(1995)}]{P95}%
  \BibitemOpen
  \bibfield  {author} {\bibinfo {author} {\bibfnamefont {M.}~\bibnamefont {{de
  Wild Propitius}}},\ }\emph {\bibinfo {title} {{Topological interactions in
  broken gauge theories}}},\ \href
  {https://doi.org/10.48550/arXiv.hep-th/9511195} {Ph.D. thesis},\ \bibinfo
  {school} {University of Amsterdam} (\bibinfo {year} {1995})\BibitemShut
  {NoStop}%
\bibitem [{\citenamefont {Yoshida}(2016)}]{Yoshida2016}%
  \BibitemOpen
  \bibfield  {author} {\bibinfo {author} {\bibfnamefont {B.}~\bibnamefont
  {Yoshida}},\ }\bibfield  {title} {\bibinfo {title} {Topological phases with
  generalized global symmetries},\ }\href
  {https://doi.org/10.1103/PhysRevB.93.155131} {\bibfield  {journal} {\bibinfo
  {journal} {Phys. Rev. B}\ }\textbf {\bibinfo {volume} {93}},\ \bibinfo
  {pages} {155131} (\bibinfo {year} {2016})}\BibitemShut {NoStop}%
\bibitem [{\citenamefont {Shirley}\ \emph {et~al.}(2020)\citenamefont
  {Shirley}, \citenamefont {Slagle},\ and\ \citenamefont
  {Chen}}]{ShirleySlagleChen20}%
  \BibitemOpen
  \bibfield  {author} {\bibinfo {author} {\bibfnamefont {W.}~\bibnamefont
  {Shirley}}, \bibinfo {author} {\bibfnamefont {K.}~\bibnamefont {Slagle}},\
  and\ \bibinfo {author} {\bibfnamefont {X.}~\bibnamefont {Chen}},\ }\bibfield
  {title} {\bibinfo {title} {Twisted foliated fracton phases},\ }\href
  {https://doi.org/10.1103/PhysRevB.102.115103} {\bibfield  {journal} {\bibinfo
   {journal} {Phys. Rev. B}\ }\textbf {\bibinfo {volume} {102}},\ \bibinfo
  {pages} {115103} (\bibinfo {year} {2020})}\BibitemShut {NoStop}%
\bibitem [{\citenamefont {Devakul}\ \emph {et~al.}(2020)\citenamefont
  {Devakul}, \citenamefont {Shirley},\ and\ \citenamefont
  {Wang}}]{DevakulShirleyWang20}%
  \BibitemOpen
  \bibfield  {author} {\bibinfo {author} {\bibfnamefont {T.}~\bibnamefont
  {Devakul}}, \bibinfo {author} {\bibfnamefont {W.}~\bibnamefont {Shirley}},\
  and\ \bibinfo {author} {\bibfnamefont {J.}~\bibnamefont {Wang}},\ }\bibfield
  {title} {\bibinfo {title} {Strong planar subsystem symmetry-protected
  topological phases and their dual fracton orders},\ }\href
  {https://doi.org/10.1103/PhysRevResearch.2.012059} {\bibfield  {journal}
  {\bibinfo  {journal} {Phys. Rev. Research}\ }\textbf {\bibinfo {volume}
  {2}},\ \bibinfo {pages} {012059} (\bibinfo {year} {2020})}\BibitemShut
  {NoStop}%
\bibitem [{\citenamefont {Ginsparg}(1989)}]{ginsparg_applied_1988}%
  \BibitemOpen
  \bibfield  {author} {\bibinfo {author} {\bibfnamefont {P.}~\bibnamefont
  {Ginsparg}},\ }\bibfield  {title} {\bibinfo {title} {Applied conformal field
  theory},\ }in\ \href@noop {} {\emph {\bibinfo {booktitle} {Fields, Strings
  and Critical Phenomena}}},\ \bibinfo {series and number} {Les Houches 1988,
  Session 49},\ \bibinfo {editor} {edited by\ \bibinfo {editor} {\bibfnamefont
  {J.}~\bibnamefont {Zinn-Justin}}\ and\ \bibinfo {editor} {\bibfnamefont
  {E.}~\bibnamefont {Br\'{e}zin}}}\ (\bibinfo  {publisher} {North-Holland},\
  \bibinfo {year} {1989})\BibitemShut {NoStop}%
\bibitem [{\citenamefont {Brennen}\ \emph {et~al.}(2009)\citenamefont
  {Brennen}, \citenamefont {Aguado},\ and\ \citenamefont {Cirac}}]{Brennen09}%
  \BibitemOpen
  \bibfield  {author} {\bibinfo {author} {\bibfnamefont {G.}~\bibnamefont
  {Brennen}}, \bibinfo {author} {\bibfnamefont {M.}~\bibnamefont {Aguado}},\
  and\ \bibinfo {author} {\bibfnamefont {J.~I.}\ \bibnamefont {Cirac}},\
  }\bibfield  {title} {\bibinfo {title} {Simulations of quantum double
  models},\ }\href {https://doi.org/10.1088/1367-2630/11/5/053009} {\bibfield
  {journal} {\bibinfo  {journal} {New Journal of Physics}\ }\textbf {\bibinfo
  {volume} {11}},\ \bibinfo {pages} {053009} (\bibinfo {year}
  {2009})}\BibitemShut {NoStop}%
\bibitem [{\citenamefont {Brell}(2015)}]{Brell2015}%
  \BibitemOpen
  \bibfield  {author} {\bibinfo {author} {\bibfnamefont {C.~G.}\ \bibnamefont
  {Brell}},\ }\bibfield  {title} {\bibinfo {title} {Generalized cluster states
  based on finite groups},\ }\href
  {https://doi.org/10.1088/1367-2630/17/2/023029} {\bibfield  {journal}
  {\bibinfo  {journal} {New Journal of Physics}\ }\textbf {\bibinfo {volume}
  {17}},\ \bibinfo {pages} {023029} (\bibinfo {year} {2015})}\BibitemShut
  {NoStop}%
\bibitem [{\citenamefont {Haegeman}\ \emph {et~al.}(2015)\citenamefont
  {Haegeman}, \citenamefont {Van~Acoleyen}, \citenamefont {Schuch},
  \citenamefont {Cirac},\ and\ \citenamefont {Verstraete}}]{Haegeman15}%
  \BibitemOpen
  \bibfield  {author} {\bibinfo {author} {\bibfnamefont {J.}~\bibnamefont
  {Haegeman}}, \bibinfo {author} {\bibfnamefont {K.}~\bibnamefont
  {Van~Acoleyen}}, \bibinfo {author} {\bibfnamefont {N.}~\bibnamefont
  {Schuch}}, \bibinfo {author} {\bibfnamefont {J.~I.}\ \bibnamefont {Cirac}},\
  and\ \bibinfo {author} {\bibfnamefont {F.}~\bibnamefont {Verstraete}},\
  }\bibfield  {title} {\bibinfo {title} {Gauging quantum states: From global to
  local symmetries in many-body systems},\ }\href
  {https://doi.org/10.1103/PhysRevX.5.011024} {\bibfield  {journal} {\bibinfo
  {journal} {Phys. Rev. X}\ }\textbf {\bibinfo {volume} {5}},\ \bibinfo {pages}
  {011024} (\bibinfo {year} {2015})}\BibitemShut {NoStop}%
\bibitem [{\citenamefont {Mochon}(2004)}]{Mochon04}%
  \BibitemOpen
  \bibfield  {author} {\bibinfo {author} {\bibfnamefont {C.}~\bibnamefont
  {Mochon}},\ }\bibfield  {title} {\bibinfo {title} {Anyon computers with
  smaller groups},\ }\href {https://doi.org/10.1103/PhysRevA.69.032306}
  {\bibfield  {journal} {\bibinfo  {journal} {Phys. Rev. A}\ }\textbf {\bibinfo
  {volume} {69}},\ \bibinfo {pages} {032306} (\bibinfo {year}
  {2004})}\BibitemShut {NoStop}%
\bibitem [{\citenamefont {Heinrich}\ \emph {et~al.}(2016)\citenamefont
  {Heinrich}, \citenamefont {Burnell}, \citenamefont {Fidkowski},\ and\
  \citenamefont {Levin}}]{Heinrich16}%
  \BibitemOpen
  \bibfield  {author} {\bibinfo {author} {\bibfnamefont {C.}~\bibnamefont
  {Heinrich}}, \bibinfo {author} {\bibfnamefont {F.}~\bibnamefont {Burnell}},
  \bibinfo {author} {\bibfnamefont {L.}~\bibnamefont {Fidkowski}},\ and\
  \bibinfo {author} {\bibfnamefont {M.}~\bibnamefont {Levin}},\ }\bibfield
  {title} {\bibinfo {title} {Symmetry-enriched string nets: Exactly solvable
  models for set phases},\ }\href {https://doi.org/10.1103/PhysRevB.94.235136}
  {\bibfield  {journal} {\bibinfo  {journal} {Phys. Rev. B}\ }\textbf {\bibinfo
  {volume} {94}},\ \bibinfo {pages} {235136} (\bibinfo {year}
  {2016})}\BibitemShut {NoStop}%
\bibitem [{\citenamefont {Cheng}\ \emph {et~al.}(2017)\citenamefont {Cheng},
  \citenamefont {Gu}, \citenamefont {Jiang},\ and\ \citenamefont
  {Qi}}]{Cheng17}%
  \BibitemOpen
  \bibfield  {author} {\bibinfo {author} {\bibfnamefont {M.}~\bibnamefont
  {Cheng}}, \bibinfo {author} {\bibfnamefont {Z.-C.}\ \bibnamefont {Gu}},
  \bibinfo {author} {\bibfnamefont {S.}~\bibnamefont {Jiang}},\ and\ \bibinfo
  {author} {\bibfnamefont {Y.}~\bibnamefont {Qi}},\ }\bibfield  {title}
  {\bibinfo {title} {Exactly solvable models for symmetry-enriched topological
  phases},\ }\href {https://doi.org/10.1103/PhysRevB.96.115107} {\bibfield
  {journal} {\bibinfo  {journal} {Phys. Rev. B}\ }\textbf {\bibinfo {volume}
  {96}},\ \bibinfo {pages} {115107} (\bibinfo {year} {2017})}\BibitemShut
  {NoStop}%
\bibitem [{\citenamefont {Gu}\ and\ \citenamefont {Wen}(2014)}]{GuWen14}%
  \BibitemOpen
  \bibfield  {author} {\bibinfo {author} {\bibfnamefont {Z.-C.}\ \bibnamefont
  {Gu}}\ and\ \bibinfo {author} {\bibfnamefont {X.-G.}\ \bibnamefont {Wen}},\
  }\bibfield  {title} {\bibinfo {title} {Symmetry-protected topological orders
  for interacting fermions: Fermionic topological nonlinear
  $\ensuremath{\sigma}$ models and a special group supercohomology theory},\
  }\href {https://doi.org/10.1103/PhysRevB.90.115141} {\bibfield  {journal}
  {\bibinfo  {journal} {Phys. Rev. B}\ }\textbf {\bibinfo {volume} {90}},\
  \bibinfo {pages} {115141} (\bibinfo {year} {2014})}\BibitemShut {NoStop}%
\bibitem [{\citenamefont {von Keyserlingk}\ and\ \citenamefont
  {Sondhi}(2016)}]{vonKeyserlingkSondhi16}%
  \BibitemOpen
  \bibfield  {author} {\bibinfo {author} {\bibfnamefont {C.~W.}\ \bibnamefont
  {von Keyserlingk}}\ and\ \bibinfo {author} {\bibfnamefont {S.~L.}\
  \bibnamefont {Sondhi}},\ }\bibfield  {title} {\bibinfo {title} {Phase
  structure of one-dimensional interacting floquet systems. i. abelian
  symmetry-protected topological phases},\ }\href
  {https://doi.org/10.1103/PhysRevB.93.245145} {\bibfield  {journal} {\bibinfo
  {journal} {Phys. Rev. B}\ }\textbf {\bibinfo {volume} {93}},\ \bibinfo
  {pages} {245145} (\bibinfo {year} {2016})}\BibitemShut {NoStop}%
\bibitem [{\citenamefont {Tantivasadakarn}\ and\ \citenamefont
  {Vishwanath}(2018)}]{TantivasadakarnVishwanath18}%
  \BibitemOpen
  \bibfield  {author} {\bibinfo {author} {\bibfnamefont {N.}~\bibnamefont
  {Tantivasadakarn}}\ and\ \bibinfo {author} {\bibfnamefont {A.}~\bibnamefont
  {Vishwanath}},\ }\bibfield  {title} {\bibinfo {title} {Full commuting
  projector hamiltonians of interacting symmetry-protected topological phases
  of fermions},\ }\href {https://doi.org/10.1103/PhysRevB.98.165104} {\bibfield
   {journal} {\bibinfo  {journal} {Phys. Rev. B}\ }\textbf {\bibinfo {volume}
  {98}},\ \bibinfo {pages} {165104} (\bibinfo {year} {2018})}\BibitemShut
  {NoStop}%
\bibitem [{\citenamefont {Borla}\ \emph {et~al.}(2021)\citenamefont {Borla},
  \citenamefont {Verresen}, \citenamefont {Shah},\ and\ \citenamefont
  {Moroz}}]{Borla21}%
  \BibitemOpen
  \bibfield  {author} {\bibinfo {author} {\bibfnamefont {U.}~\bibnamefont
  {Borla}}, \bibinfo {author} {\bibfnamefont {R.}~\bibnamefont {Verresen}},
  \bibinfo {author} {\bibfnamefont {J.}~\bibnamefont {Shah}},\ and\ \bibinfo
  {author} {\bibfnamefont {S.}~\bibnamefont {Moroz}},\ }\bibfield  {title}
  {\bibinfo {title} {{Gauging the Kitaev chain}},\ }\href
  {https://doi.org/10.21468/SciPostPhys.10.6.148} {\bibfield  {journal}
  {\bibinfo  {journal} {SciPost Phys.}\ }\textbf {\bibinfo {volume} {10}},\
  \bibinfo {pages} {148} (\bibinfo {year} {2021})}\BibitemShut {NoStop}%
\bibitem [{\citenamefont {Gaiotto}\ and\ \citenamefont
  {Kapustin}(2016)}]{GaiottoKapustin2016}%
  \BibitemOpen
  \bibfield  {author} {\bibinfo {author} {\bibfnamefont {D.}~\bibnamefont
  {Gaiotto}}\ and\ \bibinfo {author} {\bibfnamefont {A.}~\bibnamefont
  {Kapustin}},\ }\bibfield  {title} {\bibinfo {title} {Spin tqfts and fermionic
  phases of matter},\ }\href
  {https://www.worldscientific.com/doi/abs/10.1142/S0217751X16450445}
  {\bibfield  {journal} {\bibinfo  {journal} {International Journal of Modern
  Physics A}\ }\textbf {\bibinfo {volume} {31}},\ \bibinfo {pages} {1645044}
  (\bibinfo {year} {2016})}\BibitemShut {NoStop}%
\bibitem [{\citenamefont {Kapustin}\ and\ \citenamefont
  {Thorngren}(2017{\natexlab{a}})}]{kapustin2017fermionic}%
  \BibitemOpen
  \bibfield  {author} {\bibinfo {author} {\bibfnamefont {A.}~\bibnamefont
  {Kapustin}}\ and\ \bibinfo {author} {\bibfnamefont {R.}~\bibnamefont
  {Thorngren}},\ }\bibfield  {title} {\bibinfo {title} {Fermionic spt phases in
  higher dimensions and bosonization},\ }\href
  {https://doi.org/10.1007/JHEP10(2017)080} {\bibfield  {journal} {\bibinfo
  {journal} {Journal of High Energy Physics}\ }\textbf {\bibinfo {volume}
  {2017}},\ \bibinfo {pages} {80} (\bibinfo {year}
  {2017}{\natexlab{a}})}\BibitemShut {NoStop}%
\bibitem [{\citenamefont {Shukla}\ \emph {et~al.}(2020)\citenamefont {Shukla},
  \citenamefont {Ellison},\ and\ \citenamefont {Fidkowski}}]{Shukla20}%
  \BibitemOpen
  \bibfield  {author} {\bibinfo {author} {\bibfnamefont {S.~K.}\ \bibnamefont
  {Shukla}}, \bibinfo {author} {\bibfnamefont {T.~D.}\ \bibnamefont
  {Ellison}},\ and\ \bibinfo {author} {\bibfnamefont {L.}~\bibnamefont
  {Fidkowski}},\ }\bibfield  {title} {\bibinfo {title} {Tensor network approach
  to two-dimensional bosonization},\ }\href
  {https://doi.org/10.1103/PhysRevB.101.155105} {\bibfield  {journal} {\bibinfo
   {journal} {Phys. Rev. B}\ }\textbf {\bibinfo {volume} {101}},\ \bibinfo
  {pages} {155105} (\bibinfo {year} {2020})}\BibitemShut {NoStop}%
\bibitem [{\citenamefont {Wang}\ \emph {et~al.}(2019)\citenamefont {Wang},
  \citenamefont {Qi},\ and\ \citenamefont {Gu}}]{anomalousspt}%
  \BibitemOpen
  \bibfield  {author} {\bibinfo {author} {\bibfnamefont {Q.-R.}\ \bibnamefont
  {Wang}}, \bibinfo {author} {\bibfnamefont {Y.}~\bibnamefont {Qi}},\ and\
  \bibinfo {author} {\bibfnamefont {Z.-C.}\ \bibnamefont {Gu}},\ }\bibfield
  {title} {\bibinfo {title} {Anomalous symmetry protected topological states in
  interacting fermion systems},\ }\href
  {https://doi.org/10.1103/PhysRevLett.123.207003} {\bibfield  {journal}
  {\bibinfo  {journal} {Phys. Rev. Lett.}\ }\textbf {\bibinfo {volume} {123}},\
  \bibinfo {pages} {207003} (\bibinfo {year} {2019})}\BibitemShut {NoStop}%
\bibitem [{\citenamefont {Smacchia}\ \emph {et~al.}(2011)\citenamefont
  {Smacchia}, \citenamefont {Amico}, \citenamefont {Facchi}, \citenamefont
  {Fazio}, \citenamefont {Florio}, \citenamefont {Pascazio},\ and\
  \citenamefont {Vedral}}]{Smacchia11}%
  \BibitemOpen
  \bibfield  {author} {\bibinfo {author} {\bibfnamefont {P.}~\bibnamefont
  {Smacchia}}, \bibinfo {author} {\bibfnamefont {L.}~\bibnamefont {Amico}},
  \bibinfo {author} {\bibfnamefont {P.}~\bibnamefont {Facchi}}, \bibinfo
  {author} {\bibfnamefont {R.}~\bibnamefont {Fazio}}, \bibinfo {author}
  {\bibfnamefont {G.}~\bibnamefont {Florio}}, \bibinfo {author} {\bibfnamefont
  {S.}~\bibnamefont {Pascazio}},\ and\ \bibinfo {author} {\bibfnamefont
  {V.}~\bibnamefont {Vedral}},\ }\bibfield  {title} {\bibinfo {title}
  {Statistical mechanics of the cluster ising model},\ }\href
  {https://doi.org/10.1103/PhysRevA.84.022304} {\bibfield  {journal} {\bibinfo
  {journal} {Phys. Rev. A}\ }\textbf {\bibinfo {volume} {84}},\ \bibinfo
  {pages} {022304} (\bibinfo {year} {2011})}\BibitemShut {NoStop}%
\bibitem [{\citenamefont {Pollmann}\ and\ \citenamefont
  {Turner}(2012)}]{Pollmann12}%
  \BibitemOpen
  \bibfield  {author} {\bibinfo {author} {\bibfnamefont {F.}~\bibnamefont
  {Pollmann}}\ and\ \bibinfo {author} {\bibfnamefont {A.~M.}\ \bibnamefont
  {Turner}},\ }\bibfield  {title} {\bibinfo {title} {Detection of
  symmetry-protected topological phases in one dimension},\ }\href
  {https://doi.org/10.1103/PhysRevB.86.125441} {\bibfield  {journal} {\bibinfo
  {journal} {Phys. Rev. B}\ }\textbf {\bibinfo {volume} {86}},\ \bibinfo
  {pages} {125441} (\bibinfo {year} {2012})}\BibitemShut {NoStop}%
\bibitem [{\citenamefont {Marvian}(2017)}]{Marvian17}%
  \BibitemOpen
  \bibfield  {author} {\bibinfo {author} {\bibfnamefont {I.}~\bibnamefont
  {Marvian}},\ }\bibfield  {title} {\bibinfo {title} {Symmetry-protected
  topological entanglement},\ }\href
  {https://doi.org/10.1103/PhysRevB.95.045111} {\bibfield  {journal} {\bibinfo
  {journal} {Phys. Rev. B}\ }\textbf {\bibinfo {volume} {95}},\ \bibinfo
  {pages} {045111} (\bibinfo {year} {2017})}\BibitemShut {NoStop}%
\bibitem [{\citenamefont {Gaiotto}\ \emph {et~al.}(2015)\citenamefont
  {Gaiotto}, \citenamefont {Kapustin}, \citenamefont {Seiberg},\ and\
  \citenamefont {Willett}}]{generalizedglobalsymmetries}%
  \BibitemOpen
  \bibfield  {author} {\bibinfo {author} {\bibfnamefont {D.}~\bibnamefont
  {Gaiotto}}, \bibinfo {author} {\bibfnamefont {A.}~\bibnamefont {Kapustin}},
  \bibinfo {author} {\bibfnamefont {N.}~\bibnamefont {Seiberg}},\ and\ \bibinfo
  {author} {\bibfnamefont {B.}~\bibnamefont {Willett}},\ }\bibfield  {title}
  {\bibinfo {title} {Generalized global symmetries},\ }\href
  {https://doi.org/10.1007/JHEP02(2015)172} {\bibfield  {journal} {\bibinfo
  {journal} {Journal of High Energy Physics}\ }\textbf {\bibinfo {volume}
  {2015}},\ \bibinfo {pages} {172} (\bibinfo {year} {2015})}\BibitemShut
  {NoStop}%
\bibitem [{\citenamefont {Kapustin}\ and\ \citenamefont
  {Thorngren}(2017{\natexlab{b}})}]{kapustin2015higher}%
  \BibitemOpen
  \bibfield  {author} {\bibinfo {author} {\bibfnamefont {A.}~\bibnamefont
  {Kapustin}}\ and\ \bibinfo {author} {\bibfnamefont {R.}~\bibnamefont
  {Thorngren}},\ }\bibfield  {title} {\bibinfo {title} {Higher symmetry and
  gapped phases of gauge theories},\ }in\ \href
  {https://doi.org/10.1007/978-3-319-59939-7_5} {\emph {\bibinfo {booktitle}
  {Algebra, Geometry, and Physics in the 21st Century}}}\ (\bibinfo
  {publisher} {Springer},\ \bibinfo {year} {2017})\ pp.\ \bibinfo {pages}
  {177--202}\BibitemShut {NoStop}%
\bibitem [{\citenamefont {Hauschild}\ and\ \citenamefont
  {Pollmann}(2018)}]{Hauschild18}%
  \BibitemOpen
  \bibfield  {author} {\bibinfo {author} {\bibfnamefont {J.}~\bibnamefont
  {Hauschild}}\ and\ \bibinfo {author} {\bibfnamefont {F.}~\bibnamefont
  {Pollmann}},\ }\bibfield  {title} {\bibinfo {title} {{Efficient numerical
  simulations with Tensor Networks: Tensor Network Python (TeNPy)}},\ }\href
  {https://doi.org/10.21468/SciPostPhysLectNotes.5} {\bibfield  {journal}
  {\bibinfo  {journal} {SciPost Phys. Lect. Notes}\ ,\ \bibinfo {pages} {5}}
  (\bibinfo {year} {2018})}\BibitemShut {NoStop}%
\bibitem [{\citenamefont {Hastings}(2007)}]{MPSgs}%
  \BibitemOpen
  \bibfield  {author} {\bibinfo {author} {\bibfnamefont {M.~B.}\ \bibnamefont
  {Hastings}},\ }\bibfield  {title} {\bibinfo {title} {An area law for
  one-dimensional quantum systems},\ }\href
  {https://doi.org/10.1088/1742-5468/2007/08/P08024} {\bibfield  {journal}
  {\bibinfo  {journal} {Journal of Statistical Mechanics: Theory and
  Experiment}\ }\textbf {\bibinfo {volume} {2007}},\ \bibinfo {pages} {P08024}
  (\bibinfo {year} {2007})}\BibitemShut {NoStop}%
\bibitem [{\citenamefont {Verstraete}\ and\ \citenamefont
  {Cirac}(2006)}]{MPSgs2}%
  \BibitemOpen
  \bibfield  {author} {\bibinfo {author} {\bibfnamefont {F.}~\bibnamefont
  {Verstraete}}\ and\ \bibinfo {author} {\bibfnamefont {J.~I.}\ \bibnamefont
  {Cirac}},\ }\bibfield  {title} {\bibinfo {title} {Matrix product states
  represent ground states faithfully},\ }\href
  {https://doi.org/10.1103/PhysRevB.73.094423} {\bibfield  {journal} {\bibinfo
  {journal} {Phys. Rev. B}\ }\textbf {\bibinfo {volume} {73}},\ \bibinfo
  {pages} {094423} (\bibinfo {year} {2006})}\BibitemShut {NoStop}%
\bibitem [{\citenamefont {P\'erez-Garc\'{\i}a}\ \emph
  {et~al.}(2008)\citenamefont {P\'erez-Garc\'{\i}a}, \citenamefont {Wolf},
  \citenamefont {Sanz}, \citenamefont {Verstraete},\ and\ \citenamefont
  {Cirac}}]{SymFrac08}%
  \BibitemOpen
  \bibfield  {author} {\bibinfo {author} {\bibfnamefont {D.}~\bibnamefont
  {P\'erez-Garc\'{\i}a}}, \bibinfo {author} {\bibfnamefont {M.~M.}\
  \bibnamefont {Wolf}}, \bibinfo {author} {\bibfnamefont {M.}~\bibnamefont
  {Sanz}}, \bibinfo {author} {\bibfnamefont {F.}~\bibnamefont {Verstraete}},\
  and\ \bibinfo {author} {\bibfnamefont {J.~I.}\ \bibnamefont {Cirac}},\
  }\bibfield  {title} {\bibinfo {title} {String order and symmetries in quantum
  spin lattices},\ }\href {https://doi.org/10.1103/PhysRevLett.100.167202}
  {\bibfield  {journal} {\bibinfo  {journal} {Phys. Rev. Lett.}\ }\textbf
  {\bibinfo {volume} {100}},\ \bibinfo {pages} {167202} (\bibinfo {year}
  {2008})}\BibitemShut {NoStop}%
\bibitem [{\citenamefont {Wolf}\ \emph {et~al.}(2006)\citenamefont {Wolf},
  \citenamefont {Ortiz}, \citenamefont {Verstraete},\ and\ \citenamefont
  {Cirac}}]{Wolf06}%
  \BibitemOpen
  \bibfield  {author} {\bibinfo {author} {\bibfnamefont {M.~M.}\ \bibnamefont
  {Wolf}}, \bibinfo {author} {\bibfnamefont {G.}~\bibnamefont {Ortiz}},
  \bibinfo {author} {\bibfnamefont {F.}~\bibnamefont {Verstraete}},\ and\
  \bibinfo {author} {\bibfnamefont {J.~I.}\ \bibnamefont {Cirac}},\ }\bibfield
  {title} {\bibinfo {title} {Quantum phase transitions in matrix product
  systems},\ }\href {https://doi.org/10.1103/PhysRevLett.97.110403} {\bibfield
  {journal} {\bibinfo  {journal} {Phys. Rev. Lett.}\ }\textbf {\bibinfo
  {volume} {97}},\ \bibinfo {pages} {110403} (\bibinfo {year}
  {2006})}\BibitemShut {NoStop}%
\bibitem [{\citenamefont {Affleck}\ \emph {et~al.}(1987)\citenamefont
  {Affleck}, \citenamefont {Kennedy}, \citenamefont {Lieb},\ and\ \citenamefont
  {Tasaki}}]{AKLT1987}%
  \BibitemOpen
  \bibfield  {author} {\bibinfo {author} {\bibfnamefont {I.}~\bibnamefont
  {Affleck}}, \bibinfo {author} {\bibfnamefont {T.}~\bibnamefont {Kennedy}},
  \bibinfo {author} {\bibfnamefont {E.~H.}\ \bibnamefont {Lieb}},\ and\
  \bibinfo {author} {\bibfnamefont {H.}~\bibnamefont {Tasaki}},\ }\bibfield
  {title} {\bibinfo {title} {Rigorous results on valence-bond ground states in
  antiferromagnets},\ }\href {https://doi.org/10.1103/PhysRevLett.59.799}
  {\bibfield  {journal} {\bibinfo  {journal} {Phys. Rev. Lett.}\ }\textbf
  {\bibinfo {volume} {59}},\ \bibinfo {pages} {799} (\bibinfo {year}
  {1987})}\BibitemShut {NoStop}%
\bibitem [{\citenamefont {Haldane}(1983)}]{Haldane1983}%
  \BibitemOpen
  \bibfield  {author} {\bibinfo {author} {\bibfnamefont {F.}~\bibnamefont
  {Haldane}},\ }\bibfield  {title} {\bibinfo {title} {Continuum dynamics of the
  1-d heisenberg antiferromagnet: Identification with the o(3) nonlinear sigma
  model},\ }\href {https://doi.org/10.1016/0375-9601(83)90631-X} {\bibfield
  {journal} {\bibinfo  {journal} {Physics Letters A}\ }\textbf {\bibinfo
  {volume} {93}},\ \bibinfo {pages} {464 } (\bibinfo {year}
  {1983})}\BibitemShut {NoStop}%
\bibitem [{\citenamefont {den Nijs}\ and\ \citenamefont
  {Rommelse}(1989)}]{dennijs89}%
  \BibitemOpen
  \bibfield  {author} {\bibinfo {author} {\bibfnamefont {M.}~\bibnamefont {den
  Nijs}}\ and\ \bibinfo {author} {\bibfnamefont {K.}~\bibnamefont {Rommelse}},\
  }\bibfield  {title} {\bibinfo {title} {Preroughening transitions in crystal
  surfaces and valence-bond phases in quantum spin chains},\ }\href
  {https://doi.org/10.1103/PhysRevB.40.4709} {\bibfield  {journal} {\bibinfo
  {journal} {Phys. Rev. B}\ }\textbf {\bibinfo {volume} {40}},\ \bibinfo
  {pages} {4709} (\bibinfo {year} {1989})}\BibitemShut {NoStop}%
\bibitem [{\citenamefont {Pollmann}\ \emph
  {et~al.}(2010{\natexlab{b}})\citenamefont {Pollmann}, \citenamefont {Turner},
  \citenamefont {Berg},\ and\ \citenamefont
  {Oshikawa}}]{pollmann_entanglement_2010}%
  \BibitemOpen
  \bibfield  {author} {\bibinfo {author} {\bibfnamefont {F.}~\bibnamefont
  {Pollmann}}, \bibinfo {author} {\bibfnamefont {A.~M.}\ \bibnamefont
  {Turner}}, \bibinfo {author} {\bibfnamefont {E.}~\bibnamefont {Berg}},\ and\
  \bibinfo {author} {\bibfnamefont {M.}~\bibnamefont {Oshikawa}},\ }\bibfield
  {title} {\bibinfo {title} {Entanglement spectrum of a topological phase in
  one dimension},\ }\href {https://doi.org/10.1103/PhysRevB.81.064439}
  {\bibfield  {journal} {\bibinfo  {journal} {Phys. Rev. B}\ }\textbf {\bibinfo
  {volume} {81}},\ \bibinfo {pages} {064439} (\bibinfo {year}
  {2010}{\natexlab{b}})}\BibitemShut {NoStop}%
\bibitem [{\citenamefont {White}(1992)}]{White92}%
  \BibitemOpen
  \bibfield  {author} {\bibinfo {author} {\bibfnamefont {S.~R.}\ \bibnamefont
  {White}},\ }\bibfield  {title} {\bibinfo {title} {Density matrix formulation
  for quantum renormalization groups},\ }\href
  {https://doi.org/10.1103/PhysRevLett.69.2863} {\bibfield  {journal} {\bibinfo
   {journal} {Phys. Rev. Lett.}\ }\textbf {\bibinfo {volume} {69}},\ \bibinfo
  {pages} {2863} (\bibinfo {year} {1992})}\BibitemShut {NoStop}%
\bibitem [{\citenamefont {White}(1993)}]{White93}%
  \BibitemOpen
  \bibfield  {author} {\bibinfo {author} {\bibfnamefont {S.~R.}\ \bibnamefont
  {White}},\ }\bibfield  {title} {\bibinfo {title} {Density-matrix algorithms
  for quantum renormalization groups},\ }\href
  {https://doi.org/10.1103/PhysRevB.48.10345} {\bibfield  {journal} {\bibinfo
  {journal} {Phys. Rev. B}\ }\textbf {\bibinfo {volume} {48}},\ \bibinfo
  {pages} {10345} (\bibinfo {year} {1993})}\BibitemShut {NoStop}%
\bibitem [{\citenamefont {Yoshida}(2015)}]{yoshida_topological_2015}%
  \BibitemOpen
  \bibfield  {author} {\bibinfo {author} {\bibfnamefont {B.}~\bibnamefont
  {Yoshida}},\ }\bibfield  {title} {\bibinfo {title} {Topological color code
  and symmetry-protected topological phases},\ }\href
  {https://doi.org/10.1103/PhysRevB.91.245131} {\bibfield  {journal} {\bibinfo
  {journal} {Phys. Rev. B}\ }\textbf {\bibinfo {volume} {91}},\ \bibinfo
  {pages} {245131} (\bibinfo {year} {2015})}\BibitemShut {NoStop}%
\bibitem [{\citenamefont {Chen}\ \emph {et~al.}(2014)\citenamefont {Chen},
  \citenamefont {Lu},\ and\ \citenamefont {Vishwanath}}]{decorateddomainwalls}%
  \BibitemOpen
  \bibfield  {author} {\bibinfo {author} {\bibfnamefont {X.}~\bibnamefont
  {Chen}}, \bibinfo {author} {\bibfnamefont {Y.-M.}\ \bibnamefont {Lu}},\ and\
  \bibinfo {author} {\bibfnamefont {A.}~\bibnamefont {Vishwanath}},\ }\bibfield
   {title} {\bibinfo {title} {Symmetry-protected topological phases from
  decorated domain walls},\ }\href {https://doi.org/10.1038/ncomms4507}
  {\bibfield  {journal} {\bibinfo  {journal} {Nature communications}\ }\textbf
  {\bibinfo {volume} {5}},\ \bibinfo {pages} {3507} (\bibinfo {year}
  {2014})}\BibitemShut {NoStop}%
\bibitem [{\citenamefont {Komargodski}\ \emph {et~al.}(2019)\citenamefont
  {Komargodski}, \citenamefont {Sharon}, \citenamefont {Thorngren},\ and\
  \citenamefont {Zhou}}]{Komargodski_2019}%
  \BibitemOpen
  \bibfield  {author} {\bibinfo {author} {\bibfnamefont {Z.}~\bibnamefont
  {Komargodski}}, \bibinfo {author} {\bibfnamefont {A.}~\bibnamefont {Sharon}},
  \bibinfo {author} {\bibfnamefont {R.}~\bibnamefont {Thorngren}},\ and\
  \bibinfo {author} {\bibfnamefont {X.}~\bibnamefont {Zhou}},\ }\bibfield
  {title} {\bibinfo {title} {{Comments on abelian Higgs models and persistent
  order}},\ }\href {https://doi.org/10.21468/SciPostPhys.6.1.003} {\bibfield
  {journal} {\bibinfo  {journal} {SciPost Phys.}\ }\textbf {\bibinfo {volume}
  {6}},\ \bibinfo {pages} {003} (\bibinfo {year} {2019})}\BibitemShut {NoStop}%
\bibitem [{\citenamefont {Fr\"ohlich}\ \emph {et~al.}(2004)\citenamefont
  {Fr\"ohlich}, \citenamefont {Fuchs}, \citenamefont {Runkel},\ and\
  \citenamefont {Schweigert}}]{frohlichKW}%
  \BibitemOpen
  \bibfield  {author} {\bibinfo {author} {\bibfnamefont {J.}~\bibnamefont
  {Fr\"ohlich}}, \bibinfo {author} {\bibfnamefont {J.}~\bibnamefont {Fuchs}},
  \bibinfo {author} {\bibfnamefont {I.}~\bibnamefont {Runkel}},\ and\ \bibinfo
  {author} {\bibfnamefont {C.}~\bibnamefont {Schweigert}},\ }\bibfield  {title}
  {\bibinfo {title} {Kramers-wannier duality from conformal defects},\ }\href
  {https://doi.org/10.1103/PhysRevLett.93.070601} {\bibfield  {journal}
  {\bibinfo  {journal} {Phys. Rev. Lett.}\ }\textbf {\bibinfo {volume} {93}},\
  \bibinfo {pages} {070601} (\bibinfo {year} {2004})}\BibitemShut {NoStop}%
\bibitem [{\citenamefont {Aasen}\ \emph
  {et~al.}(2016{\natexlab{b}})\citenamefont {Aasen}, \citenamefont {Mong},\
  and\ \citenamefont {Fendley}}]{aasen_topological_2016}%
  \BibitemOpen
  \bibfield  {author} {\bibinfo {author} {\bibfnamefont {D.}~\bibnamefont
  {Aasen}}, \bibinfo {author} {\bibfnamefont {R.~S.~K.}\ \bibnamefont {Mong}},\
  and\ \bibinfo {author} {\bibfnamefont {P.}~\bibnamefont {Fendley}},\
  }\bibfield  {title} {\bibinfo {title} {Topological defects on the lattice: I.
  the ising model},\ }\href {https://doi.org/10.1088/1751-8113/49/35/354001}
  {\bibfield  {journal} {\bibinfo  {journal} {Journal of Physics A:
  Mathematical and Theoretical}\ }\textbf {\bibinfo {volume} {49}},\ \bibinfo
  {pages} {354001} (\bibinfo {year} {2016}{\natexlab{b}})}\BibitemShut
  {NoStop}%
\bibitem [{\citenamefont {Chang}\ \emph {et~al.}(2019)\citenamefont {Chang},
  \citenamefont {Lin}, \citenamefont {Shao}, \citenamefont {Wang},\ and\
  \citenamefont {Yin}}]{tdl}%
  \BibitemOpen
  \bibfield  {author} {\bibinfo {author} {\bibfnamefont {C.-M.}\ \bibnamefont
  {Chang}}, \bibinfo {author} {\bibfnamefont {Y.-H.}\ \bibnamefont {Lin}},
  \bibinfo {author} {\bibfnamefont {S.-H.}\ \bibnamefont {Shao}}, \bibinfo
  {author} {\bibfnamefont {Y.}~\bibnamefont {Wang}},\ and\ \bibinfo {author}
  {\bibfnamefont {X.}~\bibnamefont {Yin}},\ }\bibfield  {title} {\bibinfo
  {title} {Topological defect lines and renormalization group flows in two
  dimensions},\ }\href {https://doi.org/10.1007/jhep01(2019)026} {\bibfield
  {journal} {\bibinfo  {journal} {Journal of High Energy Physics}\ }\textbf
  {\bibinfo {volume} {2019}},\ \bibinfo {pages} {1} (\bibinfo {year}
  {2019})}\BibitemShut {NoStop}%
\bibitem [{\citenamefont {Bennett}\ \emph {et~al.}(1993)\citenamefont
  {Bennett}, \citenamefont {Brassard}, \citenamefont {Cr\'epeau}, \citenamefont
  {Jozsa}, \citenamefont {Peres},\ and\ \citenamefont
  {Wootters}}]{Teleportation}%
  \BibitemOpen
  \bibfield  {author} {\bibinfo {author} {\bibfnamefont {C.~H.}\ \bibnamefont
  {Bennett}}, \bibinfo {author} {\bibfnamefont {G.}~\bibnamefont {Brassard}},
  \bibinfo {author} {\bibfnamefont {C.}~\bibnamefont {Cr\'epeau}}, \bibinfo
  {author} {\bibfnamefont {R.}~\bibnamefont {Jozsa}}, \bibinfo {author}
  {\bibfnamefont {A.}~\bibnamefont {Peres}},\ and\ \bibinfo {author}
  {\bibfnamefont {W.~K.}\ \bibnamefont {Wootters}},\ }\bibfield  {title}
  {\bibinfo {title} {Teleporting an unknown quantum state via dual classical
  and einstein-podolsky-rosen channels},\ }\href
  {https://doi.org/10.1103/PhysRevLett.70.1895} {\bibfield  {journal} {\bibinfo
   {journal} {Phys. Rev. Lett.}\ }\textbf {\bibinfo {volume} {70}},\ \bibinfo
  {pages} {1895} (\bibinfo {year} {1993})}\BibitemShut {NoStop}%
\bibitem [{\citenamefont {Gottesman}\ and\ \citenamefont
  {Chuang}(1999)}]{GottesmanChuang1999}%
  \BibitemOpen
  \bibfield  {author} {\bibinfo {author} {\bibfnamefont {D.}~\bibnamefont
  {Gottesman}}\ and\ \bibinfo {author} {\bibfnamefont {I.~L.}\ \bibnamefont
  {Chuang}},\ }\bibfield  {title} {\bibinfo {title} {Demonstrating the
  viability of universal quantum computation using teleportation and
  single-qubit operations},\ }\href {https://doi.org/10.1038/46503} {\bibfield
  {journal} {\bibinfo  {journal} {Nature}\ }\textbf {\bibinfo {volume} {402}},\
  \bibinfo {pages} {390} (\bibinfo {year} {1999})}\BibitemShut {NoStop}%
\bibitem [{\citenamefont {Briegel}\ and\ \citenamefont
  {Raussendorf}(2001{\natexlab{b}})}]{BriegelRaussendorf2001}%
  \BibitemOpen
  \bibfield  {author} {\bibinfo {author} {\bibfnamefont {H.~J.}\ \bibnamefont
  {Briegel}}\ and\ \bibinfo {author} {\bibfnamefont {R.}~\bibnamefont
  {Raussendorf}},\ }\bibfield  {title} {\bibinfo {title} {Persistent
  entanglement in arrays of interacting particles},\ }\href
  {https://doi.org/10.1103/PhysRevLett.86.910} {\bibfield  {journal} {\bibinfo
  {journal} {Phys. Rev. Lett.}\ }\textbf {\bibinfo {volume} {86}},\ \bibinfo
  {pages} {910} (\bibinfo {year} {2001}{\natexlab{b}})}\BibitemShut {NoStop}%
\bibitem [{\citenamefont {Raussendorf}\ and\ \citenamefont
  {Briegel}(2001)}]{RaussendorfBriegel2001}%
  \BibitemOpen
  \bibfield  {author} {\bibinfo {author} {\bibfnamefont {R.}~\bibnamefont
  {Raussendorf}}\ and\ \bibinfo {author} {\bibfnamefont {H.~J.}\ \bibnamefont
  {Briegel}},\ }\bibfield  {title} {\bibinfo {title} {A one-way quantum
  computer},\ }\href {https://doi.org/10.1103/PhysRevLett.86.5188} {\bibfield
  {journal} {\bibinfo  {journal} {Phys. Rev. Lett.}\ }\textbf {\bibinfo
  {volume} {86}},\ \bibinfo {pages} {5188} (\bibinfo {year}
  {2001})}\BibitemShut {NoStop}%
\bibitem [{\citenamefont {Raussendorf}\ \emph {et~al.}(2003)\citenamefont
  {Raussendorf}, \citenamefont {Browne},\ and\ \citenamefont
  {Briegel}}]{RaussendorfBrowneBriegel2003}%
  \BibitemOpen
  \bibfield  {author} {\bibinfo {author} {\bibfnamefont {R.}~\bibnamefont
  {Raussendorf}}, \bibinfo {author} {\bibfnamefont {D.~E.}\ \bibnamefont
  {Browne}},\ and\ \bibinfo {author} {\bibfnamefont {H.~J.}\ \bibnamefont
  {Briegel}},\ }\bibfield  {title} {\bibinfo {title} {Measurement-based quantum
  computation on cluster states},\ }\href
  {https://doi.org/10.1103/PhysRevA.68.022312} {\bibfield  {journal} {\bibinfo
  {journal} {Phys. Rev. A}\ }\textbf {\bibinfo {volume} {68}},\ \bibinfo
  {pages} {022312} (\bibinfo {year} {2003})}\BibitemShut {NoStop}%
\bibitem [{\citenamefont {Raussendorf}\ \emph {et~al.}(2017)\citenamefont
  {Raussendorf}, \citenamefont {Wang}, \citenamefont {Prakash}, \citenamefont
  {Wei},\ and\ \citenamefont {Stephen}}]{Raussendorfetal2017}%
  \BibitemOpen
  \bibfield  {author} {\bibinfo {author} {\bibfnamefont {R.}~\bibnamefont
  {Raussendorf}}, \bibinfo {author} {\bibfnamefont {D.-S.}\ \bibnamefont
  {Wang}}, \bibinfo {author} {\bibfnamefont {A.}~\bibnamefont {Prakash}},
  \bibinfo {author} {\bibfnamefont {T.-C.}\ \bibnamefont {Wei}},\ and\ \bibinfo
  {author} {\bibfnamefont {D.~T.}\ \bibnamefont {Stephen}},\ }\bibfield
  {title} {\bibinfo {title} {Symmetry-protected topological phases with uniform
  computational power in one dimension},\ }\href
  {https://doi.org/10.1103/PhysRevA.96.012302} {\bibfield  {journal} {\bibinfo
  {journal} {Phys. Rev. A}\ }\textbf {\bibinfo {volume} {96}},\ \bibinfo
  {pages} {012302} (\bibinfo {year} {2017})}\BibitemShut {NoStop}%
\bibitem [{\citenamefont {Raussendorf}\ \emph {et~al.}(2019)\citenamefont
  {Raussendorf}, \citenamefont {Okay}, \citenamefont {Wang}, \citenamefont
  {Stephen},\ and\ \citenamefont {Nautrup}}]{Raussendorfetal2018}%
  \BibitemOpen
  \bibfield  {author} {\bibinfo {author} {\bibfnamefont {R.}~\bibnamefont
  {Raussendorf}}, \bibinfo {author} {\bibfnamefont {C.}~\bibnamefont {Okay}},
  \bibinfo {author} {\bibfnamefont {D.-S.}\ \bibnamefont {Wang}}, \bibinfo
  {author} {\bibfnamefont {D.~T.}\ \bibnamefont {Stephen}},\ and\ \bibinfo
  {author} {\bibfnamefont {H.~P.}\ \bibnamefont {Nautrup}},\ }\bibfield
  {title} {\bibinfo {title} {Computationally universal phase of quantum
  matter},\ }\href {https://doi.org/10.1103/PhysRevLett.122.090501} {\bibfield
  {journal} {\bibinfo  {journal} {Phys. Rev. Lett.}\ }\textbf {\bibinfo
  {volume} {122}},\ \bibinfo {pages} {090501} (\bibinfo {year}
  {2019})}\BibitemShut {NoStop}%
\bibitem [{\citenamefont {Stephen}\ \emph {et~al.}(2017)\citenamefont
  {Stephen}, \citenamefont {Wang}, \citenamefont {Prakash}, \citenamefont
  {Wei},\ and\ \citenamefont {Raussendorf}}]{Stephenetal2017}%
  \BibitemOpen
  \bibfield  {author} {\bibinfo {author} {\bibfnamefont {D.~T.}\ \bibnamefont
  {Stephen}}, \bibinfo {author} {\bibfnamefont {D.-S.}\ \bibnamefont {Wang}},
  \bibinfo {author} {\bibfnamefont {A.}~\bibnamefont {Prakash}}, \bibinfo
  {author} {\bibfnamefont {T.-C.}\ \bibnamefont {Wei}},\ and\ \bibinfo {author}
  {\bibfnamefont {R.}~\bibnamefont {Raussendorf}},\ }\bibfield  {title}
  {\bibinfo {title} {Computational power of symmetry-protected topological
  phases},\ }\href {https://doi.org/10.1103/PhysRevLett.119.010504} {\bibfield
  {journal} {\bibinfo  {journal} {Phys. Rev. Lett.}\ }\textbf {\bibinfo
  {volume} {119}},\ \bibinfo {pages} {010504} (\bibinfo {year}
  {2017})}\BibitemShut {NoStop}%
\bibitem [{\citenamefont {Devakul}\ and\ \citenamefont
  {Williamson}(2018)}]{DevakulWilliamson2018}%
  \BibitemOpen
  \bibfield  {author} {\bibinfo {author} {\bibfnamefont {T.}~\bibnamefont
  {Devakul}}\ and\ \bibinfo {author} {\bibfnamefont {D.~J.}\ \bibnamefont
  {Williamson}},\ }\bibfield  {title} {\bibinfo {title} {Universal quantum
  computation using fractal symmetry-protected cluster phases},\ }\href
  {https://doi.org/10.1103/PhysRevA.98.022332} {\bibfield  {journal} {\bibinfo
  {journal} {Phys. Rev. A}\ }\textbf {\bibinfo {volume} {98}},\ \bibinfo
  {pages} {022332} (\bibinfo {year} {2018})}\BibitemShut {NoStop}%
\bibitem [{\citenamefont {Daniel}\ \emph {et~al.}(2020)\citenamefont {Daniel},
  \citenamefont {Alexander},\ and\ \citenamefont
  {Miyake}}]{DanielAlexanderMiyake20}%
  \BibitemOpen
  \bibfield  {author} {\bibinfo {author} {\bibfnamefont {A.~K.}\ \bibnamefont
  {Daniel}}, \bibinfo {author} {\bibfnamefont {R.~N.}\ \bibnamefont
  {Alexander}},\ and\ \bibinfo {author} {\bibfnamefont {A.}~\bibnamefont
  {Miyake}},\ }\bibfield  {title} {\bibinfo {title} {Computational universality
  of symmetry-protected topologically ordered cluster phases on 2{D}
  {A}rchimedean lattices},\ }\href {https://doi.org/10.22331/q-2020-02-10-228}
  {\bibfield  {journal} {\bibinfo  {journal} {{Quantum}}\ }\textbf {\bibinfo
  {volume} {4}},\ \bibinfo {pages} {228} (\bibinfo {year} {2020})}\BibitemShut
  {NoStop}%
\bibitem [{\citenamefont {Hsieh}\ \emph {et~al.}(2017)\citenamefont {Hsieh},
  \citenamefont {Lu},\ and\ \citenamefont {Ludwig}}]{LuHsieh}%
  \BibitemOpen
  \bibfield  {author} {\bibinfo {author} {\bibfnamefont {T.~H.}\ \bibnamefont
  {Hsieh}}, \bibinfo {author} {\bibfnamefont {Y.-M.}\ \bibnamefont {Lu}},\ and\
  \bibinfo {author} {\bibfnamefont {A.~W.}\ \bibnamefont {Ludwig}},\ }\bibfield
   {title} {\bibinfo {title} {Topological bootstrap: Fractionalization from
  kondo coupling},\ }\href {https://doi.org/10.1126/sciadv.1700729} {\bibfield
  {journal} {\bibinfo  {journal} {Science advances}\ }\textbf {\bibinfo
  {volume} {3}},\ \bibinfo {pages} {e1700729} (\bibinfo {year}
  {2017})}\BibitemShut {NoStop}%
\bibitem [{\citenamefont {Ashkenazi}\ and\ \citenamefont
  {Zohar}(2022)}]{Ashkenazi21}%
  \BibitemOpen
  \bibfield  {author} {\bibinfo {author} {\bibfnamefont {S.}~\bibnamefont
  {Ashkenazi}}\ and\ \bibinfo {author} {\bibfnamefont {E.}~\bibnamefont
  {Zohar}},\ }\bibfield  {title} {\bibinfo {title} {Duality as a feasible
  physical transformation for quantum simulation},\ }\href
  {https://doi.org/10.1103/PhysRevA.105.022431} {\bibfield  {journal} {\bibinfo
   {journal} {Phys. Rev. A}\ }\textbf {\bibinfo {volume} {105}},\ \bibinfo
  {pages} {022431} (\bibinfo {year} {2022})}\BibitemShut {NoStop}%
\bibitem [{\citenamefont {Bravyi}\ \emph {et~al.}(2022)\citenamefont {Bravyi},
  \citenamefont {Kim}, \citenamefont {Kliesch},\ and\ \citenamefont
  {Koenig}}]{Bravyi22}%
  \BibitemOpen
  \bibfield  {author} {\bibinfo {author} {\bibfnamefont {S.}~\bibnamefont
  {Bravyi}}, \bibinfo {author} {\bibfnamefont {I.}~\bibnamefont {Kim}},
  \bibinfo {author} {\bibfnamefont {A.}~\bibnamefont {Kliesch}},\ and\ \bibinfo
  {author} {\bibfnamefont {R.}~\bibnamefont {Koenig}},\ }\bibfield  {title}
  {\bibinfo {title} {Adaptive constant-depth circuits for manipulating
  non-abelian anyons},\ }\href@noop {} {\bibfield  {journal} {\bibinfo
  {journal} {arXiv preprint arXiv:2205.01933}\ } (\bibinfo {year}
  {2022})}\BibitemShut {NoStop}%
\bibitem [{\citenamefont {Devakul}\ \emph {et~al.}(2018)\citenamefont
  {Devakul}, \citenamefont {Williamson},\ and\ \citenamefont
  {You}}]{Devakul18}%
  \BibitemOpen
  \bibfield  {author} {\bibinfo {author} {\bibfnamefont {T.}~\bibnamefont
  {Devakul}}, \bibinfo {author} {\bibfnamefont {D.~J.}\ \bibnamefont
  {Williamson}},\ and\ \bibinfo {author} {\bibfnamefont {Y.}~\bibnamefont
  {You}},\ }\bibfield  {title} {\bibinfo {title} {Classification of subsystem
  symmetry-protected topological phases},\ }\href
  {https://doi.org/10.1103/PhysRevB.98.235121} {\bibfield  {journal} {\bibinfo
  {journal} {Phys. Rev. B}\ }\textbf {\bibinfo {volume} {98}},\ \bibinfo
  {pages} {235121} (\bibinfo {year} {2018})}\BibitemShut {NoStop}%
\bibitem [{\citenamefont {Tantivasadakarn}\ and\ \citenamefont
  {Vijay}(2020)}]{TantivasadakarnVijay20}%
  \BibitemOpen
  \bibfield  {author} {\bibinfo {author} {\bibfnamefont {N.}~\bibnamefont
  {Tantivasadakarn}}\ and\ \bibinfo {author} {\bibfnamefont {S.}~\bibnamefont
  {Vijay}},\ }\bibfield  {title} {\bibinfo {title} {Searching for fracton
  orders via symmetry defect condensation},\ }\href
  {https://doi.org/10.1103/PhysRevB.101.165143} {\bibfield  {journal} {\bibinfo
   {journal} {Phys. Rev. B}\ }\textbf {\bibinfo {volume} {101}},\ \bibinfo
  {pages} {165143} (\bibinfo {year} {2020})}\BibitemShut {NoStop}%
\bibitem [{\citenamefont {Kitaev}\ and\ \citenamefont
  {Kong}(2012)}]{KitaevKong12}%
  \BibitemOpen
  \bibfield  {author} {\bibinfo {author} {\bibfnamefont {A.}~\bibnamefont
  {Kitaev}}\ and\ \bibinfo {author} {\bibfnamefont {L.}~\bibnamefont {Kong}},\
  }\bibfield  {title} {\bibinfo {title} {Models for gapped boundaries and
  domain walls},\ }\href {https://doi.org/10.1007/s00220-012-1500-5} {\bibfield
   {journal} {\bibinfo  {journal} {Communications in Mathematical Physics}\
  }\textbf {\bibinfo {volume} {313}},\ \bibinfo {pages} {351} (\bibinfo {year}
  {2012})}\BibitemShut {NoStop}%
\bibitem [{\citenamefont {Burnell}\ \emph {et~al.}(2014)\citenamefont
  {Burnell}, \citenamefont {Chen}, \citenamefont {Fidkowski},\ and\
  \citenamefont {Vishwanath}}]{BCFV14}%
  \BibitemOpen
  \bibfield  {author} {\bibinfo {author} {\bibfnamefont {F.~J.}\ \bibnamefont
  {Burnell}}, \bibinfo {author} {\bibfnamefont {X.}~\bibnamefont {Chen}},
  \bibinfo {author} {\bibfnamefont {L.}~\bibnamefont {Fidkowski}},\ and\
  \bibinfo {author} {\bibfnamefont {A.}~\bibnamefont {Vishwanath}},\ }\bibfield
   {title} {\bibinfo {title} {Exactly soluble model of a three-dimensional
  symmetry-protected topological phase of bosons with surface topological
  order},\ }\href {https://doi.org/10.1103/PhysRevB.90.245122} {\bibfield
  {journal} {\bibinfo  {journal} {Phys. Rev. B}\ }\textbf {\bibinfo {volume}
  {90}},\ \bibinfo {pages} {245122} (\bibinfo {year} {2014})}\BibitemShut
  {NoStop}%
\bibitem [{\citenamefont {Haah}\ \emph {et~al.}(2023)\citenamefont {Haah},
  \citenamefont {Fidkowski},\ and\ \citenamefont {Hastings}}]{HFH18}%
  \BibitemOpen
  \bibfield  {author} {\bibinfo {author} {\bibfnamefont {J.}~\bibnamefont
  {Haah}}, \bibinfo {author} {\bibfnamefont {L.}~\bibnamefont {Fidkowski}},\
  and\ \bibinfo {author} {\bibfnamefont {M.~B.}\ \bibnamefont {Hastings}},\
  }\bibfield  {title} {\bibinfo {title} {Nontrivial quantum cellular automata
  in higher dimensions},\ }\href {https://doi.org/10.1007/s00220-022-04528-1}
  {\bibfield  {journal} {\bibinfo  {journal} {Communications in Mathematical
  Physics}\ }\textbf {\bibinfo {volume} {398}},\ \bibinfo {pages} {469}
  (\bibinfo {year} {2023})}\BibitemShut {NoStop}%
\bibitem [{\citenamefont {Walker}\ and\ \citenamefont {Wang}(2012)}]{WW12}%
  \BibitemOpen
  \bibfield  {author} {\bibinfo {author} {\bibfnamefont {K.}~\bibnamefont
  {Walker}}\ and\ \bibinfo {author} {\bibfnamefont {Z.}~\bibnamefont {Wang}},\
  }\bibfield  {title} {\bibinfo {title} {(3+1)-tqfts and topological
  insulators},\ }\href {https://doi.org/10.1007/s11467-011-0194-z} {\bibfield
  {journal} {\bibinfo  {journal} {Frontiers of Physics}\ }\textbf {\bibinfo
  {volume} {7}},\ \bibinfo {pages} {150} (\bibinfo {year} {2012})}\BibitemShut
  {NoStop}%
\bibitem [{\citenamefont {Raussendorf}\ \emph {et~al.}(2006)\citenamefont
  {Raussendorf}, \citenamefont {Harrington},\ and\ \citenamefont
  {Goyal}}]{Raussendorf06}%
  \BibitemOpen
  \bibfield  {author} {\bibinfo {author} {\bibfnamefont {R.}~\bibnamefont
  {Raussendorf}}, \bibinfo {author} {\bibfnamefont {J.}~\bibnamefont
  {Harrington}},\ and\ \bibinfo {author} {\bibfnamefont {K.}~\bibnamefont
  {Goyal}},\ }\bibfield  {title} {\bibinfo {title} {A fault-tolerant one-way
  quantum computer},\ }\href
  {https://doi.org/https://doi.org/10.1016/j.aop.2006.01.012} {\bibfield
  {journal} {\bibinfo  {journal} {Annals of Physics}\ }\textbf {\bibinfo
  {volume} {321}},\ \bibinfo {pages} {2242} (\bibinfo {year}
  {2006})}\BibitemShut {NoStop}%
\bibitem [{\citenamefont {Roberts}\ and\ \citenamefont
  {Williamson}(2024)}]{Roberts2020}%
  \BibitemOpen
  \bibfield  {author} {\bibinfo {author} {\bibfnamefont {S.}~\bibnamefont
  {Roberts}}\ and\ \bibinfo {author} {\bibfnamefont {D.~J.}\ \bibnamefont
  {Williamson}},\ }\bibfield  {title} {\bibinfo {title} {3-fermion topological
  quantum computation},\ }\href {https://doi.org/10.1103/PRXQuantum.5.010315}
  {\bibfield  {journal} {\bibinfo  {journal} {PRX Quantum}\ }\textbf {\bibinfo
  {volume} {5}},\ \bibinfo {pages} {010315} (\bibinfo {year}
  {2024})}\BibitemShut {NoStop}%
\bibitem [{\citenamefont {Chen}\ and\ \citenamefont {Tata}(2023)}]{ChenTata21}%
  \BibitemOpen
  \bibfield  {author} {\bibinfo {author} {\bibfnamefont {Y.-A.}\ \bibnamefont
  {Chen}}\ and\ \bibinfo {author} {\bibfnamefont {S.}~\bibnamefont {Tata}},\
  }\bibfield  {title} {\bibinfo {title} {{Higher cup products on hypercubic
  lattices: Application to lattice models of topological phases}},\ }\href
  {https://doi.org/10.1063/5.0095189} {\bibfield  {journal} {\bibinfo
  {journal} {Journal of Mathematical Physics}\ }\textbf {\bibinfo {volume}
  {64}},\ \bibinfo {pages} {091902} (\bibinfo {year} {2023})}\BibitemShut
  {NoStop}%
\bibitem [{\citenamefont {Haah}(2013)}]{Haah13}%
  \BibitemOpen
  \bibfield  {author} {\bibinfo {author} {\bibfnamefont {J.}~\bibnamefont
  {Haah}},\ }\bibfield  {title} {\bibinfo {title} {Commuting pauli hamiltonians
  as maps between free modules},\ }\href
  {https://doi.org/10.1007/s00220-013-1810-2} {\bibfield  {journal} {\bibinfo
  {journal} {Communications in Mathematical Physics}\ }\textbf {\bibinfo
  {volume} {324}},\ \bibinfo {pages} {351–399} (\bibinfo {year}
  {2013})}\BibitemShut {NoStop}%
\bibitem [{\citenamefont {Tsui}\ and\ \citenamefont {Wen}(2020)}]{TsuiWen20}%
  \BibitemOpen
  \bibfield  {author} {\bibinfo {author} {\bibfnamefont {L.}~\bibnamefont
  {Tsui}}\ and\ \bibinfo {author} {\bibfnamefont {X.-G.}\ \bibnamefont {Wen}},\
  }\bibfield  {title} {\bibinfo {title} {Lattice models that realize
  $\mathbb{Z}_{n}$-1 symmetry-protected topological states for even $n$},\
  }\href {https://doi.org/10.1103/PhysRevB.101.035101} {\bibfield  {journal}
  {\bibinfo  {journal} {Phys. Rev. B}\ }\textbf {\bibinfo {volume} {101}},\
  \bibinfo {pages} {035101} (\bibinfo {year} {2020})}\BibitemShut {NoStop}%
\bibitem [{\citenamefont {Raussendorf}\ \emph
  {et~al.}(2005{\natexlab{b}})\citenamefont {Raussendorf}, \citenamefont
  {Bravyi},\ and\ \citenamefont {Harrington}}]{RaussendorfBravyiHarrington05}%
  \BibitemOpen
  \bibfield  {author} {\bibinfo {author} {\bibfnamefont {R.}~\bibnamefont
  {Raussendorf}}, \bibinfo {author} {\bibfnamefont {S.}~\bibnamefont
  {Bravyi}},\ and\ \bibinfo {author} {\bibfnamefont {J.}~\bibnamefont
  {Harrington}},\ }\bibfield  {title} {\bibinfo {title} {Long-range quantum
  entanglement in noisy cluster states},\ }\href
  {https://doi.org/10.1103/PhysRevA.71.062313} {\bibfield  {journal} {\bibinfo
  {journal} {Phys. Rev. A}\ }\textbf {\bibinfo {volume} {71}},\ \bibinfo
  {pages} {062313} (\bibinfo {year} {2005}{\natexlab{b}})}\BibitemShut
  {NoStop}%
\end{thebibliography}%

\onecolumngrid
\appendix

\newpage

\begin{figure}[t!]
    \centering
    \includegraphics[scale=1.2]{SPTtoKW.pdf}
    \caption{The KW MPO is obtained by starting with the MPS of the 1D cluster state flipping the legs on the B (blue) sublattice. Generalized KW dualities can be similarly obtained by a cluster state which is a nontrivial SPT protected by the desired symmetries on a bipartite lattice.}
    \label{fig:clustertoKW}
\end{figure}

\section{Matrix product for the 1D cluster state and KW}
Consider a one-dimensional lattice of $2N$ qubits. We identify two sublattices $A$ and $B$ corresponding to the odd and even sites of the lattice, respectively. The 1D cluster state can be expressed using a MPS as
\begin{align}
    \ket{\psi} = \sum_{\{s\}} \textrm{Tr}[C^{s_1} C^{s_2} \cdots  C^{s_{2N}}]\ket{s_1,s_2,...,s_{2N}},
\end{align}
where $s_n=0,1$ are $Z$-basis states and the tensor $C$ is defined as
\begin{align}
    C = \frac{1}{\sqrt{2}}\begin{pmatrix}
   \bra{0} & \bra{0}\\
   \bra{1} & -\bra{1}\\   
   \end{pmatrix}.
\end{align}
To turn this into a matrix product operator (MPO), we first double the unit cell to get an MPS with double the physical legs,
\begin{align}
    C \otimes C = \begin{pmatrix}
   \bra{0+} & \bra{1-}\\
   \bra{0-} & \bra{1+}\\   
   \end{pmatrix}.
\end{align}
Flipping the leg of the first entry upwards (see Fig.~\ref{fig:clustertoKW}) yields the MPO
\begin{align}
    \sigma = \begin{pmatrix}  
   \outer{0}{+} & \outer{1}{-} \\
   \outer{0}{-}  &\outer{1}{+}
   \end{pmatrix}.
\end{align}
This expression is exactly the Kramers-Wannier duality. For example, if we plug in the $\ket{+}$ product state, then we get the MPS for the GHZ state
\begin{align}
\begin{pmatrix}  
   \bra{0} & 0 \\
   0  &\bra{1}
   \end{pmatrix}
\end{align}

\section{More examples}
\subsection{Wen plaquette model} \label{app:Wenplaquette}
Consider the following cluster state given by stabilizers,
    \begin{align}
     \raisebox{-.5\height}{\begin{tikzpicture}[scale=0.5]
\node[label=center:$\color{red}X$] (0) at (0,0) {};
\node[label=center:$\color{red}Z$] (1) at (2,0) {};
\node[label=center:$\color{red}Z$] (2) at (0,2) {};
\node[label=center:$\color{red}Z$] (3) at (2,2) {};
\node[label=center:$\color{red}Z$] (5) at (-2,0) {};
\node[label=center:$\color{red}Z$] (6) at (0,-2) {};
\node[label=center:$\color{red}Z$] (7) at (-2,-2) {};
\draw[-,densely dotted,color=gray] (0) -- (1)  {};
\draw[-,densely dotted,color=gray] (0) -- (2)  {};
\draw[-,densely dotted,color=gray] (1) -- (3)  {};
\draw[-,densely dotted,color=gray] (2) -- (3)  {};
\draw[-,densely dotted,color=gray] (0) -- (5)  {};
\draw[-,densely dotted,color=gray] (0) -- (6)  {};
\draw[-,densely dotted,color=gray] (5) -- (7)  {};
\draw[-,densely dotted,color=gray] (6) -- (7)  {};
\end{tikzpicture}}.
\end{align}
This state is in fact the cluster state on the triangular lattice, although we have placed it on the square lattice. This cluster state is a (strong) $\ZZ_2$ subsystem SPT protected by line symmetries, given by flipping spins along the $x$ and $y$ lines of the square lattice \cite{Devakul18}. In fact, gauging this subsystem SPT gives rise to the Wen plaquette model\cite{TantivasadakarnVijay20}.

Based on this finding, we show how to prepare the Wen plaquette model via measuring an appropriate cluster state. The $A$ and $B$ are the vertices of the square (red) and dual square (blue) sublattices, respectively. We create the cluster state given by the stabilizers
    \begin{align}
     \raisebox{-.5\height}{\begin{tikzpicture}[scale=0.5]
\node[label=center:$\color{red}X$] (0) at (0,0) {};
\node[label=center:$\color{red}Z$] (1) at (2,0) {};
\node[label=center:$\color{red}Z$] (2) at (0,2) {};
\node[label=center:$\color{red}Z$] (3) at (2,2) {};
\node[label=center:$\color{red}Z$] (5) at (-2,0) {};
\node[label=center:$\color{red}Z$] (6) at (0,-2) {};
\node[label=center:$\color{red}Z$] (7) at (-2,-2) {};
\node[label=center:$\color{blue}Z$]  at (1,1) {};
\node[label=center:$\color{blue}Z$]  at (-1,1) {};
\node[label=center:$\color{blue}Z$]  at (1,-1) {};
\node[label=center:$\color{blue}Z$]  at (-1,-1) {};
\draw[-,densely dotted,color=gray] (0) -- (1)  {};
\draw[-,densely dotted,color=gray] (0) -- (2)  {};
\draw[-,densely dotted,color=gray] (1) -- (3)  {};
\draw[-,densely dotted,color=gray] (2) -- (3)  {};
\draw[-,densely dotted,color=gray] (0) -- (5)  {};
\draw[-,densely dotted,color=gray] (0) -- (6)  {};
\draw[-,densely dotted,color=gray] (5) -- (7)  {};
\draw[-,densely dotted,color=gray] (6) -- (7)  {};
\end{tikzpicture}}, &&
     \raisebox{-.5\height}{\begin{tikzpicture}[scale=0.5]
\node[label=center:$\color{red}Z$] (0) at (0,0) {};
\node[label=center:$\color{red}Z$] (1) at (2,0) {};
\node[label=center:$\color{red}Z$] (2) at (0,2) {};
\node[label=center:$\color{red}Z$] (3) at (2,2) {};
\node[label=center:$\color{blue}X$]  at (1,1) {};
\draw[-,densely dotted,color=gray] (0) -- (1)  {};
\draw[-,densely dotted,color=gray] (0) -- (2)  {};
\draw[-,densely dotted,color=gray] (1) -- (3)  {};
\draw[-,densely dotted,color=gray] (2) -- (3)  {};
\end{tikzpicture}}.
\end{align}

Note that because of the couplings within the $A$ sublattice, this cluster state is not bipartite. Now, let us measure the $X$ operators on the $A$ sublattice. The local product of stabilizers that commute with the measurements is
    \begin{align}
    - \raisebox{-.5\height}{\begin{tikzpicture}[scale=0.5]
\node[label=center:$\color{red}X$] (0) at (0,0) {};
\node[label=center:$\color{blue}Y$]  at (1,1) {};
\node[label=center:$\color{blue}Z$]  at (-1,1) {};
\node[label=center:$\color{blue}Z$]  at (1,-1) {};
\node[label=center:$\color{blue}Y$]  at (-1,-1) {};
\draw[-,densely dotted,color=gray] (0) -- (1)  {};
\draw[-,densely dotted,color=gray] (0) -- (2)  {};
\draw[-,densely dotted,color=gray] (0) -- (5)  {};
\draw[-,densely dotted,color=gray] (0) -- (6)  {};
\end{tikzpicture}}
\end{align}
and the non-local products are $\prod \color{blue}X$ along each $x$ and $y$ lines.

Thus, with measurement outcomes $X=(-1)^{s_v}$ we have the stabilizers
    \begin{align}
    (-1)^{s_v+1} \raisebox{-.5\height}{\begin{tikzpicture}[scale=0.5]
\coordinate (0) at (0,0) {};
\node[label=center:$\color{blue}Y$]  at (1,1) {};
\node[label=center:$\color{blue}Z$]  at (-1,1) {};
\node[label=center:$\color{blue}Z$]  at (1,-1) {};
\node[label=center:$\color{blue}Y$]  at (-1,-1) {};
\draw[-,densely dotted,color=gray] (0) -- (1)  {};
\draw[-,densely dotted,color=gray] (0) -- (2)  {};
\draw[-,densely dotted,color=gray] (0) -- (5)  {};
\draw[-,densely dotted,color=gray] (0) -- (6)  {};
\end{tikzpicture}}
\end{align}
which, up to single site rotations, are the stabilizers of the Wen plaquette model.

Although the Wen plaquette model is in the same topological phase as the toric code, it has the advantage of treating the $e$ and $m$ anyons on equal footing. In particular, it naturally has a dislocation defect which permutes the $e$ and $m$ anyons that encircles the defect \cite{KitaevKong12}. In other words, the dislocation hosts a Majorana zero mode. Consider the cluster state given by the graph
\begin{center}
    \includegraphics[scale=0.7]{Wenplaquettecluster.pdf}
\end{center}
which features a dislocation on the $B$ sublattice (dotted lines). Here the black lines connect $AB$ sites, while the red lines connect $AA$ sites. Performing measurements on the $A$ sublattice, the stabilizers for each plaquette on the blue sites are given by
\begin{center}
    \includegraphics[scale=0.7]{Wenplaquettestabilizers.pdf}
\end{center}

\subsection{Three-Fermion Walker-Wang model} \label{sec:3FWW}
It is argued that the three-Fermion Walker-Wang (3FWW) model\cite{BCFV14} cannot be created from a circuit; it requires a quantum cellular automaton\cite{HFH18}. Here, we argue that we can alternatively create this state by measuring an appropriate 3D cluster state. The preparation of such a state can prove useful for measurement-based quantum computation using such Walker-Wang models\cite{WW12} by effectively evolving the two-dimensional topological order on the boundary using measurements\cite{Raussendorf06,Roberts2020}. 

The 3FWW model can be obtained by gauging a $\ZZ_2^2$ 1-form SPT\cite{ChenTata21}. The response of this SPT to background $\ZZ_2$ 2-form gauge fields $B_1$ and $B_2$ is given by $B_1^2+ B_2^2 +B_1B_2$. The physical interpretation of the three terms is that they statistically transmute the anyons on the boundary to become that of fermions.

Conveniently, the above SPT phase is itself a cluster state. Therefore, combining with the cluster state that implements the KW duality on each sublattice, the cluster state we would like to perform measurements on to obtain the 3FWW is a $\ZZ_2^4$ 1-form SPT. Its response to background gauge fields $B_i$ for $i=1,2,3,4$ is $B_1^2+ B_2^2 +B_1B_2+B_1B_4 +B_2B_3$. The 3FWW is obtained by measuring the $1$ and $2$ sublattices.

Because it is a 1-form SPT, we define the cluster state on the edges of a cubic lattice, with four qubits placed per edge (i.e. 12 sites per unit cell). It is convenient to describe the cluster state using polynomials\cite{Haah13}, which denote the connectivity of this cluster state. 

As a stepping stone, we describe the stabilizers for the $B^2$ SPT,
\begin{align}
    \begin{pmatrix}
 0&(y+\bar z\bar x)(1+ z) &(z+\bar x\bar y)(1+ y) \\
 (x+\bar y\bar z)(1+z)&0 &(z+\bar x\bar y)(1+ x)  \\
 (x+\bar y\bar z)(1+y) &(y+\bar z\bar x)(1+ x)&0  \\
 \hline
1&0&0 \\
0&1&0 \\
0&0&1 \\
\end{pmatrix}
\end{align}
Here, each column denotes a stabilizer, and the top and bottom rows denote the positions of the Pauli-$Z$ and Pauli-$X$'s, respectively. Similarly, the $B_1B_2$ SPT (RBH cluster state) \cite{TsuiWen20,RaussendorfBravyiHarrington05,Yoshida2016} has stabilizers 
\begin{align}
    \begin{pmatrix}
0&0&0& 0&\bar x (1+\bar z) &\bar x (1+\bar y)     \\
0&0&0&   \bar y(1+\bar z)&0 &\bar y (1+\bar x)    \\
0&0&0&   \bar z (1+\bar y) &\bar z (1+\bar x)&0 \\
0&y (1+ z) &z (1+ y) &0&0&0\\
 x(1+z)&0 &z (1+ x) &0&0&0\\
 x (1+ y) &y (1+ x) &0&0&0&0\\
  \hline
1&0&0 & 0&0&0\\
0&1&0 & 0&0&0\\
0&0&1 & 0&0&0\\
0&0&0   & 1&0&0\\
0&0&0   & 0&1&0\\
0&0&0   & 0&0&1\\
\end{pmatrix}
\end{align}

Therefore, our desired cluster state is the +1 eigenstate of the stabilizers
\begin{align*}
\tiny
    \begin{pmatrix}
 0&(y+\bar z\bar x)(1+ z) &(z+\bar x\bar y)(1+ y)& 0&\bar x (1+\bar z) &\bar x (1+\bar y)  & 0&0&0&0&\bar x (1+\bar z) &\bar x (1+\bar y)  \\
 (x+\bar y\bar z)(1+z)&0 &(z+\bar x\bar y)(1+ x) &   \bar y(1+\bar z)&0 &\bar y (1+\bar x) & 0&0&0&  \bar y(1+\bar z)&0 &\bar y (1+\bar x) \\
 (x+\bar y\bar z)(1+y) &(y+\bar z\bar x)(1+ x)&0 &   \bar z (1+\bar y) &\bar z (1+\bar x)&0& 0&0&0&  \bar z (1+\bar y) &\bar z (1+\bar x)&0\\
0&y (1+ z) &z (1+ y)&  0&(y+\bar z\bar x)(1+ z) &(z+\bar x\bar y)(1+ y)                  & 0&\bar x (1+\bar z) &\bar x (1+\bar y)  & 0&0&0\\
 x(1+z)&0 &z (1+ x)&   (x+\bar y\bar z)(1+z)&0 &(z+\bar x\bar y)(1+ x)                   &   \bar y(1+\bar z)&0 &\bar y (1+\bar x) & 0&0&0\\
 x (1+ y) &y (1+ x)&0&  (x+\bar y\bar z)(1+y) &(y+\bar z\bar x)(1+ x)&0                  &   \bar z (1+\bar y) &\bar z (1+\bar x)&0& 0&0&0\\
0&0&0&0&y (1+ z) &z (1+ y)& 0&0&0& 0&0&0\\
0&0&0& x(1+z)&0 &z (1+ x) & 0&0&0& 0&0&0\\
0&0&0& x (1+ y) &y (1+ x)&0& 0&0&0& 0&0&0\\
 0&y (1+ z) &z (1+ y) &0&0&0&  0&0&0& 0&0&0\\
  x(1+z)&0 &z (1+ x)  &0&0&0&  0&0&0& 0&0&0\\
  x (1+ y) &y (1+ x)&0&0&0&0&   0&0&0& 0&0&0\\
   \hline
1&0&0 & 0&0&0&  0&0&0& 0&0&0\\
0&1&0 & 0&0&0&  0&0&0& 0&0&0\\
0&0&1 & 0&0&0&  0&0&0& 0&0&0\\
0&0&0   & 1&0&0& 0&0&0& 0&0&0\\
0&0&0   & 0&1&0& 0&0&0& 0&0&0\\
0&0&0   & 0&0&1& 0&0&0& 0&0&0\\
0&0&0& 0&0&0& 1&0&0 & 0&0&0  \\
0&0&0& 0&0&0& 0&1&0 & 0&0&0  \\
0&0&0& 0&0&0& 0&0&1 & 0&0&0  \\
0&0&0& 0&0&0& 0&0&0   & 1&0&0\\
0&0&0& 0&0&0& 0&0&0   & 0&1&0\\
0&0&0& 0&0&0& 0&0&0   & 0&0&1\\
\end{pmatrix}
\end{align*}

\section{Equality of long-range order for measurement outcomes in 1D \label{app:string}}

In Sec.~\ref{sec:intuitive} in the main text, we claimed that $\langle \psi_0|Z_{2m} Z_{2n}|\psi_0\rangle = - \langle \psi_1 | Z_{2m} Z_{2n} |\psi_1\rangle$. This claim can be derived using the notion of symmetry fractionalization \cite{Turner11}. In particular, since we have a gapped phase with $\prod_k X_{2k}$ symmetry, one can argue that $X_{2p} X_{2p+2} \cdots X_{2q} |\psi\rangle = U_L U_R |\psi\rangle$, where $U_{L,R}$ are exponentially localized near the end points of the original string operator. Equivalently, if we define $\tilde{\mathcal S}_{2p,2q} = X_{2p} X_{2p+2} \cdots X_{2q} $, then our state $|\psi\rangle$ is an eigenstate of $U_L \tilde{\mathcal S}_{2p,2q} U_R$. Since we are in a nontrivial SPT phase, $U_{L,R}$ will anticommute with the other $\mathbb Z_2$ symmetry $\prod_k X_{2k-1}$. Let us now revisit the situation studied in the main text, where $n$ and $m$ are separated far away from one another. Then we can choose $m \ll p \ll n \ll q$ such that $\mathcal S_{2m,2n} \times U_L \tilde{\mathcal S}_{2p,2q} U_R = - U_L \tilde{\mathcal S}_{2p,2q} U_R \times \mathcal S_{2m,2n}$. Note that since this operator leaves $|\psi\rangle$ invariant and toggles $\mathcal S_{2m,2n}$, we have that $U_L \tilde{\mathcal S}_{2p,2q} U_R |\psi_0 \rangle = e^{i\alpha} |\psi_1\rangle$. Thus,
\begin{equation}
\langle \psi_0|Z_{2m} Z_{2n}|\psi_0\rangle = e^{-i\alpha}\langle \psi_0| Z_{2m} Z_{2n}U_L \tilde{\mathcal S}_{2p,2q} U_R  |\psi_1\rangle
= -e^{-i\alpha}\langle \psi_0| U_L \tilde{\mathcal S}_{2p,2q} U_R Z_{2m} Z_{2n} |\psi_1\rangle = -\langle \psi_1| Z_{2m} Z_{2n} |\psi_1\rangle,
\end{equation}
where we used the fact that $Z_{2n}$ is odd under the spin-flip symmetry on the even sites, and since $m \ll p \ll n \ll q$ it is thus odd under $\tilde{\mathcal S}_{2p,2q}$ (whereas $Z_{2m}$ is not).

\section{Symmetry charge of push-through operator \label{app:symcharge}}

We consider the MPS-based arguments in Sec.~\ref{subsec:MPS}. There, we encountered the projective group relations $V_g V_{g'} = \omega(g,g') V_{gg'}$, which together with the Abelian symmetry relations $gg' = g'g$, imply that $V_g V_{g'} = \frac{\omega(g,g')}{\omega(g',g)} V_{g'} V_{g}$. Let us introduce $\alpha_{g,h} = \frac{\omega(g,g')}{\omega(g',g)} \in U(1)$ as a convenient shorthand notation. We now prove that Eq.~\eqref{eq:Og} implies that $U_h \mathcal O_g U_h^\dagger = \alpha_{g,h} \mathcal O_g$ or, equivalently, $U_h^\dagger \mathcal O_g U_h = \alpha_{g,h}^* \mathcal O_g$:

\begin{align*}
\raisebox{-0.9\height}{
    \begin{tikzpicture}[inner sep=1mm]
    \node[R] (R) at (0, 0) {$B$};
    \draw[-] (R) -- (0, -3.7);
    \node[U] at (0, -1) {$U_h$};
    \node[U] at (0, -2) {$\mathcal O_g$};
    \node[U] at (0, -3) {$U_h^\dagger$};
    \draw[-] (R) -- (0.7,0);
    \draw[-] (R) -- (-0.7,0);
    \end{tikzpicture}}
    &=
    \raisebox{-0.83\height}{
    \begin{tikzpicture}[inner sep=1mm]
    \node[R] (R) at (0, 0) {$B$};
    \draw[-] (R) -- (0, -2.7);
    \node[U] at (0, -1) {$\mathcal O_g$};
    \node[U] at (0, -2) {$U_h^\dagger$};
    \node[V] (Vr) at (1, 0) {$V_h^\dagger$};
    \node[V] (Vl) at (-1, 0) {$V_h$};
    \draw[-] (R) -- (Vr)-- (1.7,0);
    \draw[-] (R) -- (Vl)-- (-1.7,0);
    \end{tikzpicture}}
&&=
    \raisebox{-0.76\height}{
    \begin{tikzpicture}[inner sep=1mm]
    \node[R] (R) at (0, 0) {$B$};
    \draw[-] (R) -- (0, -1.7);
    \draw[-] (R) -- (Vr)-- (1.7,0);
    \draw[-] (R) -- (-2.7,0);
    \node[U] at (0, -1) {$U_h^\dagger$};
    \node[V] (Vr) at (1, 0) {$V_h^\dagger$};
    \node[V] (Vl) at (-1, 0) {$V_g$};
    \node[V] (Vl) at (-2, 0) {$V_h$};
    \end{tikzpicture}}
&&=\alpha_{g,h}^*
    \raisebox{-0.76\height}{
    \begin{tikzpicture}[inner sep=1mm]
    \node[R] (R) at (0, 0) {$B$};
    \draw[-] (R) -- (0, -1.7);
    \draw[-] (R) -- (Vr)-- (1.7,0);
    \draw[-] (R) -- (-2.7,0);
    \node[U] at (0, -1) {$U_h^\dagger$};
    \node[V] (Vr) at (1, 0) {$V_h^\dagger$};
    \node[V] (Vl) at (-1, 0) {$V_h$};
    \node[V] (Vl) at (-2, 0) {$V_g$};
    \end{tikzpicture}}\\
&=\alpha_{g,h}^*
    \raisebox{-0.84\height}{
    \begin{tikzpicture}[inner sep=1mm]
    \node[R] (R) at (0, 0) {$B$};
    \draw[-] (R) -- (0, -2.7);
    \draw[-] (R) -- (0.7,0);
    \draw[-] (R) -- (-1.7,0);
    \node[U] at (0, -1) {$U_h$};
    \node[U] at (0, -2) {$U_h^\dagger$};
    \node[V] (Vl) at (-1, 0) {$V_g$};
    \end{tikzpicture}}
&&=\alpha_{g,h}^*
    \raisebox{-0.56\height}{
    \begin{tikzpicture}[inner sep=1mm]
    \node[R] (R) at (0, 0) {$B$};
    \draw[-] (R) -- (0, -0.7);
    \draw[-] (R) -- (0.7,0);
    \draw[-] (R) -- (-1.7,0);
    \node[V] (Vl) at (-1, 0) {$V_g$};
    \end{tikzpicture}}
  && = \alpha_{g,h}^*
       \raisebox{-0.78\height}{
    \begin{tikzpicture}[inner sep=1mm]
    \node[R] (R) at (0, 0) {$B$};
    \draw[-] (R) -- (0, -1.7);
    \node[U] at (0, -1) {$\mathcal O_g$};
    \draw[-] (R) -- (0.7,0);
    \draw[-] (R) -- (-0.7,0);
    \end{tikzpicture}}
\end{align*}

Here, we used the fact that $V_h V_g = \alpha_{g,h}^* V_g V_h$.

Note that the above also carries through for the $A$ tensor, such that $U_{g'}^\dagger \mathcal O_g U_{g'} = \alpha_{g,g'} \mathcal O_g$ for any $g,g' \in G$. Since we are considering a mixed $G \times H$ SPT phase, the SPT must, by definition, be trivial if we restrict to just $G$ symmetry. This implies that $\alpha_{g,g'} = 1$. Hence, $U_{g'}^\dagger \mathcal O_g U_{g'} = \mathcal O_g$ for any $g,g' \in G$.

\end{document}